\documentclass[11pt]{article}
	\pdfoutput=1
\usepackage{amsmath,amssymb,booktabs,array}
\usepackage{float}
\usepackage{textcomp}
\usepackage{upgreek}
\usepackage{xstring}
 
\numberwithin{equation}{section}

\usepackage[T1]{fontenc}
\usepackage{listings}
\usepackage{datetime}
\usepackage{comment}					
\usepackage[usenames,dvipsnames]{xcolor}
\usepackage[setpagesize=false,pagebackref=false, 
linktocpage, bookmarksopen=true, colorlinks=true, 
linkcolor=Maroon,citecolor=Maroon,urlcolor=Maroon]{hyperref}

\usepackage[parsep]{collref}	
\newcommand{\rf}[1]{(\ref{#1})}
\newcommand{\ci}[1]{\cite{#1} {\small\ttfamily[\StrSubstitute{#1}{,}{,\allowbreak }]}}

\newcommand{\ZZ}{\mathbb Z}
\newcommand{\CC}{\mathbb C}
\newcommand{\RR}{\mathbb R}
\newcommand{\AdS}{{\rm AdS}}
\newcommand{\Hf}{{\rm H}^4}
\newcommand{\Str}{{\rm Str}}
\newcommand{\Tr}{{\rm Tr}}
\newcommand{\tr}{{\rm tr}}

\newcommand{\vol}{{\rm vol}}

\newcommand{\half}{\tfrac12}
\newcommand{\X}{{M_7}}
\newcommand{\cD}{\mathcal D}

\newcommand{\cK}{\mathcal K}
\newcommand{\cR}{\mathcal R}
\newcommand{\cN}{\mathcal N}
\newcommand{\cM}{\mathcal M}
\newcommand{\cC}{\mathcal C}
\newcommand{\cI}{\mathcal I}
\newcommand{\cJ}{\mathcal J}

\newcommand{\dd}{{\rm d}}
\newcommand{\hD}{\Delta}
\newcommand{\cm}[3]{[#1]^{(#2)}_{#3}}

\def\ba#1\ea{\begin{align}#1\end{align}}		        
\newcommand{\be}{\begin{equation}}
\newcommand{\ee}{\end{equation}}
\newcommand{\bea}{\begin{equation} \begin{aligned}} 
\newcommand{\eea}{\end{aligned} \end{equation}}

\newcommand{\la}{\label}

\def \lpp {\ell_{\rm p}}
\def \np {\newpage}

\def \ed {\np \small
\baselineskip 11pt
\bibliography{Ak}
\small
\bibliographystyle{JHEP-v2_9}
\end{document}
}

\def \iffa  {\iffalse}
\def \te {\textstyle}
\def\ov{\over}
\def \ci {\cite}
\def \foot {\footnote}

\def\la{\label}
\def\foot{\footnote}

\def \adsz  {AdS$_{4}\times S^{7}/\mathbb{Z}_{k}$ }
\def \OO {{\cal O}}
\def \no {\nonumber}
 \def \n {\nu} 
 
\newcommand{\al}[1]{\begin{align}{#1}\end{align}}
  \def \td {\tilde }

\def \rZ {{\rm Z}}   \def \rT {{\rm F}}

\begin{document}
\begin{titlepage}
\begin{tabbing}
\hspace*{10.5cm} \=  \kill 
\>  Imperial--TP--2026--SK--04 \\
\> 
\end{tabbing}


\vspace*{5mm}
\begin{center}
{\Large\sc  \bf
 1-loop partition function of  11d  supergravity on $\AdS_4\times S^7/\ZZ_k$:\\[6pt]
matching  $A(k)$ in  ABJM free energy   }
\vspace*{6mm}

Stefan A. Kurlyand\footnote{E-mail: \texttt{s.kurlyand23@imperial.ac.uk}     }     and   Arkady A. Tseytlin\footnote{Also at  ITMP of MSU  and Lebedev Inst. \ \  E-mail: \texttt{tseytlin@ic.ac.uk}}

\vspace*{4mm}
{\small {\it
Abdus Salam Centre for Theoretical Physics,\\ Imperial College London,  SW7 2AZ, U.K.}
}
\vspace*{0.7cm}
\end{center}
\begin{abstract}
The  function  $A(k)$ appearing  in the  large $N$ expansion of  free energy of ABJM theory
on $S^3$ is known exactly from localization. An earlier calculation
of the   1-loop partition function of 11d  supergravity
on ${\rm AdS}_4\times S^7/\mathbb Z_k$, based on the standard KK spectrum
on $S^7$, did not reproduce it. We identify a missing  1-loop  supergravity 
contribution as that of the  3-form field  fluctuations 
built from square-integrable harmonic
2-forms on Euclidean ${\rm AdS}_4$ tensored  with $S^7$ modes, which are  not captured by the 
KK spectrum. Adding  it to 
 the KK  tower contribution  expressed in terms of superconformal characters   and applying the 
 $\mathbb Z_k$ projection, the total  expression  simplifies,
and its evaluation in a  zeta-function prescription
reproduces the  expression for  $A(k)$. 
With the zero-mode measure normalized using the M2-brane charge quantization, the logarithmic and
constant terms of the localization result are reproduced as well.
We also apply a  similar approach  to the 3d ADHM  theory whose   M-theory dual  is   the orbifold 
${\rm AdS}_4\times S^7/\mathbb Z_p$, 
which has   singularity along the  fixed locus  $\AdS_4\times S^3$, supporting  a 7d  $SU(p)$ vector multiplet.
Adding the
localized KK modes of this multiplet to the $\ZZ_p$-projected supergravity spectrum gives a 1-loop free
energy that reproduces the ADHM localization counterpart of $A(k)$.
\end{abstract}
\end{titlepage}


{\small
\makeatletter
\newcommand*{\toccontents}{\@starttoc{toc}}
\makeatother
\toccontents
}

\np 

\def \CP {{\mathbb {CP}}}
\def \rms  {{\rm s}}  \def \rmv {{\rm v}} \def \F {{\rm F}}
\def \adssz  {$\AdS_4 \times S^7/\ZZ_k$ }
\def \te {\textstyle} 
\setcounter{footnote}{0}
\def \ha {\half}  \def \rH {\mathcal H}  \def \del {\partial}
\def \N {{\cal N}}  \def \rN  {{\rm N}}
\def \cF {{\cal F}}\def \rF{{\rm F}}
\def \rr {{\rm r}}
\def \rbr {{\rm b}}
\def \rZ {{\rm Z}}   \def \rT {{\rm F}}
\def \fF {{\rm f}}
\def \n {\nu}
\def \cN {{\cal N}}
\newcommand{\lab}[1]{{\small\ttfamily[#1]}}

\def \sss  {{\rm s}}

\section{Introduction and summary}\la{1}

{\bf  1.} \ 
11d  supergravity \ci{Cremmer:1978km}   being an odd-dimensional  covariant theory 
has no 1-loop logarithmic   UV divergences  (on a manifold without a boundary). 
Its  leading  power   $\Lambda^{11},\Lambda^9,\Lambda^7$ and $\Lambda^5$ divergences also cancel \ci{Fradkin:1982kf}, so 
the first  local divergence is cubic, $\Lambda^3\int\dd^{11}x\sqrt g\,(R^4+ ...)$.\foot{This  cubic divergence  regularized  by Planck scale   plays a role in the   1-loop four-graviton amplitude of 11d 
supergravity compactified on a circle \ci{Green:1997as,Russo:1997mk}. It is also related to quadratic $R^4$ 1-loop divergence 
in type II   10d supergravity regularized  by string scale \ci{Metsaev:1987ju}, see Appendix \ref{241}  below.}
Defined  using $\zeta$-function regularization the 1-loop partition function is thus finite. 

The field  equations of 11d supergravity 
admit the  maximally supersymmetric  $\AdS_4\times S^7$  solution \ci{Freund:1980xh}.
The Kaluza--Klein 
spectrum on $S^7$ can be   organized in $OSp(8|4)$ multiplets
\ci{Biran:1983iy,Casher:1984ym,Gunaydin:1985tc}   and one  can use this fact  to  study  1-loop quantum corrections on this background. 
It was found in  \ci{Gibbons:1984dg,Inami:1984vp}  that 
the Casimir energy (1-loop global $\AdS_4$ energy)  
 cancels  at every KK level.  This does not   imply that 
the Euclidean 1-loop   partition function  $Z_1$ on $\Hf\times S^7$   
 should be  equal to 1  (see  Appendix \ref{247}). 
 $Z_1$  is  a constant  that may depend  on normalization of the path integral   measure
 (see Section \ref{155}). 

 The question   about the value  of the  1-loop partition function becomes  more meaningful  if there is an additional parameter, like in the case of the   orbifold $\AdS_4\times S^7/\ZZ_k$   ($k=1,2,3,...$).  
 M-theory on $\AdS_4\times S^7/\ZZ_k$  was   suggested to  be dual to 
 a particular $\cN=6$  (for $k>2$)  supersymmetric  3d $U(N)_k \times U(N)_{-k}$ Chern-Simons-matter theory -- the ABJM theory \ci{Aharony:2008ug}. 
 Its free energy
$\rF=-\log\rZ$ on $S^3$  can be  
computed as  a function of $N$ at fixed Chern--Simons level $k$ using localization  and the Fermi-gas representation \ci{Kapustin:2009kz,Drukker:2010nc,Fuji:2011km,Marino:2011eh}.
At fixed $k$ and large $N$, this gives,    up to exponentially small corrections 
\al{
&\rZ(N,k)=\rZ_{\rm pert}(N,k)\Big[1+\OO\big(e^{-c\sqrt N}\big)\Big]\,,\qquad\qquad \ \ \ 
\rZ_{\rm pert}=C^{-1/3}\,e^{A(k)}\,{\rm Ai}\big[C^{-1/3}\hat N \big]\,,
\la{2}\\ 
&\qquad \hat N = N-B \ , \qquad \ \ \  \te C=\frac{2}{\pi^2k}\,,\qquad B=\frac{k}{24}+\frac{1}{3k}\,,\qquad c=\sqrt{8\pi^2\ov k}\,.
\la{3}
}
The constant $A(k)$ was determined  to be  \ci{Hanada:2012si} (see also
\ci{Hatsuda:2014vsa}) 
\begin{equation}
A(k)=\tfrac{2\zeta(3)}{\pi^2k}\big(1-\tfrac{k^3}{16}\big)
+\tfrac{k^2}{\pi^2}\int_0^\infty\dd x\,\frac{x\,\log(1-e^{-2x})}{e^{kx}-1}\,,
\la{4}
\ee
with,  e.g., 
$
A(1)=\frac{\log2}{4}-\frac{\zeta(3)}{8\pi^2}\,,\  A(2)=-\frac{\zeta(3)}{2\pi^2}.$
At large $k$ it has the expansion 
\begin{equation}
A(k)=-\tfrac{\zeta(3)}{8\pi^2}k^2-\tfrac16\log\tfrac{k}{4\pi}+2\zeta'(-1)+\sum_{g\ge1}\tfrac{1}{k^{2g}}\mathcal B_g\,, 
\la{6}
\end{equation}
where $\mathcal B _g$   are  rational multiples of $\pi^{2g}$. 
  $Z_{\rm pert}$ in \rf{2}   depends on $N$ only through $\hat N\equiv N-B$  so that  at large $\hat N$
  \al{\rF=-\log \rZ_{\rm pert}  &=
  \tfrac23\,C^{-1/2}\,\hat N^{3/2}  + \tfrac14\log\big( 16 \pi^2  C\hat N \big)-  A(k)+\OO\big(C^{1/2}\hat N^{-3/2}\big)\no \\
 &
=   \tfrac{\pi\sqrt{2k}}{3}\,\hat N^{3/2}+\tfrac14\log\hat N - A(k)
- \tfrac14\log \tfrac{k}{32}  
 +\OO\big(\hat N^{-3/2}\big)\,.
\la{7}
}
On the  dual  M-theory  side  $k$   is the parameter in the metric of \adssz  and 
the  counterpart of $N$  is the radius $L$ of $S^7$ (the $\AdS_4$ radius is $L/2$) in
units of the 11d Planck length $\lpp $,   with the  leading-order form 
$
L^6=32\pi^2kN\,\lpp ^6  + \OO(N^0)$.
The $C_3\wedge  R^4 $ term of the M-theory effective action shifts the M2-brane charge to
$\rN=N-\frac1{24}(k-k^{-1})$ \ci{Bergman:2009zh} suggesting  to define   
 the dimensionless radius ${\rm L}$   as  
\begin{equation}
{\rm L}^6\equiv\frac{L^6}{\lpp ^6}=32\pi^2k\,\rN\,,\qquad \qquad \ \ \ \rN=N-\tfrac{1}{24}\big(k-k^{-1} \big)\,.
\la{9}
\end{equation}
In terms of ${\rm L}$ the  large $N$  expansion of   \rf{7} then  takes the  form \ci{Beccaria:2023hhi}
\begin{align}
\rF&=\te {\rm a}_0\,\frac{{\rm L}^9}{k}+  {\rm a}_1\,\frac{{\rm L}^3}{k} + \cF_1 
+{\rm a}_2\,\frac{{\rm L}^{-3}}{k}+{\rm a}_3\,{\rm L}^{-6}+O\big({\rm L}^{-9}\big)\,,\la{01}\\
\cF_1 & = \te \half\log\frac{{\rm L}^3}{\pi k}-A(k) =  \tfrac32\log(\lpp^{-1} { L})-\tfrac12\log(\pi k)  -A(k)  \ , \la{00}\\
\te {\rm a}_0 &\te =\frac{1}{384\pi^2}\,,\qquad {\rm a}_1 =-\frac{3}{64}\,,\qquad {\rm a}_2=\frac{9\pi^2}{64}\,,\qquad {\rm a}_3=-3\pi^2\,.
\la{10}
\end{align}
Eq.\rf{01}  looks like an expansion in terms of the  large M2-brane tension ${\rm T}_2= {1\ov 4 \pi^2} {\rm L}^3$  \ci{Beccaria:2023hhi}.

As discussed in \ci{Beccaria:2023hhi}, the simple $1/k$ dependence of the ${\rm L}^9$, ${\rm L}^3$ and
${\rm L}^{-3}$ terms suggests that they originate  from local terms of the M-theory effective action, which on
the homogeneous background depend on $k$ only through the volume ${\rm vol}(S^7/\ZZ_k) = {\pi^4\ov 3k} $. The ${\rm L}^9$
term matches  the  value of the classical 11d supergravity action on $\AdS_4 \times S^7/\ZZ_k$. 
The ${\rm L}^3$ term should come from the 1-loop $R^4+...$
term in the  M-theory effective action (evaluated on $\AdS_4 \times S^7/\ZZ_k$).  This 
term  is cubically divergent in 11d supergravity    but is finite   if  one uses  a  cutoff
$\Lambda\sim\lpp ^{-1}$ \ci{Green:1997as,Russo:1997mk}. 

It is natural to    conjecture that the  terms in $\cF_1$   in \rf{00} 
 which are 
of zeroth order in the M2-brane tension  should come from the 1-loop partition
function of 11d supergravity modes or point-like M2  brane  states.

The logarithmic term in $\cF_1$ in  \rf{00}  was indeed  reproduced in \ci{Bhattacharyya:2012ye} from the 1-loop partition
function of 11d supergravity  as the 0-mode   contribution (see also \ci{Arrighi:2026xmn}).
 In  the present odd-dimensional  case  there is   no  bulk log UV divergence, but a $\log\frac{ L}{\lpp }$  term
 may  come from  a  normalization of the 0-modes in the path integral. The scale $\lpp $ enters here
 through the coupling   $2\kappa_{11}^2=(2\pi)^8\lpp ^9$  in  the 11d  supergravity action, 
  which is the only
 dimensionful parameter of the theory. For the  canonically normalized fluctuations,  the measure, and
   the volume of the zero-mode space,  may depend  on $L$ and $\kappa_{11}$ only through $L/\lpp $.

The  relevant zero modes  turned out  to be  those of the 2-form ghost of the  3-form  supergravity field.
 They are built  out of  the square-integrable harmonic 2-forms on Euclidean $\AdS_4\equiv \Hf$
 tensored with the constant scalar mode on $S^7/\ZZ_k$.  Their number $n_2$  is IR divergent,  and  after a  natural 
  regularization  (same as for  $\vol(\Hf)= {4\pi^2\ov  3} $) 
  it   is  equal to  the renormalized Euler characteristic of
   $\Hf$, i.e.  $n_2=\chi(\Hf)=1$.

The problem of  computation of  the non-zero-mode part of the 11d supergravity  free energy on \adssz was addressed  in
 \ci{Liu:2016dau}, using the KK spectrum on $S^7/\ZZ_k$, i.e. the $\ZZ_k$-invariant part of the  KK multiplet 
 spectrum on $S^7$ in 
\ci{Biran:1983iy,Nilsson:1984bj,Gunaydin:1985tc}. 
The resulting  expression  was found to have   the large-$k$ behaviour
\be 
F_{\rm LZ} =\te \frac{25\zeta(7)}{1024\pi^6}k^6-\frac{3\zeta(5)}{128\pi^4}k^4
-\frac{\zeta(3)}{18\pi^2}k^2-\frac{389}{945}\log k+\OO(1)\,,
\la{13}
\end{equation}
which  is in obvious disagreement with \rf{6}. 
In \ci{Liu:2016dau} the mismatch was attributed to possible contributions of  other M-theory  states  beyond 
supergravity  or  an ambiguity in  regularization of the sum over KK levels. 
It was also mentioned  there that the massive  KK 
spectrum on $\AdS_4$   does not contain modes associated with the harmonic 2-forms on $\AdS_4$, but this remark was not elaborated on.

In  \ci{Beccaria:2023hhi} the large-$k$ expansion of $A(k)$ was compared with that of  the 1-loop
four-graviton amplitude  in  11d supergravity on a circle. It was suggested  that $A(k)$ may   be put in a 
 form  resembling a 1-loop proper-time integral with a $1/\sinh^2  $ measure
 as in  \ci{Gopakumar:1998jq,Dedushenko:2014nya}. Starting from the integral
representation of \ci{Hanada:2012si} one finds    that formally (modulo regularization)    \ci{Beccaria:2023hhi}
\begin{equation}
A(k)=\int_0^\infty\frac{\dd t'}{t'}\,\frac{1}{e^{kt'}-1}\,\frac{1}{\sinh^2t'}
=\sum_{n=1}^\infty\int_0^\infty\frac{\dd t'}{t'}\,\frac{e^{-knt'}}{\sinh^2t'}\,,
\la{15}
\end{equation}
which is suggestive of   contributions of  a tower of massive particles  propagating in the loop. 
As we shall see below,   $t'$  in this integral representation \rf{15} 
 should  not actually be interpreted as a proper-time parameter  so this ``particle in the loop'' analogy 
 is not a  useful  analogy. 

\

{\bf 2.} \
 The   aim of the present paper is   to  point out that  the computation of \ci{Liu:2016dau}
  missed one crucial contribution --  that of    the ``harmonic  sector''  of the 3-form field fluctuations. 
  Once it is included,   the growing terms in \rf{13}  are cancelled 
   and  we reproduce  exactly the   $-A(k)$ term  in \rf{4},\rf{6}   as the 1-loop 
    supergravity   free energy. 
   
   To compute the 1-loop  partition  function of 11d   supergravity on  a  given  background  one is to  expand  the action to quadratic  fluctuation terms, fix  gauges,  introduce ghosts, etc. The result is then expressed   in  terms of determinants of differential operators, which, in the case of a symmetric   space like $\AdS_4 \times S^7/\ZZ_k$, 
    are  in principle straightforward to compute.
   In practice, due to large number of dimensions, the issues   like 
   finding  the spectrum   and regularization  of the  sum over  states 
     complicate the task.\foot{In addition,   some modes   may mix  and one may  need to deal with higher-derivative operators  (cf. \ci{Arrighi:2026xmn}). 
      Here,   in fact,   these operators   can be  factorized into 2nd  order ones modulo 
 a  multiplicative anomaly  \ci{Kontsevich:1994nc,Elizalde:1997nd,Dowker:1998tb}. 
In this work, we regularise the spectral determinants using the character zeta function (see  Section~\ref{155}), and the multiplicative anomaly may appear as the difference between this and the spectral zeta function regularisation \cite{Sun:2020ame}. In the character zeta function calculation, the anomaly terms are controlled by the coefficients in the Laurent expansion of the character integrand. These cancel separately for each supermultiplet in the KK mode
expansion (see footnote  \ref{fa}).}

  
   The  spectrum of  quadratic fluctuations near $\AdS_4 \times S^7$   can be expressed  in terms of a tower of  $OSp(8|4)$
    KK multiplets \ci{Biran:1983iy,Nilsson:1984bj,Gunaydin:1985tc}, while the $\ZZ_k$ factorization  can be implemented 
     by  summing  heat kernels over $k$ images   of the   $U(1)_m\subset SO(8)$  (Hopf  fiber) action  (see, e.g., \ci{David:2009xg}). 
   The key technical  steps of our computation will be: 
    
  (i) Express the  1-loop partition function   on $\AdS_4 \times S^7$  as  a sum
   of  KK  characters   with $U(1)_m$ 
   fugacity. 
   
      (ii)  Add    the contribution  of the   harmonic 2-form   on $\AdS_4$  written in the same character form. 
   
   (iii)  Implement the $\ZZ_k$ quotient on the  total expression.
   
   (iv)  Regularize the resulting spectral integral  using the $\zeta$-function prescription. 
   
   \noindent
  The 1-loop  supergravity   free energy will   be denoted as $F_1$:
  \begin{equation}
F_1=-\log Z_1=\ha \sum_{i}(-1)^{{\rm f}_i} \log{\det}\,(\mu^{-2}  \mathcal O_i)\ , 
\la{16}
\end{equation}
where $\OO_i$  are  the gauge-fixed second-order fluctuation operators of all fields and ghosts. We will use   prime to indicate removal of 0-modes.
Since in 11d   the top Seeley coefficient ${\rm B}_{11}$ is zero, $F_1$ in \rf{16} does not  actually depend on normalization mass  scale  $\mu$. 
At the same time, since ${\rm B}_{11} = \zeta(0) + n_0$  where $n_0$ is the total (graded) 
 number of 0-modes,  the non-zero-mode  part 
 $F_1' $ will depend on  $\mu$   via the term $n_0 \log (L\mu)$.\foot{As discussed    in \ci{Bhattacharyya:2012ye}   and below, 
  the only zero modes are 
 the   2-form ghost modes of harmonic type  $\omega_2 \otimes1$, entering with ghost weight $+1$ in the $(\det\Delta_2)^{+1}$ factor. The graded count is $n_0=-2n_2=-2$, which gives the term $-2\log(\frac12 \mu L)$ in $F_1'$.}

As in \ci{Liu:2016dau},  one may   evaluate \rf{16} by summing over the KK multiplets.
What the KK multiplets do not contain are special  ``harmonic''  
normalizable  Euclidean modes   of  3-form  fluctuation field $\td C_3$:
\begin{equation}
\td C_3=\omega_2\wedge v_1\,,\qquad\qquad  \omega_2\in\mathcal H^2(\Hf)\,,\quad \ \ \ \ v_1\in\Omega^1(S^7/\ZZ_k)\,.
\la{18}
\end{equation}
Here  $\omega_2$ is a square-integrable harmonic 2-form on $\Hf$ and $v_1$ is  a 1-form on the internal space   $S^7/\ZZ_k$.
The determinants of the  KK  tower of  operators on $\Hf$  contain   contributions
of  only the continuous spectrum on $\Hf$, while the harmonic 2-forms belong to the discrete series representations 
of $SO(4,1)$ \ci{Camp}. 
The space of such  forms is infinite-dimensional \ci{Dodziuk:1979},
but its regularized dimension is equal to one \ci{Bhattacharyya:2012ye}.

 We will  show that this  ``harmonic''  3-form fluctuation sector
decouples from the rest of the quadratic  fluctuations  
and its contribution $Z_H$   to the 1-loop 11d supergravity 
 partition function  is  given just  by the  free abelian gauge field  partition function on $S^7/\ZZ_k$, i.e.  
 by the  following combination of determinants of   vector and scalar Laplacians
\be 
\F_H=-\log Z_H= \half\log{\det}'_\X\Delta_{1}-\log{\det}'_\X\Delta_{0}\,, \qquad \ \ \ \ \ \qquad M_7\equiv S^7/\ZZ_k \ . 
\la{19} 
\end{equation}

To  compute the  free energy  contribution of the KK modes 
 we  will  use  the character representation of 1-loop determinants on $\Hf$
\ci{Bae:2016rgm,Pang:2016ofv,Basile:2018zoy,Sun:2020ame}. Consider the $\AdS_4$  field dual to the conformal
multiplet $[j]_E$ of spin $j$ and dimension $E$. The  corresponding 
 second-order operator $\cD_{E,j}$  has 
its eigenvalues as  $\lambda^2+(E-\frac32)^2$, with $\lambda$ being the $\Hf$ spectral parameter. Its contribution to
the 1-loop free energy can be written as an integral over a variable $\tau$, conjugate to the global
$\AdS_4$ energy $E$  
\al{
&F_{E,j}=\pm \half\log\det\cD_{E,j}
=-\half\,\!\int_0^\infty\frac{\dd\tau}{\tau}\,(-1)^{2j}\kappa_{B/F}(q)\,
\Big( 2+ \tfrac{(1-q)^2}{8q}    \big[ \delta -\,  (x \del_x)^2 \big]\Big)\,\cC_{E,j}(q,x)\Big|_{x=1}\ , 
\la{14}\\
&\qquad \qquad \qquad \te  \kappa_B=\frac{1+q}{1-q}\ , \qquad \ \  \kappa_F=\frac{2q^{1/2}}{1-q}\ , \qquad \qquad \qquad q=e^{-\tau}
   \ . \la{333}
  }
Here the upper sign   refers to bosons ($j\in\ZZ$), and the lower sign  to fermions ($j\in\ZZ+\tfrac12$). 
The conformal   character  is given by 
$\cC_{E,j}(q,x)={\rm tr}\big(q^D x^{2J_3}\big)$  where  $D$  is dilatation operator (with  eigenvalue $E$)
and $J_3$ is the    $SO(3)$  spin  projection generator (with the highest eigenvalue $j$).   $\delta$ equals one for fermions and zero for bosons.\foot{No prime is needed on $F_{E,j}$   in \rf{14}: the operators $\cD_{E,j}$ have a purely continuous spectrum on $\Hf$ and no
zero modes. The only zero modes are those of the harmonic sector \rf{18}, which is not part of the KK
spectrum.}

 Since the operator $ (x \del_x)^2 $  in \rf{14}  acts linearly on the characters, 
summing  over all fields of the KK spectrum we get  the expression   written in terms of  the same operator acting on the
superconformal   $OSp(8|4)$ characters
of the KK supermultiplets $\cM_n$, for which closed formulae are known \ci{Dolan:2008vc}. 
Introducing also the $U(1)_m$  fugacity $v$, 
i.e.  considering $\chi_{_\cM}(q,x,v)=\tr_{_\cM}\big(q^{D}\,x^{2J_3}\,v^{R_m}\big)$, 
  allows  then to 
implement the $\ZZ_k$ projection  in the integrand of the $\tau$-integral  \rf{14}
 representing the  1-loop   free energy in \rf{16}.
 
The non-zero-mode part of  free energy is then the sum over the KK spectrum weighted by the $so(8)$ multiplicities,
plus the harmonic sector contribution,
\al{
&F_1'=\frac1k\sum_{p=0}^{k-1}\ \sum_{[j]^{(a)}_E}\chi^{(a)}(\omega^p)\,F_{E,j}+n_2\,\F_H\,,\qquad \ \ \ \ \  \  \ n_2=1\ , \la{775}\\
&F'_1(k) = -\ha \int^\infty_0 { \dd \tau\ov \tau} \,  \cI_k(\tau)  \ , \qquad \qquad      \cI_k(\tau)\equiv\frac1k\sum_{p=0}^{k-1}\cI(q,v)\Big|_{q=e^{-\tau},\ v=\omega^p}\,,\qquad
\omega=e^{2\pi i/k}\,, 
\la{1577}
}
where $\chi^{(a)}(v)$ is the $so(8)$ character of the representation $(a)$ 
and the average over $v=\omega^p$ implements the $\ZZ_k$ projection. 

Adding   the  contribution  \rf{19}   of the harmonic  sector, 
the  integrand in \rf{1577} simplifies   to ($k\ge2$)
\begin{equation}
\cI_k(\tau)=\cC_k (\tau) +\cR(\tau)\,,\qquad\qquad   \cC_k =\frac{\coth(\tfrac{1}{4}k\tau)}{\sinh^2(\ha \tau)} \ , \qquad \ \ \ 
\cR(\tau)=e^{-2\tau}+e^{-3\tau}-{4}({e^\tau-1})^{-1}\,,
\la{20}
\end{equation}
and the non-zero-mode part of the 1-loop  free energy in \rf{16} is then found to be\foot{The last term restores    dependence on
the normalization  mass  $\mu$   and $\AdS_4$ scale $\ha L$; $\mu$ is an arbitrary normalization scale of the determinants, which drops out of the total $F_1$, see below.}
\begin{equation}
F_1'=-\half\!\int_0^\infty\frac{\dd\tau}{\tau}\,\ \frac{\coth(\tfrac{1}{4}k\tau)}{\sinh^2(\ha \tau)}  -\half\!\int_0^\infty\frac{\dd\tau}{\tau}\,
\cR(\tau)
-2\log(\ha \mu L)\,.
\la{21}
\end{equation}
The $\tau$-integral in \rf{21}    is  to be defined   using the zeta-function  regularization (see 
 \rf{171}).
  The first term in \rf{21}  then gives exactly $-A(k)$  in \rf{00} (cf.  \rf{4} and  \rf{15}).\foot{Comparing \rf{15} to the 
  first term in \rf{21}, using
$\frac{\coth(\frac{1}{4} k\tau)}{\sinh^2(\frac{1}{2}\tau)}=\frac{2}{(e^{kt'}-1)\sinh^2t'}+\frac{1}{\sinh^2t'}$
with $t'=\frac12\tau$, we see that the two integrands differ by a $k$-independent term. In the
$\zeta$-function regularization that term does not vanish: it contributes $2\zeta'(-1)+\frac16\log2$. 
Thus \rf{21} reproduces $A(k)$ exactly, whereas \rf{15} holds only in a regularization in which the singular terms at $t'\to0$ are
subtracted together with this constant. Note  that the  variable $t'$ in \rf{15}   being  related to $\tau$
is thus not a proper time: $\tau$ is
conjugate to the global $\AdS_4$ energy and it  arises from the proper-time representation \rf{65} after  the  integration 
 over the  $\AdS_4$ spectral parameter (see \rf{82},\rf{84}).}  
  
     The  $k$-independent  second term in \rf{21} gives
$\log\frac{\sqrt6}{2\pi}$, so that\foot{The  expression \rf{22}  is valid   also for $k=1$ 
after  the free $\cN=8$ singleton  multiplet contribution  is included.}  
\begin{equation}
F_1'=-A(k)+\log\tfrac{\sqrt6}{2\pi}-2\log(\ha \mu L)  \ . 
\la{22}
\end{equation}
 The $\mu$-dependence is cancelled  in $Z_1$   against   the integral over
the zero modes.  That integral also brings a dependence on
the $\AdS_4$  radius  through the normalization of the zero modes
(involving the  volume  of $S^7/\ZZ_k$)
 and on $\kappa_{11}$, i.e. on $\lpp $  (which is here the  normalization scale and not a UV cutoff). 
 
 The total 1-loop  free energy  given by the sum of $F_1'$ in \rf{21} 
 and the zero-mode contribution $F_{\rm zm}$ is  independent of $\mu$ and depends on \adssz    scale 
 $L$ only through $L/\lpp $,  i.e. 
\begin{equation}
F_{1}=F_1'+ F_{\rm zm}=\tfrac32\log(\lpp^{-1} {L})  -A(k)-\half\log k+c_0\,,
\la{23}
\end{equation} 
where $c_0$ is a $k$-independent constant, which depends on the normalization of the zero-mode measure.
We fix this normalization as in the case of the Maxwell field on a sphere \ci{Anninos:2020hfj,Law:2020cpj},
using the compactness of the group of harmonic gauge transformations of the 3-form implied by the
quantization of the M2-brane charge. This gives $c_0=-\tfrac12\log\pi$ (see the end of Section \ref{155}),
so that \rf{23} matches the localization expression in \rf{00} including the constant term.

\iffa where $c_0$ is a 
 $k$-independent constant. 
  This  matches the  localization expression in \rf{00} 
 under the single condition $c_0=-\tfrac12\log\pi$ (see  \rf{201},\rf{202})
   on the  normalization of the  measure in 11d supergravity path integral.
\fi

We   also  demonstrate   (in Appendix \ref{258}) a similar  matching   between the  1-loop 11d 
  supergravity   result   and  the localization free energy  in the  case of 3d  
  ADHM  gauge theory whose M-theory dual is a  different (singular)  orbifold 
$\AdS_4\times S^7/\ZZ_p$.\foot{The ADHM
 theory is (an unfortunate name for) the 3d $\cN=4$ $U(N)$ gauge
theory with one adjoint and $N_f=p$ fundamental hypermultiplets. 
Its Higgs branch is the moduli space of
$N$ instantons of four-dimensional $U(p)$ YM theory, described by the construction of
\ci{Atiyah:1978ri}. The theory is represented  by  $N$ M2-branes probing $\CC^2\times\CC^2/\ZZ_{p}$, with M-theory
dual $\AdS_4\times S^7/\ZZ_p$, where $\ZZ_{p}$ acts on two of the four complex coordinates and fixes a
3-sphere for $p>1$ \ci{Imamura:2010ei,Hatsuda:2014vsa}. 
This differs from the ABJM quotient
 $\ZZ_k$, which  acts freely on $S^7$  and preserves $\cN=6$
rather than $\cN=4$. For $p=1$ there is no orbifold, and the ADHM theory flows to the same $\cN=8$
fixed point as ABJM at $k=1$ \ci{Crew:2020psc}.}  
For $p>1$, the quotient has
a transverse $A_{p-1}$ singularity along $\AdS_4\times S^3$,
supporting a 7d  $SU(p)$ vector multiplet.
Including its localized KK modes in addition to the projected
supergravity spectrum reproduces the ADHM counterpart of $A(k)$.

\

{\bf 3.} \
Let us now   comment on some recent related work. 

$\bullet$ \ 
It was proposed in \ci{Gautason:2025plx} (see also  \ci{vanMuiden:2026nsp}) that the M-theory
(supergravity) partition function on $\AdS_4\times S^7/\ZZ_k$  defined  with fixed boundary value of $C_3$ 
should be matched to its ABJM localization
counterpart not in the canonical ensemble with fixed $N$, as in \rf{2}, but in the grand-canonical
ensemble in which the chemical potential $\upmu$ conjugate to $N$ is fixed.  Let us  call it the
$\upmu$-ensemble (or fixed-$C_3$ ensemble).  $\rZ_{\rm pert}(N,k)$
in \rf{2} is the inverse Laplace transform of the grand-canonical partition function $e^{J_{\rm pert}(\upmu)}$,
where $J_{\rm pert}(\upmu,k)=\frac13C\upmu^3+B\upmu+A(k)$ is the perturbative grand potential. 

Further tests of this proposal were given in
\ci{Bobev:2026gir}, where 
the leading and subleading terms in the grand potential,
including the coefficient $B$,  were   computed for several supersymmetric
backgrounds using the two- and higher-derivative
supergravity actions. This work  also analyzed the logarithmic
contribution and its dependence on the choice of ensemble.

 The grand potential
$J_{\rm pert}$ contains no logarithms, so in this ensemble both the logarithmic term $\frac32\log\frac{L}{\ell_p}$
and the term $-\frac12\log k$ in \rf{23} should be absent, and the 1-loop free energy should be given by $-A(k)$.

The zero-mode part in \rf{23}, $\frac32\log{L\ov \lpp}-\tfrac12\log k-\half\log\pi$, is equal to
$\frac14\log N-\frac14\log k+\frac54\log2$. This is exactly the Gaussian factor
$\tfrac12\log\big[2\pi J''_{\rm pert}(\upmu_s,k)\big]$ of the Laplace transform from $e^{J_{\rm pert}}$ to
$\rZ_{\rm pert}$, evaluated at its saddle point $\upmu_s$.
This factor combines the prefactor $C^{-1/3}$ in \rf{2} with the $C$-dependence of the large-argument form
of the Airy function, giving $\frac14\log\hat N+\frac14\log C+\frac12\log(4\pi)$.
 The non-zero modes give the constant $-A(k)$ of the
grand potential. Thus the zero modes account for the change from the ensemble with fixed $\upmu$
to the one with fixed $N$.

In \ci{Bobev:2026gir} the $\upmu$-ensemble was implemented on the gravity side by relative boundary
conditions\foot{The relative boundary conditions for a $p$-form $\omega$ mean the Dirichlet-type conditions
$\iota^*\omega=0$, where $\iota^*$ is the pull-back to the boundary, i.e. the tangential components of
$\omega$ are set to vanish. For the fluctuations of $C_3$ this keeps the boundary value of $C_3$ fixed, and
this boundary value is the chemical potential for the M2-brane charge $N$, which is the flux of $\ast F_4$
through $S^7/\ZZ_k$. Fixing the boundary value leaves the charge free as in the grand-canonical ensemble,
while fixing the flux gives the canonical ensemble. The two are related by a Legendre transform of the
action, i.e. a Laplace transform of the partition function.
}
 on the 3-form and on its 2-form ghost.  These conditions are meant to apply only to the part of $C_3$ and of its ghosts
that is constant on the internal space, i.e. to the boundary holonomy of $C_3$ averaged over $S^7/\ZZ_k$,
which is the variable conjugate to $N$.\foot{We thank  J. van Muiden  for this clarification.}
 With these conditions the harmonic ghost zero modes
$\omega\otimes1$, which have non-vanishing pull-back to the boundary of $\Hf$, are absent, and with them the
logarithmic term, in agreement with the absence of logarithms in grand-canonical $J_{\rm pert}$.

The modes with non-zero eigenvalue of the internal Laplacian are not affected by the choice of ensemble.
Since $\upmu$ is conjugate to the total charge $N$, fixing it constrains only the internal average of the
holonomy. This is a choice of which of the  global quantities -- the flux or the averaged holonomy --  is held
fixed, and not a local boundary condition on the 11d field. For the massive KK fields the condition is the
absence of sources, i.e. square integrability, in both ensembles. It keeps the harmonic-sector modes
$\omega\wedge v_1$ in \rf{18} and the corresponding ghost modes built from non-constant scalar harmonics,
which give the contribution \rf{19}. Had the relative conditions been imposed on all KK modes, these modes
would be excluded, since $\iota^*(\omega\wedge v_1)=(\iota^*\omega)\wedge v_1\ne0$.

Our computation  is a formal 1-loop computation in 11d supergravity in which all fluctuations are assumed to
be square-integrable (Section \ref{44}), and as such it does not refer to a choice of ensemble. With this
assumption made also in the sector constant on the internal space, the ghost zero modes are present
\ci{Bhattacharyya:2012ye}, and the total result \rf{23} is to be compared with the canonical-ensemble
expressions \rf{7},\rf{01}. The non-zero-mode part $F_1'$, which contains $-A(k)$, is the same in the two
ensembles. Thus $A(k)$, the $\upmu$-independent term of $J_{\rm pert}$, may equally be viewed as computed in
the $\upmu$-ensemble. What remains to be done there is to evaluate the contribution of the constant sector
with the relative boundary conditions, which should contain no logarithmic terms and should cancel the
dependence of $F_1'$ in \rf{189} on the normalization scale $\mu$.

\iffa  
In \ci{Bobev:2026gir} the $\upmu$-ensemble was implemented on the gravity side by relative boundary
conditions\foot{The relative boundary conditions for a $p$-form $\omega$ mean the Dirichlet-type conditions
$\iota^*\omega=0$, where $\iota^*$ is the pull-back to the boundary, i.e. the tangential components of
$\omega$ are set to vanish. For the fluctuations of $C_3$ this keeps the boundary value of $C_3$ fixed, and
this boundary value is the chemical potential for the M2-brane charge $N$, which is the flux of $\ast F_4$
through $S^7/\ZZ_k$. Fixing the boundary value leaves the charge free as in the grand-canonical ensemble,
while fixing the flux gives the canonical ensemble. The two are related by a Legendre transform of the
action, i.e. a Laplace transform of the partition function.
}
on the 3-form and on the 2-form ghost. With these conditions the harmonic
ghost zero modes, which have non-vanishing pull-back to the boundary of $\Hf$, are absent, and with them the
logarithmic term, in agreement with the absence of logarithms in $J_{\rm pert}$. Taken literally, the same
conditions also exclude the modes $\omega\wedge v_1$ of the harmonic sector of the $\td C_3$ fluctuations in
\rf{18}, since $\iota^*(\omega\wedge v_1)=(\iota^*\omega)\wedge v_1\ne0$ for any non-zero internal 1-form $v_1$.

In  our    approach  when    comparing with the  canonical ensemble   expressions in \rf{7} and   \rf{01} 
 we  assume   the square-integrability condition discussed in Section \ref{44}. It keeps the whole harmonic sector,
including the ghost zero modes that produce the logarithmic term \ci{Bhattacharyya:2012ye}.

The constant $A(k)$ is the $\upmu$-independent term of $J_{\rm pert}$ and is therefore present in both
ensembles. In our computation it arises only after the harmonic-sector  modes \rf{18}, which have a  non-zero internal
eigenvalue, are added to the standard KK spectrum.
 In the $\upmu$-ensemble the ghost zero modes must be
absent, while the  harmonic 3-form  modes must  therefore be kept. 
The relative conditions of \ci{Bobev:2026gir} remove both. A
condition that separates  them would have to act differently on the constant internal mode and on the modes
with non-zero internal eigenvalue, i.e. it would be non-local in 11d.
 How $A(k)$ arises in the grand-canonical computation with the boundary conditions of
\ci{Bobev:2026gir} is therefore an open question.
A boundary problem implementing this distinction, consistently with the variational principle, gauge
symmetry and supersymmetry, remains to be specified.
\fi

\

$\bullet$ \   
A gravitational description of the ABJM Airy-function
coefficients $C$ and $B$ in \rf{3}, based on 4d 
higher-derivative supergravity, was given in
\ci{Hristov:2022lcw}.
This framework was subsequently extended to a general class
of M2-brane theories associated with toric Calabi--Yau
fourfolds, including arbitrary sphere squashing and mesonic
mass deformations, using equivariant volumes
\ci{Cassia:2025jkr}.
These constructions determine the Airy-function argument
while leaving the $N$-independent prefactor undetermined.

The program of relative equivariant localization in supergravity \ci{BenettiGenolini:2023kxp}
  was applied  \ci{BenettiGenolini:2026cyc,BenettiGenolini:2026kjy}  to the 11d supergravity action in order to
reproduce the Airy function structure found from localization on the 3d ABJM theory side.
These references assumed, following  \ci{Gautason:2025plx}, that the  supergravity description   should  correspond to the grand canonical ensemble description on the ABJM side. 
 In
\ci{BenettiGenolini:2026cyc} the coefficients $C$ and $B$  in  \rf{3} 
were obtained from the  $C_3\wedge F_4 \wedge F_4$ 
and  $C_3\wedge  R^4$  terms, and the result was  stated ``up to the prefactor'', so, in particular,  $A(k)$ was  not computed. 

In
\ci{BenettiGenolini:2026kjy} the 1-loop factor was conjectured to be given by  $\log Z_{\text{1-loop}}=\zeta'_{\mathcal K}(0)$,
with $\zeta_{\mathcal K}$  being 
the Mellin transform of the equivariant index of $\bar\partial_W\otimes(TW-T^*W)$ on a
12d  filling $W$ of the  11d space. At an isolated fixed point with weights $w_\alpha$ the
index is $\sum_\alpha(q_\alpha^{-1}-q_\alpha)/\prod_\alpha(1-q_\alpha)$, with $q_\alpha=e^{-{\rm t}w_\alpha}$ and ${\rm t}$ being 
the equivariant parameter. 
For the  ABJM  theory on $S^3$  it gives 
\ci{BenettiGenolini:2026kjy}
\begin{equation}
\mathcal K_k({\rm t})=\frac{\coth(\tfrac{1}{4} k{\rm t})}{2\sinh^2(\tfrac{1}{2}{\rm t})}\,,\qquad\qquad 
\zeta'_{\mathcal K_k}(0)=A(k)\,, 
\la{25}
\end{equation}
which matches our expression in \rf{20},\rf{21}  ($\mathcal K_k=\frac12\cC_k$).
Note that  here the parameter ${\rm t}$ is conjugate to the weights of the
Killing vector, not to the $\AdS_4$ energy $E$   as $\tau$  in \rf{14},\rf{20}. 

The proposal of \ci{BenettiGenolini:2026kjy} gives
analytic predictions for the ABJM Airy constant at
general squashing and for arbitrary supersymmetric
mass deformations, reproducing the available independent
field-theory results at special values of these parameters.
A first-principles derivation of the relevant gauge-fixed
supersymmetry complex, including its boundary conditions
and determinant normalization, remains open.
Our explicit  calculation  for \adssz spectrum  is for  the round $S^3$ 
 case and provides a spectral test of
the proposed character in this setting. 
Our total result \rf{23} includes the ghost zero modes and corresponds to the canonical ensemble, whereas the
conjectured 1-loop formula of \cite{BenettiGenolini:2026kjy} assumes matching to the grand-canonical
ensemble. Since $A(k)$ comes from the non-zero-mode part $F_1'$, which is common to the two ensembles, the
comparison in \rf{25}--\rf{27} is not affected by this difference.

\iffa 
We stress, however, that our calculation  corresponds   to  the 
canonical ensemble, whereas the conjectured 1-loop formula
of \cite{BenettiGenolini:2026kjy} assumes matching to  the grand-canonical
ensemble.
\fi

Before the $\ZZ_k$  projection, the
character  corresponding to the index  in  \ci{BenettiGenolini:2026kjy}  (with the  $so(8)$  fugacities as in  Appendix \ref{258} below) is
\begin{equation}
\mathcal K_W(q;u,v)=\frac{2(q^{-1}-q)+\sum_{y}\big(q^{1/2}y^{-1}-q^{-1/2}y\big)}
{(1-q)^2\prod_{y}\big(1-q^{-1/2}y\big)}\,,\qquad y=\big\{uv,\tfrac vu,\tfrac1{uv},\tfrac uv\big\}\,,
\la{26}
\end{equation}
and we find the exact identity
\begin{equation}
2\,\mathcal K_W(q;u,v)=\cI(q,u,v)+\cI_{\cM_1}(q,u,v)-\cR(q)\,,
\la{27}
\end{equation}
where $\cI$ is our total integrand in \rf{154} (KK tower plus harmonic sector, for $u\ne1$ in the closed
form of Appendix \ref{258}), $\cI_{\cM_1}$ is the singleton integrand and $\cR$ is the $k$-independent
function in \rf{20}.
Thus the proposed character corresponds to the sum of the 11d supergravity KK tower, of the harmonic 2-form
sector and of the singleton, up to $\cR$. 

The identity \rf{27}, including the remainder $\cR$,  provides  a spectral
interpretation of the character of \ci{BenettiGenolini:2026kjy}.
It relates that character to
the standard KK contribution, the harmonic 2-form sector
and the singleton contribution. 

The prescription of \ci{BenettiGenolini:2026kjy}
identifies the function $A(k)$, including its
$k$-independent part, without an adjustable additive
constant. In our calculation, the fugacity-independent
remainder $\cR$ contributes both a finite term and the
determinant-scale dependence displayed below.
The comparison of the complete 1-loop free energies therefore also involves the normalization of the
zero-mode measure, represented by $c_0$ in our result, which is fixed at the end of Section \ref{155}. The
identity \rf{27} establishes the relation between the characters. Identifying the fully normalized
partition functions requires this additional input.

\iffa 
The comparison of the complete 1-loop free energies
therefore also involves the functional-measure
normalization, represented by $c_0$ in our result.
The identity \rf{27} establishes the relation between
the characters; identifying the fully normalized
partition functions requires this additional input.
\fi 
\iffa 
 It  gives  the full $A(k)$, including the $k$-independent part, without an undetermined 
additive constant, so that its prediction is unambiguous even though its derivation from first principles
remains open. This is distinct from our calculation, in which $\cR$ contributes a $k$-independent constant
to $F_1'$ and the total $k$-independent constant also depends on the normalization of the zero-mode
measure, $c_{\rm zm}$ in \rf{201}.  
\fi

For the ADHM quotient, the finite-group projection
of the  character \rf{26} alone gives
$\frac12[A(p)+A(1)]$ for the Airy constant.
The additional contribution
$\frac12(p^2-1)A(1)$ required by the localization
result \ci{Hatsuda:2014vsa} is supplied by the
contribution of the 7d $SU(p)$ vector multiplet localized
on the singular locus $\AdS_4\times S^3$,  which we include in Appendix \ref{258}.
This provides a concrete example of the additional
contributions associated with orbifold singularities.


\

{\bf 4.} \
The rest of the paper is organized as follows. In Section \ref{28} we set up the 1-loop computation:
the 11d action and the $\AdS_4\times S^7$ background, the definition of $F_1$ through $\zeta$-function regularized determinants, and the $\ZZ_k$ quotient. 

Section \ref{44} is devoted to the 3-form field. We recall the properties of the square-integrable
harmonic 2-forms on $\Hf$, the computation of their regularized number $n_2=1$, show that the harmonic form sector \rf{18} decouples from the
metric and gravitino fluctuations, and derive its contribution \rf{19} to the free energy.

Section \ref{61} presents  the character representation of the 1-loop free energy. Section \ref{68}
introduces the $SO(3,2)\times SO(8)$ characters of the conformal and KK multiplets, 
derives \rf{14} starting from the heat kernel on $\Hf$, defines   the Plancherel measure and the
transformation to the variable $\tau$, and gives the resulting integrand of the whole KK tower in
closed form. Section \ref{140} rewrites the harmonic contribution \rf{19} in the same variables, and
Section \ref{150} combines the two into the total integrand.

Section \ref{155} implements   the $\ZZ_k$ projection to  obtain \rf{20}, evaluates its contribution to \rf{1577}, 
  and finally adds the zero mode  part  leading to  the final result in  \rf{23}.
Some  concluding remarks are made in  Section \ref{203}.

Appendix \ref{206} lists the KK spectrum of 11d supergravity on $\AdS_4\times S^7$ in terms of
$OSp(8|4)$ supermultiplets. Appendix
\ref{235} presents   the closed form of the characters of the $\frac12$-BPS multiplets and their sum over the
tower. Appendix \ref{214} derives the contribution of the harmonic
2-form sector from the Hodge decomposition of the 3-form heat kernel.
Appendix \ref{252}   comments on the 
properties of the discrete series of representations of $SO(4,1)$ 
 and of the harmonic 2-forms on $\Hf$. 

Appendix \ref{241} reviews the 1-loop  UV  divergences of 11d supergravity.
 Appendix \ref{247} discusses  the 1-loop Casimir energy in global $\AdS_4$  (defined by the
Lorentzian spectrum on $\RR\times S^2$). 
It   shows that the vanishing of the Casimir energy found in
\ci{Gibbons:1984dg,Inami:1984vp} for $\AdS_4\times S^7$ holds level by level also for
$\AdS_4\times S^7/\ZZ_k$, in agreement with the vanishing of the vacuum energy of the dual   conformal 
theory on
$\RR\times S^2$.

 Appendix \ref{280} discusses the relation between the 1-loop computation in \adssz and 
  in its  reduction to type IIA on $\AdS_4\times\CP^3$. We note that the   IIA  supergravity 
1-loop  contribution   corresponds to  the $U(1)_m$-neutral part of the 11d result, 
and that  the large-$k$ limit does not commute with the $\tau$-integral.


Appendix \ref{258} extends the construction to the ADHM quotient with $p$ flavours, which requires a second
fugacity and the $SU(p)$ vector multiplets localized on the fixed   3-sphere. The projected bulk modes and
the $p^2-1$ localized multiplets give the combination $\frac12\big[A(p)+p^2A(1)\big]$, which reproduces the
known relation between the ADHM and ABJM   $A$-constants \ci{Hatsuda:2014vsa}. 

\section{1-loop supergravity on $\AdS_4\times S^7/\ZZ_k$}\la{28}


In Euclidean signature the bosonic action of 11d  supergravity \ci{Cremmer:1978km} is
\begin{equation}
2\kappa_{11}^2S=\int\dd^{11}x\sqrt g\,\big(-R+\half|F_4|^2\big)+\tfrac{i}{6}\int C_3\wedge F_4\wedge F_4\,,\qquad \qquad 
2\kappa_{11}^2= (2\pi)^8 \lpp ^9 \ , 
\la{30}
\end{equation}
with $F_4=\dd C_3$ and $|F_4|^2=\frac1{4!}F_{ABCD}F^{ABCD}$. The  $\Hf\times S^7$   background
\ci{Freund:1980xh}   with the $S^7$ radius $L$ and the $\AdS_4$ radius $L/2$ is given by 
\begin{equation}
\dd\hat s^2=L^2\big(\tfrac14\,\dd s^2_{\Hf}+\dd s^2_{S^7}\big)\,,\qquad\qquad \ \ \ 
\hat F_4=\tfrac{3i}{8}\,L^3\,{\rm vol}_{\Hf}\,,
\la{31}
\end{equation}
where $\dd s^2_{\Hf}$, $\dd s^2_{S^7}$ and  the volume form ${\rm vol}_{\Hf}$ refer to the unit-radius metrics
($\int   {\rm vol}_{\Hf} = \vol(\Hf)=\frac{4\pi^2}{3}$).\foot{In what follows we will often set
$L=1$. The dependence on $L$  can be restored by dimensional analysis.} 

Expanding the fields around the background, $g=\hat g+h$, $C_3=\hat C_3+\td C_3$, $\Psi_A$, and fixing
the gauges on fluctuations
the 1-loop partition function is formally given  by
\begin{equation}
Z_1=\big({\rm Sdet}\,\cD\big)^{-1/2}\,,
\la{32}
\end{equation}
where $\cD$ is the  second-order  operator acting on $(h_{AB},\td C_3,\Psi_A)$ and on the
ghosts. 
The metric and 3-form fluctuations are coupled through the background flux and the
3-form to itself through the Chern--Simons term in \rf{30} (see Section \ref{44}), so that $\cD$ is not block diagonal in
the fields. 

Let $\mathcal O$ be one of the
 operators in \rf{16}, with $n_0$ zero modes. The
regularized heat trace has the small-$t$  expansion
\begin{equation}
\Tr\,e^{-t\mathcal O}\simeq\sum_{n\ge0}{\rm B}_n(\mathcal O)\,t^{(n-D)/2}\,,\qquad\ \ \ \ \  {\rm B}_n ={\te{1\ov (4\pi)^{D/2}}  }
\int d^{D} x \sqrt g \, {\rm b}_n\ , 
\ \ \ \ \ \  \ \  D=11\,,
\la{35}
\end{equation}
where ${\rm B}_D(\mathcal O)$ is  the coefficient of $t^0$. 
In the  standard proper-time   representation 
\be 
\log \det \OO = - \int^\infty_\varepsilon  {\dd t\ov t }\,   \Tr\,e^{-t\mathcal O} \ , 
\qquad \qquad \varepsilon = \Lambda^{-2} \to 0   \ . \la{305}
\ee
To remove the resulting power divergences   one may use 
the spectral  zeta function  prescription 
\al{
&\zeta_{\mathcal O}(z)=\frac{1}{\Gamma(z)}\int_0^\infty\dd t\,t^{z-1}\big(\Tr\,e^{-t\mathcal O}-n_0\big)\,,\qquad \qquad 
\zeta_{\mathcal O}(0)={\rm B}_{D}-n_0 \ ,
\la{36}
\\
& \ha \log{\det}'\big(\mu^{-2} \mathcal O\big)=-\ha \zeta'_{\mathcal O}(0)-\zeta_{\mathcal O}(0)\log\mu\,. 
\la{37}
}
On a compact  manifold without  boundary the coefficients ${\rm b}_n$ with odd $n$
vanish, so  that $\rm b_{11}=0$ and $\zeta_{\mathcal O}(0)=-n_0$. Thus the $\mu$ 
dependence of a primed determinant is entirely due to the one of the zero mode contribution. Since all
eigenvalues are  $\sim L^{-2}$, the same  applies to  the dependence on  $L$. 

The  non-zero-mode part of  free energy
\begin{equation}
F_1'=\half\,{\Str}'\log\big(\mu^{-2} \cD\big)=\ha \sum_{i}(-1)^{{\rm f}_i} \log{\det}'\big(\mu^{-2} \mathcal O_i \big)\,,
\la{38}
\end{equation}
 thus depends on $\mu$ and $L$ only through $n_0\log(\mu L)$, with $n_0$ being  the graded
number of zero modes.  

The integral over the collective coordinates of the zero modes  
cancels the $\log\mu$ dependence of
$F_1'$.  For canonically
normalized fluctuations,  the zero-mode measure, i.e. the norm of the zero modes 
depends on $L$ and on $\kappa_{11}$, and hence on $L/\lpp $. 
The normalization of this measure is fixed at the end of Section \ref{155}.

All operators in \rf{38}  here are built from the Laplacians and Dirac operators on
the symmetric spaces $\Hf$ and $S^7$, and, e.g.,  the    Laplacian on the  direct product space is  additive,
$
\Delta=\Delta_{\Hf}\otimes1+1\otimes\Delta_{S^7}$.
The spectrum on $S^7$ is discrete and organized by $SO(8)$ representations. The spectrum on $\Hf$
consists of the continuous (principal) series, with spectral parameter $\lambda$ and the
Plancherel measure of  \ci{Camporesi:1993mz,Camp}, plus, for the 
2-forms only, the discrete series of square-integrable harmonic forms.


The fluctuations on $\AdS_4\times S^7$  can be organized into unitary
representations of $OSp(8|4)$ \ci{Freedman:1983na,Gunaydin:1985tc}.
In the case of \adsz    space  the $\ZZ_k$ projection acts
within each supermultiplet through the restriction on  $U(1)_m$ charges. This is what makes the  use of the character representation
of Section \ref{61} natural.

To implement   the $\ZZ_k$  quotient   let us introduce the   embedding coordinates $w_i$ on 
 $S^7\subset\CC^4$. The Hopf  fiber $U(1)$  group acts as $w_i\to e^{i\theta}w_i$, and
$\ZZ_k\subset U(1)$ is generated by $\theta=2\pi/k$. It acts freely, so 
\begin{equation}
M_7\equiv S^7/\ZZ_k
\la{41}
\end{equation}
is a smooth manifold for all $k\ge1$. 
 
 Let the    subgroup 
$U(1)_m\subset SO(8)$  represent  the isometry along the M-theory circle. It is  generated by an element $R_m$ of the
Cartan subalgebra of $so(8)$ defined in \rf{212}.
 A mode survives the quotient when its charge $q_m$ is
a multiple of $k$. The
projector onto $\ZZ_k$ 
invariant modes is
\begin{equation}
\Pi_k=\frac1k\sum_{p=0}^{k-1}g^p\,,\qquad \qquad \ \ \ g=e^{2\pi iR_m/k}\,,
\la{42}
\end{equation}
so that for any operator $\mathcal O$ commuting with $R_m$ 
\begin{equation}
\Tr_{M_7}\,e^{-t\mathcal O}=\frac1k\sum_{p=0}^{k-1}\Tr_{S^7}\big(g^p\,e^{-t\mathcal O}\big)\,.
\la{43}
\end{equation}
Eq. \rf{43} is  the statement that the fluctuation  operator on $M_7$ is the
operator on $S^7$ restricted to the $\ZZ_k$-invariant modes: the average over the group is the
projector onto them. 

The $p=0$ term  in \rf{43}  is $1/k$ times the covering-space
trace. For the unprimed heat trace of a local elliptic operator and a freely acting $g^p$, $p\ne0$, the
$p$-th term has no fixed-point contribution and is exponentially small as $t\to0$, so its contribution
to a zeta-function regularized determinant is finite and independent of the regularization.
 This observation  is used in Section \ref{155}.  

\section{3-form fluctuations  and  harmonic 2-forms}\la{44}

Here we shall  first    write  down  the quadratic action for the 3-form fluctuations and their couplings to the
metric  and  then identify the sectors that decouple  from the rest.
Let us   define the   notation  for the scalar product  
and the Hodge Laplacian on $p$-forms,
\begin{equation}
\langle\alpha,\beta\rangle=\int\alpha\wedge\ast\beta\,,\qquad\qquad \ \ \ 
\Delta_p=\dd\dd^\dagger+\dd^\dagger\dd\,,
\la{45}
\end{equation}
where  all contractions are  done  with the background  metric $\hat g$. The part of the quadratic fluctuation action
that depends on $\td C_3$ is
\begin{equation}
2\kappa_{11}^2 S_2(\td C_3)=\half\big\langle \td C_3,\dd^\dagger\dd \td C_3+\ast(i\hat F_4\wedge\dd \td C_3)\big\rangle
+\big\langle\half(\hat g\cdot h)\hat F_4-h\cdot\hat F_4,\,\dd \td C_3\big\rangle\,,
\la{46}
\end{equation}
where 
$ (h\cdot\hat F_4)_{B_1\dots B_4}=4\,h^A{}_{[B_1}\hat F_{|A|B_2B_3B_4]}\,,\  \hat g\cdot h=h^A{}_A\,, $
i.e. 
\begin{equation}
h\cdot\hat F_4=\tfrac1{3!}\,h^A{}_{B_1}\hat F_{AB_2B_3B_4}\,e^{B_1}\wedge\dots\wedge e^{B_4}\,.
\la{48}
\end{equation}
The second term in the first bracket in \rf{46} comes from the Chern--Simons term in \rf{30}, the second bracket from the
expansion of $\sqrt g|F_4|^2$. Thus  generic 3-form fluctuations mix with the metric fluctuations. 
The
linearized gauge symmetries combine diffeomorphisms with the 3-form gauge transformations,
\begin{equation}
\delta h_{AB}=\mathcal L_\xi\hat g_{AB}\,,\qquad \delta \td C_3=\iota_\xi\hat F_4+\dd\Lambda_2\,,\qquad
\delta\Lambda_r=\dd\Lambda_{r-1}\ , \ \ \ \ \ \ \ r=1,2\,,
\la{49}
\end{equation}
so that the background flux also mixes diffeomorphisms with the gauge symmetry of $\td C_3$.

  If the  mixing  were   absent    the  $C_3$  part of the 1-loop partition function 
   would  take   the familiar form \ci{Schwarz:1978cn,Siegel:1980jj,Obukhov:1982dt}
\begin{equation}
Z_C=(\det\Delta_3)^{-1/2}(\det\Delta_2)^{+1}(\det\Delta_1)^{-3/2}(\det\Delta_0)^{+2}
\la{33}
\end{equation}
with $\Delta_p$  being  the Hodge Laplacians on $p$-forms. 

There  are   two   decoupled sectors  where mixing terms   drop out:
(i) the 3-form  harmonic sector
\rf{57}, and   (ii)  the zero modes of the 2-form
ghost (cf.  the explicit zero-mode analysis in  \ci{Arrighi:2026xmn}).  

Before discussing the  3-form  decoupled sector  let us make some preliminary remarks.
First,  $\Hf$ is a complete Riemannian manifold:
every geodesic can be continued indefinitely, and the conformal boundary is at  an  infinite distance. On a
complete manifold the operators $\dd$ and $\dd^\dagger$ are adjoint to each other on the space of
square-integrable forms whose $\dd$ and $\dd^\dagger$ are also square-integrable, with no boundary
terms \ci{Gaffney:1955}. This 
makes the
integrations by parts in  this section legitimate. Second, for a Laplace-type operator $\Delta$ acting
on a bundle over  $\Hf$ the  heat kernel density has the local small-$t$ expansion (cf. \rf{35}) 
\begin{equation}
\tr\,e^{-t\Delta}(x,x)\simeq\frac{1}{(4\pi t)^2}\sum_{n\ge0}{\rm b}_n\,t^{n/2}\,, \qquad \ \ \  {\rm b}_{2n+1}=0 \ . 
\la{51}
\end{equation}
The Seeley coefficients ${\rm b}_n$ are local invariants built from the curvature  
and are constant on the homogeneous
space $\Hf$. 

On ${\rm H}^{2d}$ there are square-integrable harmonic forms of  the middle degree $d$, i.e. 2-forms on $\Hf$.  The space
$\mathcal H^2(\Hf)$ of  square integrable  harmonic 2-forms is infinite-dimensional \ci{Dodziuk:1979}, and  its 
explicit orthonormal basis $\omega^{(l)}$ \ ($l=1,2,\dots$)  was constructed in \ci{Camp}. These
forms are closed and co-closed, and they are not part of the continuous spectrum. 

The Plancherel measure of the continuous (principal) series of  2-form  eigenvalues $h$, integrated
against $e^{-t { h  } }$, gives a heat kernel trace   whose small-$t$ expansion in \rf{51} reproduces ${\rm b}_0$
and ${\rm b}_2$ of the Hodge Laplacian on 2-forms but not ${\rm b}_4$.
The deficit is a $t$-independent
constant, precisely the contribution of the harmonic forms: the diagonal of the projector onto
$\mathcal H^2(\Hf)$ is the constant $\varrho={3\ov 4\pi^2}$ for  unit-radius  $\Hf$ 
\ci{Camp}, which equals ${1 \ov 32 \pi^2} \mathcal E_4 $ where 
$
\mathcal E_4=R_{abcd}R^{abcd}-4R_{ab}R^{ab}+R^2={24}
$
is the density of the   Euler  number $\chi$ of $\Hf$. The regularized number of the harmonic 2-forms is therefore
\begin{equation}
n_2=\varrho\,\vol (\Hf)=\chi (\Hf)=1\,.
\la{53}
\end{equation}
Here the standard IR regularization is applied to  the volume  and thus to the Euler number of $\Hf$. 
This is the  number used for the  harmonic 2-form  zero modes in \ci{Bhattacharyya:2012ye}.\foot{\la{100}One may illustrate 
  the construction of  the harmonic 2-form as follows.  Consider $\Hf$  with the  metric  ($ |x|<1$) 
$
\dd s^2=\frac{4dx^a dx^a }{(1-|x|^2)^2}. 
$
The Hodge star on 2-forms in 4d is conformally invariant, so that  the flat 2-form
$\omega=\dd x^1\wedge\dd x^2$ is closed and co-closed in the hyperbolic space metric, and its norm,
$
\|\omega\|^2=\int_{|x|<1}\dd^4x=\frac{\pi^2}{2}\,,
$
is finite. As a smooth form it is exact, $\omega=\dd \alpha, \  \alpha=x^1\dd x^2 $, but $\alpha$  has infinite
1-form norm in the hyperbolic metric,  and no square-integrable  analog of it exists.
 Thus  the harmonic
2-forms cannot be removed by a gauge
transformation.   This example
is one element of the infinite-dimensional space. The two constant flat forms
$\dd x^1\wedge\dd x^2\pm\dd x^3\wedge\dd x^4$ are representatives of the lowest  
types of the two representations
$\pi_\pm$ of Appendix \ref{252}. The complete infinite family is parameterized explicitly in
\ci{Camp}.} 

For  all  other  operators   that  appear  in $Z_1$ in \rf{32}   (scalars, vectors,
symmetric traceless tensors, spinors, vector-spinors) the Plancherel theorem on $\Hf$ 
\ci{Camporesi:1993mz,Camp}  implies  no discrete part of the spectrum, so
 the continuous measure reproduces the
local coefficients ${\rm b}_n$  in \rf{51}  by itself. 

Let us  now   show the decoupling   of the harmonic form sectors   in the  ${\td C}_{3}$  and in the 2-form ghost (cf. \rf{33}) 
actions. 
Consider the special  3-form fluctuations of the form \rf{18},
\begin{equation}
{\td C}_{3,H}=\omega^{(l)}\wedge v_{(l)}\,,\qquad\qquad \ \ \  v_{(l)}\in\Omega^1(\X)\,,
\la{57}
\end{equation}
where $\omega^{(l)}$  is a basis of harmonic 2-forms on $\Hf$ (we assume summation over $l$).
 The 
corresponding gauge parameters  are  $\Lambda_2=\phi_{(l)}\omega^{(l)}$,  where 
$\phi_{(l)}\in\Omega^0(\X)$. 
The  special  subset of  2-form ghosts  have the same form $b_{(l)}\omega^{(l)}$.
 Since the exterior differential   splits as 
$\dd=\dd_4+\dd_7$ and the $\omega^{(l)}$ are closed and co-closed, 
we have 
$\dd {\td C}_{3,H}=\omega^{(l)}\wedge\dd_7v_{(l)}$.

The kinetic term in the (gauge-fixed)  fluctuation action \rf{46} contains  the  Hodge Laplacian, which  acts on \rf{57} as
$\omega^{(l)}\wedge\hD_7v_{(l)}$.
The Chern--Simons term $\td C_3\wedge\dd \td C_3\wedge\hat F_4$   vanishes on  the harmonic part 
in \rf{57}. 
The  mixing term  with $h$ needs $\dd \td C_3$ with at least three $\Hf$ indices  as 
$\hat F_4\sim{\rm vol}(\Hf)$, so it also  vanishes.

Turning to the 2-form ghost action, we observe that 
the only possible mixing between the 2-form ghost $\bar b_{(l)}\omega^{(l)}$ and the
diffeomorphism ghost $\zeta$ (from  the variation of the gauge condition $\dd^\dagger \td C_3$) 
 is the
pairing of $\omega^{(l)}\wedge\dd_7\bar b_{(l)}$ with $\iota_\zeta\hat F_4$. Here  the first form has two
and the second  has three $\Hf$ indices, so it also vanishes.

We  conclude that the special harmonic  fluctuations  \rf{57}  
 are  thus a  decoupled   sector  of the   action  \rf{46}   on  which
 the   quadratic  fluctuation operator  reduces to $1\otimes\hD_7$ acting  on the 
 vector fields  on $\X$:\foot{The same conclusion can be  reached starting from the quadratic action in  \ci{Arrighi:2026xmn}.}
\begin{equation}
2\kappa_{11}^2 S^{(H)}_2=\sum_l\Big[\half\langle v_{(l)},\hD_7v_{(l)}\rangle_7
+\langle\bar b_{(l)},\hD_7b_{(l)}\rangle_7\Big]\, . 
\la{58}
\end{equation}
The  gauge symmetry $v\to v+\dd_7\phi$ is fixed by
the Feynman gauge with  the corresponding ghosts.

As a result,   the  non-zero-mode contribution of the  harmonic  $\td C_3$ 
sector \rf{57} to the 1-loop partition function is  given by 
\begin{equation}
Z^{(H)}_1=\Big[\frac{{\det}'_\X \Delta_0}{\big({\det}'_\X\Delta_1\big)^{1/2}}\Big]^{n_2}= e^{-n_2 \F_H} \,,\qquad\qquad 
\F_H=\half\log{\det}'_\X \Delta_1-\log{\det}'_\X\Delta_0\,,
\la{59}
\end{equation}
where $n_2$  is the regularized number \rf{53} of harmonic
2-forms.
 Since the first Betti number of $\X$ vanishes, $b_1(\X)=\dim H^1(\X,\RR)=0$, there are no vector  zero modes.
  Splitting 1-forms into  co-exact plus exact we have 
$\Delta_1=\Delta_{1,\perp}\oplus\Delta_0'$  so that  \rf{59} is equivalent to
\begin{equation}
\F_H=\half\log{\det}_\X\Delta_{1,\perp}-\half\log{\det}'_\X\Delta_0\,.
\la{60}
\end{equation}
Once the decoupling of the harmonic sector   is established,  its contribution  \rf{59}   to the free energy 
can also be read off directly from the 3-form complex \rf{33}, by a Hodge decomposition of  the corresponding  heat kernel 
trace  (see  Appendix \ref{214}).

 
Let us make a comment on boundary conditions.
 The 1-loop determinants are defined by the square-integrable spectral problem on
the complete space $\Hf\times M_7$. On a complete manifold the Hodge Laplacians are essentially
self-adjoint on smooth compactly supported forms \ci{Gaffney:1955},  so their  domain is fixed by square
integrability and no condition at the conformal boundary is imposed.
 The only exception is the $35_s$
scalars of the $n=2$ multiplet, whose mass lies in the window where two quantizations are possible, and for
which we use the supersymmetric value $E=1$ (see discussion in Section \ref{68}). 

The  square integrable  harmonic 2-forms belong to this
domain. Their boundary values are non-zero distributions on $S^3$  
\ci{Lott:2000}, and they satisfy neither the relative (Dirichlet) condition $\iota^*\omega=0$ nor the absolute (Neumann) 
condition
$\iota^*(\ast\omega)=0$. Indeed, $\ast$ acts with opposite signs on the two chiralities  in \rf{254}, whose
boundary values lie in complementary subspaces (see Appendix \ref{252}), so either condition forces
$\iota^*\omega_\pm=0$ and hence $\omega=0$. The  condition of square integrability 
 is therefore not one of the two local
conditions of a space with boundary. 
It is the definition  that was  used in the zero-mode analysis of
\ci{Bhattacharyya:2012ye}.

Square integrability is an assumption on the space of fluctuations in the formal 1-loop computation,
independent of any interpretation in terms of ensembles. For the modes with non-zero internal eigenvalue it
is the condition that no sources of the massive KK fields are turned on. For the part of the 3-form gauge
complex that is constant on $M_7$ it corresponds to fixed flux $N$. The relative conditions of
\ci{Bobev:2026gir}, which fix instead the averaged boundary holonomy of $C_3$, apply to this part only.

 \iffa 
 The harmonic forms of \ci{Camp} have non-vanishing
pull-back to the conformal boundary of $\Hf$ (cf. Appendix \ref{252}). As was already  mentioned in the Introduction, 
in  \ci{Bobev:2026gir} the 
special  boundary conditions are imposed on the 3-form
fluctuations and on the 2-form ghost (vanishing pull-back to the boundary), which exclude
 the  harmonic ghost zero modes. With that choice the harmonic 3-form sector \rf{57} would be excluded as
well. 

The  contribution   of the  constant 
 mode of the 2-form ghost  field 
$\omega^{(l)}\otimes1$,  i.e.  the zero mode of \ci{Bhattacharyya:2012ye},  
 will be  discussed   in Section \ref{155}.
\fi

\section{Superconformal characters and 1-loop partition function}\la{61}

In this section we construct the 1-loop partition function from the KK spectrum, organized in
multiplets of $OSp(8|4)$, as was also done in \ci{Liu:2016dau}. The KK spectrum, together with the
notation for the multiplets, their $so(8)$ labels, their $U(1)_m$ charges and the superconformal characters, is collected in Appendices \ref{206} and  \ref{235}. 

As discussed  in the Introduction, the KK modes determine the spectrum of the quadratic fluctuation operators. We first write the contribution of each conformal multiplet, and then express the sum over the KK tower in terms of the superconformal characters (Section \ref{68}). We then add the contribution of the harmonic sector in the same character-like form (Section \ref{140}). Remarkably, the contributions of both sectors combine into a simple expression (Section \ref{150}). 


\subsection{Contribution of the KK modes}\label{68}

For an operator $\mathcal O$ describing a  supergravity fluctuation  mode, the spectral determinant 
 can be formally written as
\begin{equation}
\log{\det}'\mathcal O=-\int_0^\infty\frac{\dd t}{t}\,\Tr'\,e^{-t\mathcal O}\,.
\la{69}
\end{equation}
To obtain the total  non-zero-mode 1-loop partition function associated
with the operator $\cD$ in \rf{32} or \rf{16} one  is to sum over all fields
and account for their statistics.

In the case of $M_{7}=S^7/\mathbb{Z}_k$, the modes on the quotient can be obtained by projecting to the $\mathbb{Z}_k$-invariant modes on $S^7$. To implement this projection, it is convenient to introduce the  fugacity $v$ corresponding to
 $U(1)_m\subset SO(8) $ with the   generator $R_m$ 
 and consider the ``refined'' heat kernel
\begin{equation}
K(t,v)=\Str\,(e^{-t\cD}\,v^{R_m})=\sum_{}(-1)^{\fF}\,\Tr\, (e^{-t\cD}\,v^{R_m})\,,
\la{70}
\end{equation}
where $(-1)^\fF$ is $+1$ for bosons and $-1$ for fermions. The corresponding ``refined'' free energy of non-zero modes is 
\begin{align}
    F_1'(v) =-\tfrac12\int_0^\infty\frac{\dd t}{t}\,K'(t,v)\, . \la{65}
\end{align}
As  in \rf{42},\rf{43} (see also  Section \ref{155})  the non-zero-mode 1-loop partition function on $\AdS_4\times S^7/\mathbb{Z}_k$ can then be expressed  as  the finite group average 
\begin{equation}
\log Z_1'=-F_1'=-\frac{1}{k}\sum_{p=0}^{k-1}F_1'(\omega^{p}) \,,\qquad\qquad 
\omega=e^{2\pi i/k}\,.
\la{71}
\end{equation}
To understand how the modes transform under the $\mathbb{Z}_k$ action, one can keep track of the $so(8)$ weights appearing in the KK expansion of the supergravity modes. As we discuss below, the refined heat kernel \rf{70} for the KK modes is related to the superconformal character
\begin{align}\label{64}
    \chi_{_\cM}(q,x,v) &= \tr_{_{\mathcal{M}}}(q^{D}x^{2J_{3}}v^{R_m})\,,
    \qquad \qquad
    \cM = \bigoplus_{n\geq2}\cM_n\,,
\end{align}
where $\mathcal{M}_n$ are the $OSp(8|4)$ multiplets appearing in the KK expansion of the supergravity modes  (see Appendix~\ref{206}). Here $D$ is the dilatation generator of the boundary conformal algebra,\footnote{In Lorentzian global AdS$_4$ it is the Hamiltonian conjugate to global time, so that its eigenvalue $E$ is the global energy of a single-particle state
 (in units of the inverse AdS$_4$ radius), equal to the scaling dimension of the dual  conformal operator.} and $J_3$ is the Cartan generator of the $so(3)$ spin algebra of $so(3,2)$, with eigenvalues $m\in\{-j,-j+1,\ldots,j\}$  for  a spin-$j$ multiplet. Note that $x^{2J_3}$ has even powers of $x$ for bosons and odd powers for fermions.

The calculation of the superconformal character is rather straightforward and allows us to obtain an explicit expression for the contribution of the KK modes \rf{71}.

In Lorentzian $\AdS_4$, after KK compactification, each 4d bulk field $\varphi$ describes a conformal multiplet $\cm{j}{\mathrm{r}}{E}$, where $j$ is its $\AdS_4$ spin and $\mathrm{r}$ labels an irreducible representation of $so(8)$.
The equations of motion for the field $\varphi$   can be written in terms of 
 the quadratic Casimir operator $C_2$ of $so(3,2)$:
\begin{equation}
\big[C_2-C_2(E,j)\big]\varphi=0\,,\qquad\qquad 
C_2(E,j)=E(E-3)+j(j+1)\,.
\la{72}
\end{equation}
Here $C_2$ is realized on $\varphi$ by the Killing vectors of $\AdS_4$ acting on the spin part as well as on the argument, and $C_2(E,j)$ is its value on $[j]_E$ (see, for instance, \ci{deWit:2002vz}). Eq. \rf{72} just expresses the fact that the quadratic Casimir acts on the representation $[j]_E$ as multiplication by the constant $C_2(E,j)$. The quadratic fluctuation operator acting on $\varphi$ can therefore be expressed as a Laplace-type operator:
\begin{equation}
\cD_{E,j}=4\big[-C_2+C_2(E,j)\big]\,,
\la{73}
\end{equation}
where the factor $4$    is the square  of the inverse   $\AdS_4$ radius $\tfrac12$ in units of $L$ in  \rf{31}.\footnote{For instance, for a  scalar field  with  $j=0$ and conformal dimension $E$, once expressed in terms of derivatives, the Casimir operator $C_2$ becomes  the d'Alembert operator on  $\AdS_4$, while the effective mass  satisfies $m^2=C_2(E,0)=E(E-3)$ (in units of $\AdS_4$ radius). 
For bosonic fields, after Euclidean continuation and imposing the transversality and tracelessness constraints, $\cD_{E,j}=4\big(-\nabla^2+m^2_{E,j}\big)$ with $m^2_{E,j}=E(E-3)-j$. Here $\nabla^2$ is the Laplacian of the unit-radius $\Hf$ acting on symmetric transverse traceless rank-$j$ tensors. 
For fermions, $\cD_{E,j}$ denotes the corresponding squared Dirac or Rarita--Schwinger kinetic operator.}

The Euclidean continuation replaces $so(3,2)$ by $so(4,1)$ and the Lorentzian metric by that of $\Hf=SO_0(4, 1)/SO(4)$. The differential operator $\cD_{E,j}$ is unchanged in form, with $C_2$ now the Casimir of $SO(4,1)$ acting on fields on $\Hf$. On square-integrable spin-$j$ fields on $\Hf$ the operator \rf{73} is self-adjoint and, for the propagating fields of the KK spectrum, its spectrum is purely continuous. Its generalised eigenfunctions are the spin-$j$ harmonics of the principal series, $\psi_{\lambda,j}(x)$, labelled by $\lambda\in\RR_+$. On these harmonics \ci{Camporesi:1993mz,Camp}
\begin{equation}
C_2\,\psi_{\lambda,j}
=\big[-\lambda^2-\tfrac94+j(j+1)\big]\psi_{\lambda,j}\,.
\la{75}
\end{equation}
The Casimir of a mode consists of a universal part $-\lambda^2-\frac94$  (the same for all spins)
 and of the spin part $j(j+1)$  (the Casimir of the $SO(3)$ representation carried by the spin components of the field). The same spin part appears in $C_2(E,j)$ and therefore cancels in \rf{73},
\begin{equation}
\cD_{E,j}\,\psi_{\lambda,j}=e(\lambda)\,\psi_{\lambda,j}\,,\qquad\qquad 
e(\lambda)=4\big(\lambda^2+\nu^2\big)\,,\qquad\qquad 
\nu\equiv E-\tfrac32\,.
\la{76}
\end{equation} 
The eigenvalue $e(\lambda)$ depends on the multiplet only through the energy $E$ and on the mode only through~$\lambda$.\foot{\la{fhh}Let us note that the Euclidean continuation of global $\AdS_4$ with
noncompact time (i.e. with boundary $\mathbb R\times S^2$)  is the same hyperbolic space $\Hf$
that admits the  foliation with $S^3$ boundary.
Its $\mathbb R\times S^2$  boundary metric is conformal to the
round $S^3$ metric with two poles removed. 
 Below the determinant   of  the operator $\cD_{E,j}$   is defined on complete $\Hf$ assuming 
 the $S^3$  boundary  when  regularizing the AdS$_4$  volume.
Thus the use of the  global-energy representations does not imply that  it corresponds to 
the  partition function  regularized   under the assumption that   the boundary is  the 
 cylinder  $\mathbb R\times S^2$  (see also footnote \ref{fgg}).}

The heat-kernel trace density on $\Hf$ is given by an integral over the spectral parameter $\lambda$ with the spin-dependent Plancherel measure $\mu_j(\lambda)$. The Plancherel measure on $\Hf$ for a spin-$j$ field has the form \ci{Camporesi:1993mz}
\begin{align}
\mu_j(\lambda)&= \tfrac{2j+1}{8\pi^2} 
\lambda\big[\lambda^2+(j+\tfrac12)^2\big]\, T_j(\lambda)\,,
\qquad\qquad \ \ \ 
T_j(\lambda)=
\begin{cases}
\tanh(\pi\lambda)\,,&j\in\ZZ_{\ge0}\,,\\[1mm]
\coth(\pi\lambda)\,,&j\in\ZZ_{\ge0}+\tfrac12\,.
\end{cases}
\la{78}
\end{align}
Therefore, the contribution of one unshortened conformal multiplet to the refined heat kernel \rf{70} is
\begin{equation}
K'_{(E,j,\mathrm{r})}(t,v)
=\chi^{(\mathrm{r})}(v)\;\cK_{E,j}(t)\,,
\la{79}
\end{equation}
where $\chi^{(\mathrm{r})}(v)$ is the $so(8)$ character (see Appendix~\ref{235}) and $\cK_{E,j}(t)$ is the normalised trace over $\Hf$ with spin-$j$ degeneracy and statistics included (see also \ci{Liu:2016dau})
\begin{equation}
\cK_{E,j}(t)
=\tfrac{1}{6}c_j
\int_0^\infty\dd\lambda\,
\lambda\, \big[\lambda^2+(j+\tfrac12)^2\big]\,
T_j(\lambda)\,e^{-4t(\lambda^2+\nu^2)}\,, \qquad\qquad \ \  c_{j}=(-1)^{2j}(2j+1)\, ,
\la{80}
\end{equation}
where we have used  the regularized  expression ${4\pi^2\ov 3} $  for the volume of unit-radius $\Hf$.

The contribution of the conformal multiplet $[j]_E^{(\mathrm{r})}$ to the free energy \rf{65} can thus be written  as
\begin{align}
F'_{1,(E,j,\mathrm{r})}(v)
&=-\tfrac12\int_0^\infty\frac{\dd t}{t}\,K'_{(E,j,\mathrm{r})}(t,v)
=-\tfrac12\chi^{(\mathrm{r})}(v)\int_0^\infty\frac{\dd t}{t}\,\cK_{E,j}(t)\,.
\la{88}
\end{align}
Following \cite{Bae:2016rgm,Pang:2016ofv,Basile:2018zoy,Sun:2020ame}, the 1-loop partition function resulting from the heat kernel \rf{80} can be expressed in terms of  the $so(3,2)$ character of the conformal multiplet.
To obtain this relation, one can use  that 
\begin{equation}
e^{-4t\lambda^2}
=\frac{1}{\sqrt{16\pi t}}
\int_{-\infty}^\infty\dd\tau\,
e^{-{1\ov 16 t} \tau^2}\ e^{i\tau\lambda}\,, 
\la{82}
\end{equation}
which  converts the integral over the spectral parameter $\lambda$ in \rf{80} into a Fourier transform. 
Evaluating the proper-time integral,
the free energy contribution \rf{88} can be represented  formally
as\foot{For the multiplets appearing in Appendix~\ref{206}, the
superconformal primary is given by the scalar field with conformal
dimension $E=\frac12 n$. Recall that the unitarity bound for the scalar
field in $\AdS_4$ requires $E\geq\frac12 $. The supergravity fields for
$n>2$ satisfy the standard quantization in the bulk with $E\geq \frac 32 $.
For $n=2$ the $35_s$ scalars have $m^2=-2$ and two admissible boundary
conditions, corresponding to
$E=1$ and $E=2$, with the same spectrum $\lambda^2+\frac14$ on $\Hf$.
The $\cN=8$ organisation of the KK multiplets selects $E=1$. The
character expression is the same in both cases,
$\cC_{E,j}(q,x)=q^E\chi_{2j}(x)P(q,x)$  (where $P$ is defined in \rf{2231}) and the quantization enters
only through the value of the conformal dimension $E$ of the dual
primary \cite{Sun:2020ame}. For $E<\frac32$ the right-hand side of \rf{84}
is defined by analytic continuation in $E$ from $E>\frac32$, which
corresponds to the chosen boundary condition. The two differ by $\frac{1}{8\pi^2}\zeta(3)$ term
per real scalar \cite{Klebanov:2011gs,Giombi:2013fka}. One can also
formally extend \rf{84} to the $n=1$ singleton multiplet, with its
null states removed.
}
\begin{align}
F'_{1,(E,j,\mathrm{r})}(v)
=-\tfrac12\chi^{(\mathrm{r})}(v)
\int_0^\infty\frac{\dd\tau}{\tau}\,\cJ_{E,j}(\tau)\,.
\la{84}
\end{align}
Using the expression for $\cK_{E,j}(t)$ in \rf{80} and the Plancherel measure for bosonic (integer spin) and fermionic (half-integer spin) fields in \rf{78}, we get
\begin{align}
\cJ^B_{E,j}
&=c_j\,\frac{1+q}{1-q}
\Big[
\frac{q^E}{(1-q)^3}
-\tfrac16 {j(j+1)}\,\frac{q^{E-1}}{1-q}
\Big]\,,
\qquad\ \ \ \ \ \ \  j\in\ZZ_{\ge0}\,,
\la{86}\\
\cJ^F_{E,j}
&=c_j\,\frac{2q^{1/2}}{1-q}
\Big[
\frac{q^E}{(1-q)^3}
-\tfrac{1}{6} {(j-\tfrac12)(j+\tfrac32)}\,
\frac{q^{E-1}}{1-q}
\Big]\,,
\qquad j\in\ZZ_{\ge0}+\tfrac12\,,
\la{87}
\end{align}
where $c_j$ was defined in \rf{80} 
and  we use the variable
\begin{align}\la{63}
q\equiv e^{-\tau}\,.
\end{align}
The relation between \rf{86} and \rf{87} and the $so(3,2)$ characters is explicit. The $so(3,2)$ character $\cC_{E,j}(q,x)$ of a conformal multiplet can be expressed in terms of the spin-$j$ character $\chi_{2j}(x)$ as\foot{\la{fgg}
Let us  stress that no Lorentzian spectrum is used as an input here. The starting point is the  square-integrable 
spectral decomposition on $\Hf$, with the Plancherel density \rf{78} and the eigenvalues
$\lambda^2+\nu^2$, which define the determinant with the $S^3$ boundary. The energies  of
global $\AdS_4$ appear as the output of the transformation to $\tau$, through the poles of 
$\tanh\pi\lambda$ at $\lambda=\pm i(n+\frac12)$ and $\coth\pi\lambda$ at $\lambda=\pm i n$, which generate the series
$q^E/(1-q)^3=\sum_nd_nq^{E+n}$. This series is the 
$so(3,2)$ scalar character with lowest energy $E$,
which enters the character integral representation of
\ci{Sun:2020ame,Anninos:2020hfj}. The character thus enters as a generating function of the
representation labelled by $E$, which is also what fixes the boundary condition at $S^3$, and
not as a sum over states of the theory on $\RR\times S^2$.} 
\al{
\cC_{E,j}(q,x)\equiv \tr (q^D x^{2J_3})
=&\frac{q^E\ \chi_{2j}(x)}
{(1-q)(1-qx^2)(1-qx^{-2})}\,,\qquad \ \   \la{7766} \\
\qquad
\chi_{2j}(x)=&\sum_{m=-j}^{j}x^{2m}
=\frac{x^{2j+1}-x^{-2j-1}}{x-x^{-1}}\,.  \la{7777}
}
Using character representations of 1-loop zeta functions in AdS given  in \ci{Bae:2016rgm,Pang:2016ofv,Basile:2018zoy}, $\cJ_{E,j}(\tau)$ can be represented in terms of  a differential operator acting on the conformal character,
\begin{equation}
\cJ_{E,j}(\tau)=
\begin{cases}
\displaystyle
\kappa_B(q)\, 
\Big[2-\tfrac{(1-q)^2}{8q}(x\partial_x)^2\Big]
\cC_{E,j}(q,x)
\Big|_{x=1}\,,
&\qquad j\in\ZZ_{\ge0}\,,\\[3mm]
\displaystyle
-\kappa_F(q)\, 
\Big[2+\tfrac{(1-q)^2}{8q}
-\tfrac{(1-q)^2}{8q}(x\partial_x)^2\Big]
\cC_{E,j}(q,x)
\Big|_{x=1}\,,
&\qquad j\in\ZZ_{\ge0}+\tfrac12\,,
\end{cases}
\la{101}
\end{equation}
where we have introduced
\begin{equation}
\kappa_B(q)=\frac{1+q}{1-q}\,,
\qquad\qquad \ \ \ 
\kappa_F(q)=\frac{2q^{1/2}}{1-q}\,.
\la{89}
\end{equation}
The total contribution of the KK modes is the sum of the per-field integrands in \rf{84} over all conformal multiplets of the KK spectrum, with the null-state subtractions of the conserved-current multiplets included,
\footnote{\label{fa}In applying the character representation \eqref{84}, one can worry
about the multiplicative anomaly (finite difference  between determinant of a product and product of determinants). 
 The finite multiplicative anomaly
term results from the $1/\tau$ terms in the Laurent expansion of
$\cJ_{E,j}/(2\tau)$ (see,  e.g., \cite{Sun:2020ame}   where  ``multiplicative anomaly''
refers to  a finite difference between the spectral zeta-function regularized 
 determinant on Euclidean AdS and the result 
 obtained by zeta-regularising the character integral after dropping its UV regulator, cf. $\lambda^2 + \nu^2 \to (\lambda + i \nu) (\lambda-i \nu)$). 
Concretely, if $\cJ_{E,j}^{(0)}$ denotes the $\tau$-independent part of $\cJ_{E,j}$ and
$\cJ_{E,j}^{(0)}=a_0(j)+a_2(j)\nu^2+a_4(j)\nu^4$
(with $\nu=E-\tfrac32$), the finite multiplicative anomaly results
from the $a_2(j)\nu^2$ and $a_4(j)\nu^4$ terms.
Using the multiplets in Appendix~\ref{206}, it is straightforward
to check explicitly that, for each $\cM_n$,
$\sum_{[j]_{E}^{(\mathrm{r})}\in\cM_n}
\chi^{\mathrm{r}}(v)\,a_2(j)\nu^2=0$
and
$\sum_{[j]_{E}^{(\mathrm{r})}\in\cM_n}
\chi^{\mathrm{r}}(v)\,a_4(j)\nu^4=0$.
There is thus no multiplicative anomaly for the refined free
energy \eqref{65}.
}
\begin{equation}
\cI_{\rm KK}(\tau,v)
\equiv
\sum_{\cm{j}{\mathrm{r}}{E}\in\cM}
\chi^{(\mathrm{r})}(v)\,\cJ_{E,j}(\tau)\,,\qquad \ \ 
\qquad
\cM=\bigoplus_{n\ge2}\cM_n\,.
\la{93}
\end{equation}
Here the sum in \rf{93} runs over the conformal multiplets of Table \ref{211}, for all $n\ge2$. To express \rf{93} in terms of the superconformal character \rf{64} (see also \rf{237}),
\begin{equation}
\chi_{\rm KK}(q,x,v)
\equiv
\sum_{n\ge2}\chi_{_{\cM_n}}(q,x,v)
=
\sum_{\cm{j}{\mathrm{r}}{E}\in\cM}
\chi^{(\mathrm{r})}(v)\,\cC_{E,j}(q,x)\,,
\la{103}
\end{equation}
one can extract the bosonic and fermionic contributions from the even and odd parts in $x$ of $\chi_{\rm KK}(q,x,v)$:
\begin{equation}
\chi^\pm_{\rm KK}(q,x,v)
\equiv
\half\big[
\chi_{\rm KK}(q,x,v)\pm\chi_{\rm KK}(q,-x,v)
\big]\,.
\la{106}
\end{equation}
For \rf{93}, using \rf{101},\rf{103} and \rf{106}, we finally obtain
\begin{align}
\cI_{\rm KK}(\tau,v)
={}&
\kappa_B(q)
\, 
\Big[2-\tfrac{(1-q)^2}{8q}(x\partial_x)^2\Big]
\chi^+_{\rm KK}(q,x,v)
\Big|_{x=1}
\nonumber\\
&-\kappa_F(q)
\,
\Big[2+\tfrac{(1-q)^2}{8q}
-\tfrac{(1-q)^2}{8q}(x\partial_x)^2\Big]
\chi^-_{\rm KK}(q,x,v)
\Big|_{x=1}\,.
\la{107}
\end{align}
The superconformal character \rf{103} involves the standard $so(3,2)$ and $so(8)$ characters and, with the states of the superconformal multiplet specified in Appendix~\ref{206}, can be computed explicitly. Taking  derivatives in \rf{107}, we get
\begin{align}
\cI_{\rm KK}(\tau,v)
&=
\frac{4{\rm Q}-6(1+q)+4q{\rm Q}^{-1}}{1-q}
+\frac{(1-q)\big[
        4(1+q^2){\rm Q}
        -(1+q)(2q^2+q+2)
        \big]}{{\rm Q}^4} \, , \la{138}\\
&\ \ \ \ {\rm Q}(q,v)\equiv 1+q-(v+v^{-1})q^{1/2}\, , \ \ \ \qquad \ \ \ \ q= e^{-\tau} \, . \la{111}
\end{align}
The same construction can be formally applied to the $n=1$ singleton multiplet, which is absent from the KK tower. This gives the $\cN=8$ singleton contribution discussed  in Section \ref{155}
\begin{equation}
\cI_{\cM_1}=4(v+v^{-1}) \frac{\,q^{1/2}}{1-q}\,.
\la{112}
\end{equation}

\iffa 
Let   us finish with a remark  that the boundary conditions are part of the definition of the 1-loop determinants. 
In particular, the $35_s$ scalars of the massless multiplet obey the
supersymmetric (dimension $E=1$) boundary condition.  Their contribution is obtained
by analytic continuation in the   parameter $\nu=E-\frac32$ (cf. \ci{Giombi:2013fka,Sun:2020ame}). 
The  spectral integral depends only on $\nu^2$
and instead gives the $E=2$ value.  The scalar   boundary condition  needs to be specified separately from the one in 
 the harmonic sector  below. 
\fi

\subsection{Contribution of the harmonic 3-form  sector}\la{140}

To obtain the total 1-loop partition function, one must also include
the harmonic  sector contribution \rf{19},\rf{60} 
(see  Appendix \ref{214}). It  can be written in the same integral form \rf{84}  
 so that its integrand can be added to $\cI_{\rm KK}$ in \rf{107}. The corresponding eigenvalues follow by decomposing
the forms into irreducible representations of  
${so}(8)$
and expressing the Hodge Laplacian through the quadratic Casimir.
With the $S^7$ radius set to one  we have \ci{Ikeda:1978}:
\begin{align}
    \Delta_p\, \phi_{p\, (\mathrm r)}
    =
    \big[\sum_{a=1}^{4}r_a(r_a+8-2a)\big]\, \phi_{p\, (\mathrm r)}\,,
\end{align}
where $\phi_{p\, (\mathrm r)}$ is a $p$-form mode in representation $({\rm r})=(r_1,r_2,r_3,r_4)$.
 With the conventions
of Appendices \ref{206} and \ref{235}, the non-constant scalar modes
decompose as \ci{Ikeda:1978}
\begin{align}
    \Omega^0(S^7)'
    =
    \bigoplus_{n=1}^{\infty}\te 
    \big(\frac n2,\frac n2,\frac n2,-\frac n2\big)\,,
\end{align}
while the 1-forms decompose into exact and co-exact parts
(recall that $b_1(S^7)=0$)
\begin{align}
    \Omega^1(S^7)
    =
    \bigoplus_{n=1}^{\infty}{\te 
    \big(\frac n2,\frac n2,\frac n2,-\frac n2\big)}
    \oplus
    \bigoplus_{n=2}^{\infty}
    {\te \big(\frac n2,\frac n2,\frac n2-1,1-\frac n2\big)\,.}
\end{align}
The corresponding eigenvalues are
\begin{align}
    n(n+2\gamma)=(n+\gamma)^2-\gamma^2\,,\la{141}
\end{align}
with $\gamma=3,\, n\geq1$ for the scalar and exact 1-form modes,
and $\gamma=2,\, n\geq2$ for the co-exact 1-form modes.

With  the regularized number $n_2=1$ of the harmonic 2-forms on
$\Hf$ in  \rf{53}, the ``refined'' analog of the  harmonic  sector contribution  ${\rm F}_H$ in \rf{59}   may be written as 
\begin{align}
    \F_H(v)
    &=-\tfrac12\int_0^\infty\frac{\dd t}{t}\,
    \cK_{H}(t,v)\,,
    \\
    \cK_{H}(t,v)
    &=
    \Tr_{\Omega^1(S^7)}
    \big(v^{R_m}e^{-t\Delta_1}\big)
    -2\Tr'_{\Omega^0(S^7)}
    \big(v^{R_m}e^{-t\Delta_0}\big)\,. \la{434}
\end{align}
As in  \rf{82}  we  may write ($\gamma=2,3$) 
\begin{align}
    \int_0^\infty\frac{\dd t}{t}\,
    e^{-tn(n+2\gamma)}
    =
    \tfrac{1}{4\sqrt{\pi}}
    \int_{-\infty}^{\infty}\dd\tau\,
    e^{{1\ov 2} \gamma\tau}
    \int_0^\infty{\dd t\ov t^{3/2}}\,
    e^{-{1\ov 16t} \tau^2-t(n+\gamma)^2}
    =
    \int_0^\infty\frac{\dd\tau}{\tau}\,
    e^{-{1\ov 2} n\tau}\big(1+e^{-\gamma\tau}\big)\,. \la{435} 
\end{align}
This identity relates two divergent integrals, and for their regularized values it holds up to a
scale-dependent  term. In the $\zeta$-function prescription \rf{171} the right-hand side gives
$-\log\frac n2-\log\frac{n+2\gamma}{2}=-\log\frac{n(n+2\gamma)}{4}$, i.e. it represents the eigenvalue
divided by $4$ (in units of the $\AdS_4$ radius  equal to $\frac12$), exactly as the KK integrand of
Section \ref{68} represents $\frac14 e(\lambda)=\lambda^2+\nu^2$. The two sectors are thus treated in the same
convention, and the conversion to the physical scale $\mu$ is made for the total integrand in
eq. \rf{189} below.

Applying this transformation and using the
${so}(8)$ characters \rf{236}, we obtain
\begin{align}
    \F_H(v)
    &=
    -\tfrac12\int_0^\infty\frac{\dd\tau}{\tau}\,
    \cI_{H}(\tau,v)\,,
    \\
    \cI_{H}(\tau,v)
    &=
    (1+q^2)\sum_{n=2}^{\infty}q^{n\ov 2}\,
    \chi^{(\frac n2,\frac n2,\frac n2-1,1-\frac n2)}(v)
    -(1+q^3)\sum_{n=1}^{\infty}q^{n\ov 2}\,
    \chi^{(\frac n2,\frac n2,\frac n2,-\frac n2)}(v)\,.\la{148}
\end{align}
Here the exact 1-form contribution cancelled one of
the two scalar contributions. Summing over $n$ gives
\begin{align}\la{149}
    \cI_H(\tau,v)
    =
    2+q^2+q^3
    -
    \frac{
        (1-q)\big[
        4(1+q^2){\rm Q}
        -(1+q)(2q^2+q+2)
        \big]
    }{{\rm Q}^4}\,,
\end{align}
where ${\rm Q}$ was  defined in \rf{111}.

\subsection{The combined contribution}\la{150}

Adding together   \rf{138} and \rf{149} we  notice that the
${\rm Q}^{-3}$ and ${\rm Q}^{-4}$ terms cancel in the combined expression, so that we get 
\begin{equation}
\cI(q,v)\equiv\cI_{\rm KK}+\cI_H
=q^2+q^3+\frac{4q}{(1-q)(1-q^{1/2}v)(1-q^{1/2}v^{-1})}
-\frac{4q^{1/2}(v+v^{-1})+4q}{1-q}\,,
\la{154}
\end{equation}
where we used  that ${\rm Q}=1+q-(v+v^{-1})q^{1/2}$.
 We obtain a similar cancellation when the second
fugacity for the $U(1)$  generator $2R_1$ is  introduced  in Appendix~\ref{258}.

The  total ``refined'' non-zero-mode part of the 1-loop  free energy  is thus given by 
\begin{align}\la{666}
    F_1'(v)
    =-\tfrac12\int_0^\infty\frac{\dd\tau}{\tau}\,
    \cI(e^{-\tau},v)\,.
\end{align}
Note that the identity \rf{154} between the summed character
integrands at fixed $\tau$ holds before any projection or
regularization.
To evaluate the free energy it remains to project \rf{154}
onto the $\ZZ_k$-invariant states and to perform the regularized
$\tau$-integral.

\section{$\ZZ_k$ projection and  $A(k)$}\la{155}

The non-zero-mode part of free energy for the $\mathbb{Z}_k$  quotient  of $S^7$ 
is obtained from \rf{154}, \rf{666}  using  a finite group
average (Molien projection)   as in \rf{42},\rf{43}. 
One can define the projected
integrand $\cI_k(\tau)$ as the average of the total integrand \rf{154}
over the group elements,
\begin{equation}
\cI_k(\tau)\equiv\frac1k\sum_{p=0}^{k-1}
\cI(q,v)\Big|_{q=e^{-\tau},\ v=\omega^p}\,,
\qquad\qquad 
\omega=e^{2\pi i/k}\,.
\la{157}
\end{equation}
$\mathbb{Z}_k$-quotient acts freely on $S^7$ and there are no additional ``orbifold degrees of freedom''.\footnote{This 
changes in the case of the ADHM quotient discussed in Appendix~\ref{258}, where one has
to add an additional vector multiplet contribution localised on
$\AdS_4\times S^3$.}
With \rf{157}, the free energy of the non-zero supergravity 
modes on $S^7/\mathbb{Z}_k$ becomes
\begin{equation}
F_1'=-\half\int_0^\infty\frac{\dd\tau}{\tau}\,\cI_k(\tau)\,.
\la{158}
\end{equation}
Using that
\begin{align}
\frac{1}{k}\sum_{p=0}^{k-1}\omega^{pl} = 
 \left\{
\begin{array}{ll}
1, &\qquad  l\in k\mathbb{Z}\, ,\\
0, & \qquad  l\not\in k\mathbb{Z}\, , 
\end{array}
\right.
\la{159}
\end{align}
 the projected integrand $\cI_k(\tau)$
becomes\foot{Here, we have used that one can expand \rf{154}
in powers of $q^{1/2}$ and apply \rf{159} at each order.
In particular, one  finds that 
$
\frac1k\sum_{p=0}^{k-1}
\frac{1}{(1-q^{1/2}\omega^p)(1-q^{1/2}\omega^{-p})}
=\frac{1+q^{k/2}}{(1-q)(1-q^{k/2})}\,.
$
}
\begin{gather}
\cI_k(\tau)
=\cC_k(\tau)+\cR(\tau)
-\delta_{k,1}\,\frac{8e^{-\frac12 \tau} }{1-e^{-\tau}}\,,
\la{161}\\
\cC_k(\tau)
=\frac{\coth(\frac{1}{4}k\tau)}{\sinh^2(\frac{1}{2}\tau)}\,,
\qquad\qquad
\cR(\tau)
=e^{-2\tau}+e^{-3\tau}-\frac{4}{e^\tau-1}\,.
\la{163}
\end{gather}
For $k\geq2$, the only $k$-dependence comes from $\cC_k(\tau)$.
The last term in \rf{161}, relevant only for $k=1$, can be removed
by formally adding the singleton contribution to the one of the 
bulk supergravity
modes.  Indeed,  using \rf{112},    we have 
\begin{align}
\frac1k\sum_{p=0}^{k-1}
\cI_{\cM_1}(q,v)\Big|_{q=e^{-\tau},\ v=\omega^p}
=\frac{8q^{1/2}}{1-q}\,\delta_{k,1}\,.
\la{165}
\end{align}
For $k=1$, the singleton describes the decoupled   abelian $\cN=8$
``centre-of-mass''  sector of the $U(N)_1\times U(N)_{-1}$ ABJM theory.\foot{\la{191}Its $S^3$ free energy (i.e. that of 
8 real conformal scalars and 4  Dirac  fermions) is 
$
F_{\rm free}=\te 8\big(\frac{\log2}{8}-\frac{3\zeta(3)}{16\pi^2}\big)+4\big(\frac{\log2}{4}+\frac{3\zeta(3)}{8\pi^2}\big)=2\log2\,,
$
which is exactly $-\frac12\zeta'(0)$ of the integrand $8q^{1/2}/(1-q)$ in \rf{165}, whose Mellin transform is
$8\zeta(z,\frac12)$. The Airy function \rf{2} refers to the full $U(N)_1\times U(N)_{-1}$ theory, so the
comparison at $k=1$ must be made with the singleton added.}

The free-energy expression \rf{158} still requires a regularization
prescription. In particular, the $k$-dependent part of the integrand
$\cC_k(\tau)$ in \rf{161}  is singular in the limit  $\tau\to0$:
\begin{equation}
\cC_k(\tau)
=\frac{16}{k\,\tau^3}
+\frac{k^2-4}{3k\,\tau}
+\OO(\tau)\,.
\la{169}
\end{equation}
Then the   leading small-$\tau$ divergence in \rf{158}  is  
  cubic, $\int_\delta \dd\tau\,( k^{-1} \tau^{-4})\sim k^{-1} \delta^{-3}$.

This  reminds of  the local cubic $\int R^4$-type UV divergence  of 11d  supergravity, i.e. $\Lambda^3 I_8\sim{1\ov k} (\Lambda L)^3$ (see Appendix \ref{241})  where  $\Lambda= \varepsilon^{-1/2} $ is  related to the proper-time cutoff  $t > \varepsilon$. 
 The proper time  $t$  in \rf{69}   multiplies eigenvalues of the full 11d  operator and 
 the cutoff suppresses the contributions of  large eigenvalues. 
 The latter  are proportional to   $L^{-2}$,   where $L$ is  the scale  of \adssz in \rf{31},
 so that the  dimensionless cutoff parameter is  actually $\Lambda L$. 
 
 The   parameter  $\tau$ in the  character integral representation \rf{84},\rf{158}    multiplies 
  instead  the eigenvalue of the  dilatation 
  operator or  the global AdS  energy $E$ of a given representation 
    so that the cutoff $\tau > \delta$ suppresses its large values.  Since $E$ scales as $L^{-1}$ 
    the  dimensionless cutoff  is $ \delta^{-1} L$. 
    
    The two  cubic divergences  appearing in the two different integral representations of the same 
    1-loop  free energy should thus be related.      To   match them   to  the ${\rm L}^3$ term 
    in the ABJM free energy \rf{01}  one may assume that $\varepsilon^{1/2} \sim \delta \sim \lpp $
    (see   \ci{Beccaria:2023hhi}). 
    
     To  discard these 
 power divergences in both $t$-  and $\tau$- integral representations of $F'_1$ one may apply  the 
 corresponding zeta-function regularizations. 
Explicitly,  to regularize the integral in \rf{158} one   may  use the (Mellin)
$\zeta$-function prescription  defining the  finite part of $F_1'$  as 
\begin{align}\la{171}
F_1'=-\half\zeta'_{\cI_k}(0)\,,
\qquad\qquad
\zeta_{\cI_k}(z)
=\frac{1}{\Gamma(z)}
\int_0^\infty\dd\tau\,\tau^{z-1}\, \cI_k(\tau)\, .
\end{align}
The Mellin integral  converges for $\operatorname{Re}z>3$ and is defined near $z=0$ by an analytic continuation. 

Evaluating \rf{171}   for the $k$-dependent term $\cC_k$ in \rf{161},\rf{163}
 gives the standard integral representation of the function $A(k)$ in \rf{4}
 (as reviewed in \ci{Beccaria:2023hhi}  following \ci{Hanada:2012si,Hatsuda:2014vsa}) 
\be
A(k)
=\half\zeta'_{\cC_k}(0)
=\tfrac{2}{\pi^2k}\zeta(3) \big(1-\tfrac{k^3}{16}\big)
+\tfrac{k^2}{\pi^2}\int_0^\infty\dd x\,\frac{x\,\log(1-e^{-2x})}{e^{kx}-1}\,.
\la{182}
\ee
Here $\zeta(z)$ is the Riemann zeta function. Thus $\cC_k$ contributes $-A(k)$ to $F'_1$ in \rf{171}.\foot{The same 
 character appeared  in
\cite{BenettiGenolini:2026kjy},  ${\mathcal K}_k=\half \mathcal C_k$, see  \rf{25}. 
We have 
$
\mathcal C_k(\tau)
=4\sum_{m=1}^{\infty}m e^{-m\tau}
+8\sum_{m,r=1}^{\infty}m e^{-(m+kr/2)\tau}, 
$
so that for  $\operatorname{Re}z>3$, term-wise Mellin transform
in  \rf{171} then gives
$
\zeta_{\mathcal C_k}(z)
=4\zeta(z-1)
 +8\sum_{m,r=1}^{\infty}\frac{m}{(m+kr/2)^z}
=4\zeta(z-1)
 +8\zeta_3\big(z,1+\tfrac{k}{2}\mid1,1,\tfrac{k}{2}\big),
$
where $\zeta_3$ is the Barnes triple zeta function.
The Barnes expression is thus a compact rewriting of the same Mellin
transform, with the analytic continuation to $z=0$ understood.}
 Applying the same prescription to the $k$-independent term $\cR$ gives
\begin{align}
\zeta_{\cR}(z)
=2^{-z}+3^{-z}-4\zeta(z)\,,
\qquad\qquad
-\half\zeta'_{\cR}(0)
=\log\tfrac{\sqrt6}{2\pi}\,.
\la{188}
\end{align}
Finally, combining \rf{182} and \rf{188}, with the singleton contribution \rf{165} included for $k=1$, we get
\begin{equation}
F_{1}'=-A(k)+\log\tfrac{\sqrt6}{2\pi}-2\log(\half\mu L)\, ,
\la{189}
\end{equation}
where, in line with the discussion in  Section~\ref{28} (see  \rf{36},\rf{38}   and \rf{53}), we restored the  radius 
$L$  dependence with
 $n_{0} = -2n_{2}=-2$ and 
 introduced the $\AdS_4$ energy scale $\mu$.\foot{\la{987}One may expect that  applying our   character   computation 
  procedure  and the $\zeta$-function  regularization \rf{171}  just to the KK part of the integrand  in \rf{138}
  one  would reproduce  the large-$k$ behaviour of   \ci{Liu:2016dau} in \rf{13}.
   One can check that  this is indeed the case  (we thank M. Beccaria 
  for pointing this out).  This shows that  our procedure  and the level-sum regularization 
  used  in \ci{Liu:2016dau}   agree on the KK sector.} 

As already discussed in Section \ref{44}, ignoring the constant scalar  $S^7$ mode in \rf{59}, 
 the only  non-trivial zero modes  in the 1-loop partition function 
 originate from  the special ``harmonic''   2-form ghost modes  of the structure 
 $\omega^{(l)}\otimes1$  where   $\omega^{(l)}$ are harmonic 2-forms on $\Hf$ \ci{Bhattacharyya:2012ye}. The integral over them, performed with the same ultralocal measure that defines the
$\mu$-normalized determinants, cancels the scale dependence $-2\log(\mu L)$ of the non-zero modes in
\rf{189}, as  must happen  since ${\rm b}_{11}=0$ so that  the total 1-loop free energy has no
logarithmic UV  divergence. Also, the zero modes are normalized with respect to the
11d measure, in which  $\lpp $ entering the action in \rf{30} through $\kappa_{11}$, 
and the volume of the internal space $\X= S^7/\ZZ_k$  appear. 
This gives a dependence on $L/\lpp $, and the result   is that the
total  log term is   $\frac32\log(L/\lpp )$, matching  the $\frac14\log N$ term in  \rf{7}  \ci{Bhattacharyya:2012ye}.

The $k$-dependence of the zero-mode factor in the  1-loop partition function 
 follows from the similar scaling argument. The $\AdS_4$ norm
of a 2-form is scale invariant, so the 11d norm of $\omega^{(l)}\otimes1$ is
proportional to the internal volume, given by (up to $L^7$ factor)
\begin{equation}
\|\omega^{(l)}\otimes1\|^2\  \sim \ \vol(\X)=\tfrac1k \vol(S^7)={  \tfrac{\pi^4}{ 3k}} \,.
\la{199}
\end{equation}


The constant can be fixed by normalizing the zero-mode measure as in the Maxwell case on the sphere \ci{Anninos:2020hfj,Law:2020cpj}.\foot{We thank Z. Sun for  suggesting this argument.} The canonically normalized 3-form  field in the 11d supergravity  action is $A_{3}=(2\kappa_{11}^2)^{-1/2}C_{3}$,
with gauge transformations $\delta A_3=\dd\Lambda_2$. The measure consistent with the definition of the
determinants as $\det(\mu^{-2}\cD)$ is induced by the metrics
\begin{equation}
\dd s^2_{A}=\frac{\mu^2}{2\pi}\langle\delta A_3, \delta A_3\rangle\,,\qquad\qquad
\dd s^2_{\Lambda}=\frac{\mu^4}{2\pi}\langle\delta\Lambda_2,\delta\Lambda_2\rangle \,,
\la{301}
\end{equation}
where the invariant norm of a $p$-form is defined in \eqref{45} and the powers of $\mu$ are fixed by dimensions.

Let us consider an M2-brane with the worldvolume $\Sigma = \mathrm{H}^2\times S^1$ 
(where $\mathrm{H}^2$ is Euclidean $\AdS_2$ space)  embedded in \adssz   as in the case of 
 the dual of 
     $\ha$-BPS Wilson loop in the ABJM theory (see, for instance, \cite{Giombi:2023vzu}). 
 The M2-brane  couples to 
  the 3-form field, so its partition function contains the factor 
\begin{equation}
e^{iT_2\int_{\Sigma} C_3}=e^{iq\int_{\Sigma} A_3} \ , \qquad \qquad  q=\sqrt{2\kappa_{11}^2}\ T_2=(2\pi)^2\lpp^{3/2}\,.
\la{302}
\end{equation}
Consider gauge parameters $\Lambda_2=\omega\,\phi$, giving
$\delta A_3=\omega\wedge\dd_7\phi$, where $\omega$ is a
square-integrable harmonic 2-form on $\Hf$. 
Let us assume that $\omega$ is normalised so that $\int_{\mathrm{H}^2}\omega=1$.
\foot{In the ball coordinates of footnote~\ref{100}, for instance,
one may take
$
\omega=\frac{1}{\pi}
\big(\dd x^1\wedge\dd x^2+\dd x^3\wedge\dd x^4\big).
$
This form has $\|\omega\|^2=1$ and unit integral over the
$\mathrm{H}^2$ disc defined by $x^3=x^4=0$.}

Let us  use a compact gauge-parameter normalization in which
$\phi$ has period $2\pi/q$, set by the M2-brane charge.
For its normalized constant mode,
$\phi=\alpha_0/\sqrt{V_7}$, this gives
\begin{align}
\alpha_0\sim\alpha_0+{2\pi}{q}^{-1}\sqrt{V_7}\,.
\end{align}
Here $V_7=\vol(S^7/\ZZ_k)$ includes the radius dependence. 
Extending this normalization to the  harmonic sector,
we can formally define its effective  gauge-group $G_0$  volume by
\begin{equation}
\vol(G_0)=\big({\mu^2\sqrt{2\pi V_7}}\, q^{-1} \big)^{n_2}\,,
\la{304}
\end{equation}
where $n_{2}$ is the number of the harmonic 2-forms.  For $\Hf$ we assume that this number is replaced by its regularized value $n_2=1$ in \rf{53}, as for the other
contributions proportional to the volume of the hyperbolic space.

\iffa 
Consider the $A_{3}$ gauge transformations given by $\omega \wedge \dd_7\phi$ where $\omega$ is the harmonic 2-form on $\rm{H}^4$ normalised so that $\int_{\rm{H}^2}\omega=1$.\foot{As a 
simplest example one can consider (see footnote \ref{100}) 
 $\omega = \tfrac{1}{\pi}(dx^{1}\wedge dx^{2} + dx^{3}\wedge dx^{4}) $ so
  that $\int_{\mathrm{H}^2}\omega =1$.} 
  For the normalised constant gauge transformations $\phi =\alpha_{0}/\sqrt{V_7}$, the gauge-invariance of the M2-brane requires ($V_7= \vol(S^7/\mathbb Z_k)$)
\begin{align}
    \alpha_{0} \sim \alpha_{0}+{2\pi}{q}^{-1} \sqrt{V_{7}} \, .
\end{align}
Equivalently, with the canonical measure normalisation, the volume of the effective gauge group $G_0$  becomes
\begin{equation}
\vol(G_0)=\big({\mu^2\sqrt{2\pi V_7}}\, q^{-1} \big)^{n_2}\,,
\la{304}
\end{equation}
where $n_{2}$ is the number of the harmonic 2-forms.  For $\Hf$ we assume that this number is replaced by its regularized value $n_2=1$ in \rf{53}, as for the other
contributions proportional to the volume of the hyperbolic space.
\fi

 With
$V_7= \vol(S^7/\mathbb Z_k)=\frac{\pi^4}{3k}L^7$ we then get for the zero-mode contribution to the 1-loop free energy
\al{
F_{\rm zm} &=   \log\vol(G_0) = n_2\log\big({\mu^2\sqrt{2\pi V_7}}\, {q}^{-1}\big)\no \\
&=2\log(\mu L)+\tfrac32\log(\lpp^{-1}L)-\half\log k+\half\log\tfrac{\pi}{24}\,.
\la{200}
}
Adding \rf{189} and \rf{200}, the dependence on $\mu$ cancels, and the 1-loop free energy of 11d
supergravity on $\AdS_4\times S^7/\ZZ_k$ is  for all $k\ge1$  (with  the singleton contribution included for
$k=1$)
\begin{equation}
F_{1}=\tfrac32\log(\lpp^{-1}L)-A(k)-\half\log k-\half\log\pi\,.
\la{201}
\end{equation}
This agrees with the ABJM localization expression in \rf{00}, including the $k$-independent constant.

\iffa 
If the
zero-mode measure depends on $(L,k)$ only through this norm and the  coefficient in the action or $\lpp$, as in
\ci{Bhattacharyya:2012ye}, 
then
the zero-mode  contribution to the 1-loop free energy is 
\begin{equation}
F_{\rm zm}=2\log(\mu L)+\tfrac32\log({\ell^{-1}_p}L)-\half\log k+c_{\rm zm}\,.
\la{200}
\end{equation}
Here the term
$-\tfrac12\log k$ is the $k$-dependent part of  $\frac{1}{2}n_2\log\vol(\X)$ with $n_2=1$ as in \rf{53}.
$c_{\rm zm}$ is a $k$-independent constant, which is determined below by the normalization of the zero-mode
measure.
\

Adding \rf{189} and \rf{200} together, the dependence on $\mu$ cancels out, and  we conclude that the 1-loop
free energy of 11d supergravity on $\AdS_4\times S^7/\ZZ_k$ is given,  for
all $k\ge1$  (with the singleton contribution  included for $k=1$)  by 
\begin{equation}
F_{1}=\tfrac32\log({\ell^{-1}_p}L) -A(k)-\half\log k+c_0\,,\qquad\qquad \ \ \ 
c_0=\log\tfrac{2\sqrt6}{\pi}+c_{\rm zm}\,.
\la{201}
\end{equation}
This has  the same $k$-dependence as  the ABJM localization  expression   in \rf{00}, including the $-\tfrac12\log k$ term. It
agrees with it   exactly if 
\begin{equation}
c_0=-\half\log\pi\,.
\la{202}
\end{equation}
\fi

\section{Concluding remarks}\la{203}

The matching between the 1-loop 11d supergravity   free energy   on \adssz 
and  the constant $A(k)$   in the $N^0$  part  in free energy of the dual ABJM theory demonstrated in this paper
is an example of a  precision test of holography. 

It is more  non-trivial  than similar tests    in  odd-dimensional AdS  case     where 
 supergravity 1-loop correction   matches  $N^0$ term   in the Weyl anomaly  coefficient in dual  gauge theory 
 (see \ci{Beccaria:2014xda}  for $\AdS_5 \times S^5$,   \ci{Beccaria:2014qea}  for 
 $\AdS_7 \times S^4$ and  also \ci{ArabiArdehali:2013jiu,Liu:2017ruz}   for some less supersymmetric  cases)
 as   here  not just a number but  the   whole    function of the parameter $k$ is   matched.  
 
The structure of the integrand $\cC_k(\tau)$ in \rf{163}, which yields $A(k)$ through \rf{171},\rf{182}, was
anticipated in \ci{Beccaria:2023hhi}, where an integral representation of $A(k)$ involving a $\sinh^{-2}$
factor was discussed. Its similarity to the Schwinger-type ``proper-time'' integrals appearing in the
M-theory derivation of topological string amplitudes \ci{Gopakumar:1998jq,Dedushenko:2014nya} was noted
there.\foot{In that derivation the factor
$\sinh^{-2}$ is the proper-time propagator of a BPS particle in the constant self-dual graviphoton background,
and the sum over the momenta along the M-theory circle, i.e. over the D0-brane charges, produces the
contributions of all genera, the constant-map part having the same large-$k$ structure as $A(k)$ with $k$
in the role of $1/g_s$.
}
The computation in the present  paper gives this factor a definite interpretation
(although here $\tau$ is conjugate
to the global $\AdS_4$ energy rather than a proper time).  Writing
$\cC_k=\sinh^{-2}(\frac12\tau) \coth(\frac14k\tau) = \sinh^{-2}(\frac12\tau)\big[1+2\sum_{r\ge1}e^{-\frac12 kr\tau}\big]$, the factor $\sinh^{-2}(\frac12\tau)$ is common to
all Hopf-charge sectors, and is by itself the contribution of the neutral modes to $\cC_k$, while
$\coth(\frac14k\tau)$ is the sum over the Hopf charges $kr$ retained by the $\ZZ_k$ projection.

\iffa
The approach of the present work can also be applied to other internal spaces $M_7$. In the 1-loop
calculation of Section~\ref{61} we have relied on the organization of the KK tower into $OSp(8|4)$
multiplets. The contribution of each conformal multiplet within a KK supermultiplet is determined by its
$so(3,2)$ character, as in Section~\ref{68}, while the $so(8)$ character supplies the multiplicities and the
charges needed to implement the quotient. For a supersymmetric Freund--Rubin background $\AdS_4\times M_7$
with two real Killing spinors on $M_7$, for example a Sasaki--Einstein $M_7$, the KK tower is organized into
$OSp(2|4)$ multiplets, and the role of the $so(8)$ character is played by the characters
of the isometry group of $M_7$, which
includes the $SO(2)_R$ R-symmetry generated by the Reeb vector.

 If this multiplet spectrum, including the energies and multiplicities,
is known, the KK contribution can be computed as in Section~\ref{61}, with the sum running over the conformal
multiplets within the $OSp(2|4)$ supermultiplets.\foot{We thank J. van Muiden for suggesting this
possibility. Explicit $OSp(2|4)$ KK spectra for $M^{111}$ and $Q^{111}$ are discussed in
\cite{Fabbri:1999mk,Merlatti:2000ed}. See also \cite{Eager:2013mua} for the short-multiplet spectrum on
Sasaki--Einstein 7-manifolds.} In contrast to the  case of $S^7$, the energies of the long multiplets are,  in general, 
irrational functions of the quantum numbers, so the sum over the tower is not expected to reduce to a closed
form, but the integral representation can still be evaluated numerically.

The harmonic sector  contribution has again  to be added to the KK contribution. Since $M_7$ has positive Ricci curvature, its
first Betti number vanishes, and the harmonic sector is now given  by the determinants  \rf{19}  of the 
Laplacians  on $M_7$, with $n_2=1$. To implement a finite-group quotient as in
Section~\ref{155} one must also know the action of the quotient group on the KK modes. For a cyclic quotient group 
contained in a $U(1)$ isometry, this requires retaining the corresponding fugacity, as for the M-theory circle
action $U(1)_m$ in the $S^7/\ZZ_k$ case.   If the quotient has fixed loci, the 1-loop partition function  should 
also include the contributions of the modes localized there, as in the ADHM example of Appendix~\ref{258}.
A comparison with the dual field theory requires the knowledge of the corresponding constant in the Airy-function partition
function, which is   available  for some $\cN=2$ examples.
\fi

The approach of the present work can also be applied to other
smooth compact internal spaces $M_7$. In the 1-loop calculation
of Section~\ref{61}, we used the organization of the $S^7$ KK
tower into $OSp(8|4)$ multiplets. The contribution of each
conformal multiplet is determined by its $so(3,2)$ character,
as in Section~\ref{68}, while the $so(8)$ characters encode
the internal multiplicities and the charges needed to
implement a quotient.
For a Freund--Rubin background $\AdS_4\times M_7$ preserving
$\cN=2$ supersymmetry, as for a generic Sasaki--Einstein
$M_7$ with two real Killing spinors, the KK spectrum is
instead organized into $OSp(2|4)$ multiplets.
The $so(8)$ characters are replaced by the characters of
the isometry representations carried by the internal
modes, with the Reeb vector determining their R-charges.

If the full multiplet spectrum, including the energies
and multiplicities, is known, the KK contribution can be
computed as in Section~\ref{61}, with the sum running over
the conformal multiplets within the $OSp(2|4)$
supermultiplets.\foot{We thank J. van Muiden for suggesting
this possibility. Explicit $OSp(2|4)$ KK spectra for
$M^{111}$ and $Q^{111}$ are discussed in
\cite{Fabbri:1999mk,Merlatti:2000ed}.
See also \cite{Eager:2013mua} for the short-multiplet
spectrum on Sasaki--Einstein 7-manifolds.}
Unlike the $S^7$ case, the energies of the long multiplets
are generally nonlinear functions of the internal quantum
numbers and are generically irrational.
The sum over the KK tower need not admit a closed form,
but numerical evaluation remains possible once the
spectrum and the regularization prescription are specified.

With the same spectral prescription, the contribution
built from harmonic 2-forms on Euclidean $\AdS_4$ must
again be added to the KK contribution.
For a smooth compact $M_7$ with positive Ricci curvature,
the Bochner theorem gives $b_1(M_7)=0$.
The contribution therefore retains the determinant form
\rf{19}, with the internal Laplacians now acting on $M_7$
and with the same regularized multiplicity $n_2=1$.
To implement a supersymmetry-preserving finite-group
quotient as in Section~\ref{155}, one must also know
the action of the quotient group on the KK modes.
For a cyclic subgroup of a $U(1)$ isometry, this requires
retaining the corresponding fugacity, as for the
M-theory circle action $U(1)_m$ in the $S^7/\ZZ_k$ case.
If the quotient has fixed loci, one must also include
any additional degrees of freedom supported at the
orbifold singularities, as in the ADHM example of
Appendix~\ref{258}.\foot{
Refined characters  with extra  fugacities 
may give access to the squashed-sphere and real-mass deformations treated within the relative
equivariant localization approach
 in
\ci{BenettiGenolini:2026kjy}.}

A comparison with the dual field theory also requires
an independent determination of the coefficient $A$
in its perturbative Airy-function partition function.
Partial field-theory results are available for several
$\cN=2$ examples \cite{Geukens:2024zmt,Bobev:2025ltz}.

\section*{Acknowledgements}
We thank M. Beccaria, J. Gauntlett,  J. van Muiden and Z. Sun for   helpful  comments  on the draft. 
We also   acknowledge   K. Mkrtchyan  and 
J. Park    for useful discussions of related questions.  
SAK is  supported  by   the President's PhD Scholarship at Imperial College London.
 AAT is   supported by the STFC Consolidated Grant    ST/X000575/1 
 and  STFC Consolidated Award UKRI4230: Abdus Salam Centre for Theoretical Physics at Imperial College London.
 Some of the computations of this paper  were assisted by AI models. 

\iffa 
 Introducing the  second  internal-space fugacity  
  the construction applies to any $\ZZ_k\subset U(1)^2$
quotient of $S^7$, with the modes localized at fixed points added, as for the ADHM quotient in
Appendix
\ref{258}. For other Sasaki--Einstein spaces $X_7$ with vanishing first Betti number the $N^0$ term is fixed
by the KK spectrum of $X_7$ and the harmonic sector \rf{59}, and the $\cN=2$ examples with Airy-function
partition functions would provide further tests. 
\fi

\appendix
\section{KK spectrum in terms of  $OSp(8|4)$ supermultiplets}\la{206}

The linearised fluctuations of 11d supergravity near $\AdS_4\times S^7$
 can be organized into unitary
representations of $OSp(8|4)$ \ci{Freedman:1983na,Gunaydin:1985tc}.  Below we follow the conventions of
\ci{Dolan:2008vc} (see also \ci{Cordova:2016emh}).

The bosonic subalgebra is $so(3,2)\oplus so(8)$.
The conformal algebra $so(3,2)$ is generated by the dilatation $D$, the translations $P_{\alpha\beta}$,
the special conformal transformations $K^{\alpha\beta}$ and the Lorentz generators $M_\alpha{}^\beta$
($\alpha,\beta=1,2$ are 3d spinor indices). Its maximal compact subalgebra
$so(2)\oplus so(3)$ is generated by $D$ and by the spin
generators $J_3$ and $J_\pm$ representing $M_\alpha{}^\beta$. The $R$-symmetry $so(8)$ is generated by
$R_{rs}=-R_{sr}$ $(r,s=1,\dots,8)$. The Cartan generators $R_a\equiv R_{2a-1,2a}$  with  $a=1,\dots,4$
generate rotations in four orthogonal planes of $\RR^8$.

The odd generators are the supercharges
$Q_{r\alpha}$ (a vector of $so(8)$, with dimension $\tfrac12$ and spin $\tfrac12$) and the superconformal charges
$S_r{}^\alpha$, with $\{Q,Q\}\sim P$, $\{S,S\}\sim K$ and $\{Q,S\}\sim M+D+R$. It is convenient to
combine them into $Q_{a\alpha}=\frac{1}{\sqrt2}(Q_{2a-1,\alpha}+iQ_{2a,\alpha})$ and their conjugates
$\bar Q_{a\alpha}$ (which carry $R_a$ charge $\pm1$), and similarly for $S$-generators.

A unitary multiplet $\cM$
is generated by a highest-weight state annihilated by $K$, $S$, $J_+$ and the $so(8)$ raising operators.
It decomposes into a finite sum of conformal multiplets,  
each generated from a conformal primary
(a state annihilated by $K$) by the translations $P$. A conformal primary is labelled by its energy
$E$ (the eigenvalue of $D$, which has the interpretation of global energy in AdS$_4$),
its spin $j\in\tfrac12\ZZ_{\ge0}$ (the eigenvalue of $J_3$ for the highest state of
the spin multiplet) and its $so(8)$ weights $r_a\in\tfrac12\ZZ$ (the $R_a$ eigenvalues of the highest
weight), $r_1\ge r_2\ge r_3\ge|r_4|$. Equivalently, it carries the Dynkin labels
\begin{equation}
[a_1,a_2,a_3,a_4]=[r_1-r_2,\,r_2-r_3,\,r_3+r_4,\,r_3-r_4]\,.
\la{207}
\end{equation}
We use the notation $\cm{j}{r_{1}, r_{2}, r_{3}, r_{4}}{E}$ for the conformal multiplet with these labels
(i.e. we  use $R_{a}$ eigenvalues to denote $so(8)$ representations).

With this notation, the three eight-dimensional representations are
$8_v=(1,0,0,0)$,
$8_c=(\tfrac12,\tfrac12,\tfrac12,\tfrac12)$,
$8_s=(\tfrac12,\tfrac12,\tfrac12,-\tfrac12)$.
Like in
\ci{Gunaydin:1985tc,Dolan:2008vc} we assume that the supercharges
are in $8_v$, while the coordinates of $S^7$ transform in $8_s$.

The KK spectrum \ci{Gunaydin:1985tc} consists of one $\frac12$-BPS multiplet for each $n\ge2$,
\begin{equation}
\cM=\bigoplus_{n\ge2}\cM_n\,,\qquad\qquad  \cM_n=B_{1}\cm{0}{n/2,\, n/2,\, n/2,\, -n/2}{n/2}\,.
\la{208}
\end{equation}
The superconformal primary is a scalar of dimension $E=n/2$ in the $n$-th symmetric traceless
power of $8_s$, i.e. the $n$-th scalar harmonic on $S^7$.
In the notation of \ci{Dolan:2008vc} these $\cM_n$
are the $(4,B,-)$ representations with $r_1=r_2=r_3=-r_4=n/2$. For $n\ge4$ each $\cM_n$ is the sum of
the 15 conformal multiplets
\begin{equation}
\cM_n=\bigoplus_{0\le b_1\le b_2\le b_3\le b_4\le2}
\Big[\half\Big(2-\half\sum_{i=1}^4(-1)^{b_i}\Big)\Big]^{(n/2-b_1,\,n/2-b_2,\,n/2-b_3,\,-n/2+b_4)}_{(n+b_1+b_2+b_3+b_4)/2}\,.
\la{209}
\end{equation}
Here $b_a\in\{0,1,2\}$ is the number of components of the $a$-th supercharge
acting on the superconformal primary to produce the conformal primary. For the highest
weight the primary is annihilated by $Q_1,Q_2,Q_3$ and $\bar Q_4$ (Table 1 of \ci{Dolan:2008vc}), so the
active supercharges are $\bar Q_1,\bar Q_2,\bar Q_3$ and $Q_4$.

Each supercharge raises the energy by $\frac12$ and shifts the $so(8)$ weight by one unit, $r_a\to r_a-1$
for $a=1,2,3$ and $r_4\to r_4+1$. The spin of the resulting conformal multiplet is determined by the
number of unpaired supercharges: for a given $a$ the two components $Q_{a\alpha}$, $\alpha=1,2$, applied
together ($b_a=2$) form a spin singlet, while a single one ($b_a=1$) carries a free spinor index. The
conformal multiplet labelled by $b=(b_1,\dots,b_4)$ therefore has
\be \la{210}
E_{b}=\tfrac n2+\tfrac12\sum_{a=1}^4 b_a\, , \qquad \qquad
j_{b}=\tfrac12\,n_{\rm odd}(b)=\tfrac12\big[2-\tfrac12\sum_{a=1}^4(-1)^{b_a}\big]\,  ,
\ee
where $n_{\rm odd}(b)$ is the number of odd entries of $b$.
The
$so(8)$ weights are
\be \la{109}
\mathrm{r}_b =\te  (\frac n2-b_1,\frac n2-b_2,\frac n2-b_3,-\frac n2+b_4)\, ,
\ee which the Weyl group orders as
$b_1\le b_2\le b_3\le b_4$.

The resulting multiplets are listed in Table \ref{211}. They are the fifteen KK
towers of \ci{Biran:1983iy,Nilsson:1984bj}.\footnote{The labels there are related to the present ones by the triality transformation exchanging $a_1$ and
$a_4$, and by the shift of $n$ by two.}
For $n=3$ the terms with $b_3=b_4=2$ are absent and $\cM_3$ has
nine conformal multiplets. For $n=2$ the massless fields with $j\ge1$ correspond to conserved-current multiplets:
$\cM_2$ is the massless graviton multiplet (graviton, $8_v$ gravitini, $28$
gauge fields, $56_v$ spinors, $35_c\oplus35_s$ scalars), whose conformal multiplets with $j\ge1$ are shortened by the
conservation conditions. For $n=2$ the $35_s$ scalars have $E=1$, see Section \ref{68}.
The multiplet with $n=1$ is the $\cN=8$ singleton (one Rac in $8_s$ and one Di  in $8_c$ of 
\ci{Flato:1978qz}), which does not correspond to propagating bulk fields and is
absent from the KK spectrum. $n=0$ is the identity.
\begin{table}[h]
\centering
\begin{tabular}{cccc}
\toprule
spin $j$ & energy $E$ & $so(8)$ labels $[a_1,a_2,a_3,a_4]$ & $(b_1,b_2,b_3,b_4)$\\
\midrule
$2$ & $n/2+2$ & $[0,0,0,n-2]$ & $(1,1,1,1)$\\
$3/2$ & $n/2+3/2$ & $[1,0,0,n-2]$ & $(0,1,1,1)$\\
$3/2$ & $n/2+5/2$ & $[0,0,1,n-3]$ & $(1,1,1,2)$\\
$1$ & $n/2+1$ & $[0,1,0,n-2]$ & $(0,0,1,1)$\\
$1$ & $n/2+2$ & $[1,0,1,n-3]$ & $(0,1,1,2)$\\
$1$ & $n/2+3$ & $[0,1,0,n-4]$ & $(1,1,2,2)$\\
$1/2$ & $n/2+1/2$ & $[0,0,1,n-1]$ & $(0,0,0,1)$\\
$1/2$ & $n/2+3/2$ & $[0,1,1,n-3]$ & $(0,0,1,2)$\\
$1/2$ & $n/2+5/2$ & $[1,1,0,n-4]$ & $(0,1,2,2)$\\
$1/2$ & $n/2+7/2$ & $[1,0,0,n-4]$ & $(1,2,2,2)$\\
$0$ & $n/2$ & $[0,0,0,n]$ & $(0,0,0,0)$\\
$0$ & $n/2+1$ & $[0,0,2,n-2]$ & $(0,0,0,2)$\\
$0$ & $n/2+2$ & $[0,2,0,n-4]$ & $(0,0,2,2)$\\
$0$ & $n/2+3$ & $[2,0,0,n-4]$ & $(0,2,2,2)$\\
$0$ & $n/2+4$ & $[0,0,0,n-4]$ & $(2,2,2,2)$\\
\bottomrule
\end{tabular}
\caption{\small The conformal multiplets in $\cM_n$, $n\ge4$, in \rf{209}. The last column gives the numbers
$b_a$ of components of the supercharges ($\bar Q_1,\bar Q_2,\bar Q_3,Q_4$) acting on the
superconformal primary. Representations with a negative
label are absent for $n=2,3$.}
\la{211}
\end{table}

The $U(1)$ subgroup corresponding to the Hopf $S^1$ fiber of $S^7$ of Section \ref{28} (i.e. the M-theory circle
along which the reduction to type IIA supergravity is performed) will be denoted
by $U(1)_m$. It is generated
by the Cartan element
\begin{equation}
R_m=2R_2\,,
\la{212}
\end{equation}
so that on the highest weight state of a representation with Dynkin labels $[a_1,a_2,a_3,a_4]$ the
corresponding charge is
\begin{equation}
q_m=2r_2 = 2a_2+a_3+a_4\,.
\la{213}
\end{equation}
The weights of the $S^7$ coordinates (in the $8_s$ representation) have $q_m=\pm1$, so
that the $\ZZ_k\subset U(1)_m$ action is
free. For the supercharges, which are in the $8_v$ representation, the charges are $q_m=0$ for six of the eight components
and
$q_m=\pm2$ for the remaining two ($Q_2$ and $\bar Q_2$ in the notation of \ci{Dolan:2008vc}).

The
commutant of the algebra $u(1)_m$ in $so(8)$ is $su(4)\oplus u(1)_m$, with $su(4)\simeq so(6)$, and under it one has
$8_v\to6_0\oplus1_{2}\oplus1_{-2}$.
For $k>2$ the two charged supercharges are projected out. The six
neutral ones, with their superconformal partners, give the $12+12$ odd generators of $osp(6|4)$, corresponding to the
$\cN=6$ superconformal algebra of the ABJM theory with $SO(6)$ R-symmetry. The generator $R_m$ commutes with
$osp(6|4)$, so the symmetry that survives the quotient is $osp(6|4)\oplus u(1)_m$, where $u(1)_m$ acts as
a baryonic symmetry (the topological $U(1)$ of ABJM).
For $k=1,2$ all supercharges
survive.

Another  special choice 
of an abelian $SO(8)$ subgroup  is the $U(1)_{\widetilde m}$  
generated by
\be\la{115}
 \widetilde{R}_m \equiv  R_{1}-R_{2} \, . 
\ee
On the highest-weight state of a representation with Dynkin labels
$[a_1,a_2,a_3,a_4]$, the corresponding charge is  
\be
    \widetilde{q}_{m} =r_1-r_2 =  a_{1} \, .\la{887}
\ee
In this case, four of the eight weights of the coordinate representation $8_s$
have $\widetilde{q}_m=0$, while the remaining four weights have charges $\widetilde{q}_m=\pm1$ (two of each).

 Thus the
$U(1)_{\widetilde m}$ action has a fixed locus $S^3\subset S^7$.
For the supercharges, four of the eight $so(8)$ components are neutral,
while the remaining four ($Q_{1,2}$ and $\bar Q_{1,2}$ in the notation
of \ci{Dolan:2008vc}) have charges $\widetilde{q}_m=\pm1$.
The quotient by $\mathbb{Z}_{p}\subset U(1)_{\widetilde m}$
gives the ADHM orbifold (see 
Appendix \ref{258}).

\section{Characters of superconformal multiplets}\la{235}

\iffa This Appendix recalls the closed form of the characters of the $\frac12$-BPS multiplets $\cM_n$ from
\ci{Dolan:2008vc}, which provides an independent derivation of the KK characters of Section \ref{68}. We use the variable $s=q^{1/2}$ of \ci{Dolan:2008vc} and the fugacities
$y=(y_1,\dots,y_4)$ that correspond to the Cartan generators $R_1,\dots,R_4$ in Appendix \ref{206}.
The fugacities in the main text correspond
to $y=(u^2,v^2,1,1)$. 
\fi

This Appendix recalls the closed form of the characters of the $\frac12$-BPS multiplets $\cM_n$ from \ci{Dolan:2008vc}.

We use $q=s^2$, where $s$ is the fugacity for $2E$ in \ci{Dolan:2008vc}.
The spin fugacity $x$ corresponds to $2J_3$ as in \rf{64}, and
$y=(y_1,\dots,y_4)$ are the fugacities for the Cartan generators
$R_1,\dots,R_4$. The main text uses a fugacity $v$ for the generator $R_m=2R_2$ in \rf{212}. This corresponds to the specialization $y=(1,v^2,1,1)$. We also introduce a fugacity $u$ for the generator $R_r\equiv 2R_1$ with $y_3=y_4=1$, so that retaining both charges gives $y=(u^2,v^2,1,1)$.

For a representation labelled by $\mathrm{r}=(r_1,\dots,r_4)$, the Weyl
formula for the $so(8)$ characters is \ci{Dolan:2008vc} 
\begin{equation}
\chi^{(\mathrm{r})}(y)=\frac{\det\big[y_i^{\,r_j+4-j}+y_i^{-r_j-4+j}\big]+\det\big[y_i^{\,r_j+4-j}-y_i^{-r_j-4+j}\big]}
{2\prod_{i<j}\big(y_i+y_i^{-1}-y_j-y_j^{-1}\big)}\,.
\la{236}
\end{equation}
It is valid also for non-dominant $\mathrm{r}$, where it gives the signed characters needed for the shortened
multiplets.\foot{A weight $\rm r$ is dominant if $r_1\ge r_2\ge r_3\ge|r_4|$, i.e. if all its Dynkin labels are non-negative.
For non-dominant $\rm r$ the expression \rf{236} either vanishes, when $\mathrm{r}+\rho$ ($\rho$ is the
Weyl vector)
lies on a Weyl wall   (a hyperplane in weight space that is fixed by one of the reflections generating the Weyl group), 
or equals $\pm$ the character of the representation with highest weight $w(\mathrm{r}+\rho)-\rho$,
(where the Weyl group element $w$ makes $\mathrm{r}+\rho$ dominant). These  characters implement the shortening of the multiplets with small $n$. Thus eq.  \rf{237} holds for
all $n\ge1$, including the shortened multiplets and the supersingleton multiplet. In particular, at $n=2$ it produces the conserved-current characters, i.e. the null-state subtractions  \ci{Dolan:2008vc}.} 


For the $\frac12$-BPS multiplet $\cM_n$ with superconformal primary in $[0,0,0,n]$,
in the notation of \rf{209}, the superconformal character can be expressed as
\begin{align}
\chi_{_{\cM_n}}(q,x,u,v) &= \tr_{_{\mathcal{M}_n}}(q^{D}x^{2J_{3}}u^{2R_{1}}v^{2R_{2}}) =\sum_{0\le b_1\le b_2\le b_3\le b_4\le2}
\cC_{E_b,j_b}(q,x)\,\chi^{(\mathrm{r}_b)}(u^2,v^2,1,1)\,,
\la{237}
\end{align}
where the conformal dimension $E_b$, spin $j_b$ and $so(8)$ R-charges $\mathrm{r}_{b}$ are given in \rf{210}, \rf{109}. The $so(3,2)$ character $\cC_{E,j}(q,x)$ can be expressed in terms of the spin-$j$ character $\chi_{2j}(x)$ as
\al{
&\cC_{E,j}(q,x)=\tr(q^{D}x^{2J_{3}}) = q^{E}\, \chi_{2j}(x)\, P(q, x) 
 \ , \qquad \qquad 
\chi_{2j}(x)=\sum_{m=-j}^{j}x^{2m}
=\tfrac{x^{2j+1}-x^{-2j-1}}{x-x^{-1}}\,, \la{332}\\
& \qquad  P(q, x)=\tfrac{1}{(1-q)(1-qx^2)(1-qx^{-2})} \ , \la{2231}
}
where $P(q,x)$ is the generating function of the conformal descendants, i.e. the character of the action of the three momentum generators on a conformal primary.
For $n=1$ \rf{237} reduces to the $\mathcal{N}=8$
singleton character 
\al{
&\chi_{_{\cM_1}}(q,x,v)=\mathcal D_{\rm Rac}(q,x)\,\chi^{8_s}(v)+\mathcal D_{\rm Di}(q,x)\,\chi^{8_c}(v)\,,\la{099}\\
&\mathcal D_{\rm Rac}=q^{1/2}(1-q^2)\,P(q,x)\,,\qquad\qquad   \mathcal D_{\rm Di}=q\,(1-q)\, (x + x^{-1}) \,P(q,x)\,,
\la{139}
}
describing a free conformal scalar of dimension $\tfrac12$ in $8_s$ and a free spinor of dimension $1$ in $8_c$. 

The character of the bulk KK tower is
\begin{equation}\la{240}
\chi_{\rm KK}(q,x,u,v)
=\sum_{n=2}^{\infty}\chi_{_{\cM_n}}(q,x,u,v)\,.
\end{equation}
Since $\cC_{E,j}(q,x)=(-1)^{2j}\cC_{E,j}(q,-x)$, one can project \rf{240} onto its integer spin and
half-integer spin parts
\begin{equation}
\chi^\pm_{\rm KK}(q,x,u,v)
=\tfrac12\big[
\chi_{\rm KK}(q,x,u,v)
\pm\chi_{\rm KK}(q,-x,u,v)
\big]\,.
\la{5533}
\end{equation}
This allows us to express the bosonic and fermionic characters entering the
differential operators in \rf{107} in terms of the superconformal character.

The two $U(1)$ generator choices \rf{212} and \rf{115}  in Appendix \ref{206} correspond to different
specializations of the fugacities. For the ABJM quotient, the character related to the KK spectrum on $S^7/\mathbb{Z}_{k}$ can be obtained by reducing to one fugacity as 
\begin{align}
\chi_{\rm KK}(q,x,v)=\lim_{u\to1}\chi_{\rm KK}(q,x,u,v)\,,
\end{align}
and computing the $\ZZ_k$  group average
\begin{align}
\chi_{\rm KK}^{(k)}(q,x)
=\frac{1}{k}\sum_{p=0}^{k-1}\chi_{\rm KK}(q,x,e^{2\pi i p/k})\,.
\end{align}
In the case  of  the ADHM  11d    circle action in \rf{115}
the projection on the $\mathbb{Z}_{p}$-invariant supergravity modes
  can be obtained  in a similar way.
  In this case, one should, however, consider also an 
  additional contribution of the
   modes localised on AdS$_4\times S^3$, as we discuss further in Appendix~\ref{258}.

\section{Harmonic 2-form  sector contribution to 11d partition function } 
\la{214}

Here   we derive the  contribution \rf{59}  
of  the harmonic 2-form sector \rf{57}  of   3-form  fluctuations  to  the 1-loop 11d supergravity  partition function. 

The starting point   will be  the 3-form   partition function \rf{33}  in  the case 
 when  the flux and Chern--Simons couplings to  other fields are  formally absent.  Such decoupling indeed  takes place as was shown  in  Section \ref{44}. 

The heat  kernel   for   Hodge Laplacians on $p$-forms  on 11d space $\Hf\times M_7$   will be denoted as 
\begin{equation}
K_p(t)=\Tr_{\Omega^p(\Hf\times M_7)}\,e^{-t\Delta_p}\,.
\la{215}
\end{equation}
Then  the   $C_3$ free energy corresponding to  \rf{33}    may be written as 
\begin{equation}
F_C=-\log Z_C= -\half\int_0^\infty\frac{\dd t}{t}\,K_C(t)\,,\qquad\qquad 
  K_C=K_3-2K_2+3K_1-4K_0\,.
\la{216}
\end{equation}
The $p$-forms on the product  space decompose into $a$-forms on $\Hf$ times $b$-forms on $M_7$ with
$a+b=p$, and the heat  kernels  factorize as 
\begin{equation}
K_p=\sum_{a+b=p}K_a^{(4)}\,K_b^{(7)}\,,\qquad\ \ \ 
K_a^{(4)}=\Tr_{\Omega^a(\Hf)}\, e^{-t\Delta_a}\,,\qquad\ \  K_b^{(7)}=\Tr_{\Omega^b(M_7)}\, e^{-t\Delta_b}\,.
\la{217}
\end{equation}
Each  space of forms  is decomposed into exact, co-exact  and harmonic  parts
\begin{equation}
\Omega^a=\dd\Omega^{a-1}\oplus\dd^\dagger\Omega^{a+1}\oplus\mathcal H^a\ . 
\la{218}
\end{equation}
 Let us introduce  the   notation  for the traces on $\Hf$ (for  the scalar trace, the trace over co-exact 1-forms, and the trace over the square-integrable 
 harmonic 2-forms)  as  
\begin{equation}
S(t)=\Tr_{\Omega^0(\Hf)}\, e^{-t\Delta_0}\,,\qquad
V(t)=\Tr_{\dd^\dagger\Omega^2(\Hf)}\, e^{-t\Delta_1}\,,\qquad
H_2(t)=\Tr_{\mathcal H^2(\Hf)}\, e^{-t\Delta_2} =n_2\,.
\la{219}
\end{equation}
$H_2$  is $t$-independent and equal to the regularized number $n_2$ in  \rf{53}.
There are no square-integrable  harmonic
forms of  degrees $0,1,3,4$, and the co-exact 2-forms are isospectral to the co-exact 1-forms (the
map $\ast\dd$ intertwines them on the non-zero spectrum), so
\begin{equation}
K_0^{(4)}=S\,,\qquad\ \  K_1^{(4)}=V+S\,,\qquad\ \  K_2^{(4)}=2V+H_2\,,\qquad\ \  K_3^{(4)}=V+S\,.
\la{220}
\end{equation}
On $M_7= S^7/\ZZ_k $ (which has no real  cohomology in degrees $p=1,2,3$)  
we define the trace over non-constant scalars and the trace over co-exact $p$-forms  as 
\begin{equation}
{\rms} (t)=\Tr'_{\Omega^0(M_7)}e^{-t\Delta_0}\,,\qquad\qquad 
{\rmv}_p (t)=\Tr_{\dd^\dagger\Omega^{p+1}(M_7)}e^{-t\Delta_p}\,.
\la{221}
\end{equation}
Then  in \rf{217} we have 
\begin{equation}
K_0^{(7)}=\rms +1\,,\qquad\ \  K_1^{(7)}=\rmv_1 +\rms \,,\qquad \ \ K_2^{(7)}=\rmv_2+\rmv_1\,,\qquad \ \ K_3^{(7)}=\rmv_3+\rmv_2\,.
\la{222}
\end{equation}
Substituting \rf{217}, \rf{220} and \rf{222} into \rf{216} gives 
\begin{equation}
 K_C=\big(S\,\rmv _3+V\,\rmv _2+V\,\rmv _1+S\,\rms \big)+H_2\,(\rmv _1-\rms-2)\,.
\la{223}
\end{equation}
The first bracket is the standard KK  content of the 3-form heat kernel: internal co-exact 3-forms give
$\AdS_4$ scalars, internal 2-forms give $\AdS_4$ vectors, internal 1-forms give (dualized) $\AdS_4$ vectors and internal
scalars give (dualized) $\AdS_4$ scalars.\foot{Here ``dualized'' refers to the standard
 treatment in which the $\AdS_4$ 2-form and 3-form
components of $C_3$ are traded, by the 4d Hodge duality $\ast_4\dd_4$, for vectors and
scalars. This map is one-to-one on the continuous spectrum of $\Hf$ but  misses the harmonic
2-forms.}

The dependence on the harmonic forms is only  in the last term in \rf{223}. The $-2$ there  is
a  constant internal mode of the 2-form ghost corresponding to the zero mode of \ci{Bhattacharyya:2012ye}.  Projecting  out these zero modes  gives 
\be\la{2244}
 F'_C=-\tfrac12\int{\dd t\ov t}\  K_C'\ , \qquad \ \ 
 \qquad K_C'= K_C+2H_2 \ , \ee
 where  $ K_C'$ thus contains  the non-zero-mode harmonic contribution  $H_2(\rmv_1-\rms)$.
The latter contributes  to  the free energy  as  $n_2 \F_H$ with $\F_H$  given  in \rf{60}, i.e.  
\begin{equation}
   F'_H=-\tfrac12\int{\dd t\,\ov t}\ H_2(\rmv_1-\rms) =  n_2 \F_H \ , \qquad 
        \qquad \F_H= \half\log{\det}_{M_7}\, \Delta_{1,\perp}-\half\log {\det}'_{M_7}\, \Delta_0\,,
\la{224}
\end{equation}
 Equivalently, since $\Delta_1=\Delta_{1,\perp}\oplus\Delta_0'$, we have  
$\F_H= \tfrac12\log\det'\Delta_1-\log\det'\Delta_0$  as  in \rf{59}.


\section{Harmonic forms and discrete series representations}\la{252}

This Appendix  discusses the square integrable   harmonic 2-forms on $\Hf=SO_0(4,1)/SO(4)$  and their  interpretation. 

The spectrum of  the Hodge Laplacian $\Delta_p$ on the space of square integrable    $p$-forms 
 $\Omega^p(\Hf)$ is purely
continuous except   for $p=2$  where  there  is an infinite dimensional space   $\mathcal H^2(\Hf)$ of 
zero modes, i.e. of harmonic  2-forms \ci{Camp,CarronPedon:2004}.
In 4d   we have $\ast^2=1$ on 2-forms, and thus may split 
\begin{equation}
\mathcal H^2=\mathcal H^2_+\oplus\mathcal H^2_-\,,\qquad\qquad  \ast\omega_\pm=\pm\omega_\pm\,.
\la{254}
\end{equation}
Each $\mathcal H^2_\pm $
is an irreducible discrete-series representation $\pi_\pm$ of $SO_0(4,1)$
\ci{Pedon:1997,Prlic:2020}, with vanishing quadratic Casimir. 
Its restriction to $SO(4)$ consists of the transverse vector harmonics of $S^3$ of one chirality, with multiplicity one at each angular momentum.  Special   forms 
$\dd x^1\wedge\dd x^2\pm\dd x^3\wedge\dd x^4$ mentioned in footnote \ref{100}  
represent its lowest ``$K$-type'', i.e. the lowest $SO(4)$ representation  ($(1,0)$ or $(0,1)$ of $Spin(4)=SU(2) \times SU(2)$) of (anti-)self-dual 2-forms at the origin, in the
decomposition of $\pi_\pm$ under the maximal compact subgroup $K=SO(4)$.
 
The
multiplicity $n_2=1$ in  \rf{53} is the renormalized trace of the projector onto
$\mathcal H^2$,  $1\ov 2$ per chirality. 


 In even dimension the middle-degree
Poisson transform identifies the  harmonic forms with special exact boundary forms \ci{Lott:2000}. 
\iffa $-\tfrac12$,
\begin{equation}
\mathcal H^2(\Hf)\simeq H^{-1/2}_{\rm exact}\Omega^2(S^3)\,.
\la{255}
\end{equation} \fi 
The two chiralities in \rf{254} 
correspond to the two signs of the curl on co-closed boundary 1-forms. 
In the upper-half-space parametrization,
$\dd s^2=z^{-2}(\dd z^2+\dd x_i\dd x_i)$, an explicit family is
$F=\dd A$, where
\begin{equation}
 A=\int\frac{\dd^3p}{(2\pi)^3}
 e^{i\mathbf p\cdot\mathbf x-pz}a_i(\mathbf p)\,\dd x^i,
 \qquad p=|\mathbf p|,\qquad p_i a_i(\mathbf p)=0.\la{2010}
\end{equation}
The 2-form $F$ is closed and co-closed,  with the standard form norm,
$
 \|F\|^2=\int F\wedge\ast F
 =\int\frac{\dd^3p}{(2\pi)^3}\,p\,|a(\mathbf p)|^2.
$
We assume that the last integral is finite, as is the case for a
smooth transverse wave packet with compact Fourier support away
from $p=0$. Then $F$ is square integrable, while its potential $A$
is not square integrable: the norm of $A$ contains
$\int_0^\infty\dd z\,z^{-2}e^{-2pz}$, which diverges at the
conformal boundary (cf. footnote \ref{100}).

The algebras $so(4,1)$ and $so(3,2)$ are two real forms of
$so(5,\CC)$. The positive-energy representations of $so(3,2)$
relevant to Lorentzian $\AdS_4$ and the discrete series of
$SO_0(4,1)$ have different unitary Hilbert-space realizations.
Analytic continuation relates differential equations and
representation-theoretic expressions, but does not by itself
provide a unitary identification of these Hilbert spaces.
Their admissible boundary and normalizability conditions must
be distinguished.

When the same harmonic 2-form is regarded as a 2-form gauge
potential $B_2=\dd  V_1$, its 1-form $V_1$ is a
non-normalizable gauge parameter. This is the mechanism discussed
in \ci{Larsen:2015aia} in connection with quantum inequivalence
between dual field descriptions  in AdS$_4$ (cf. \ci{Duff:1980qv}), with the
regularized mode count as in \rf{53} related to the Euler characteristic.

In the Maxwell interpretation above, $F$ is instead a field
strength. Although it is a zero mode of $\Delta_2$, its potential
has nonzero boundary data, so it does not give an admissible zero
mode of the Maxwell fluctuation problem with homogeneous Dirichlet
data for $A$. It may describe a solution with prescribed sources,
or enter a formulation in which the boundary connection is
integrated over \ci{Marolf:2006nd,Witten:2003ya}. The latter
requires a specified boundary action and does not automatically
make these configurations zero modes of that action.

 \iffa This does not contradict their inclusion in the  1-loop  computation in this paper. 
The relevant modes of the 3-form are $\omega\wedge v_1$ with a non-constant internal 1-form $v_1$ as in  eq.
\rf{57}. After the KK reduction
they are massive modes of the Euclidean operator (see eq. \rf{257} below), and the question is only whether
the  boundary conditions  for  the 3-form and for  its ghost  field  allow them (cf. the end of this Appendix). 
\fi 

Let us now comment on the implications  of the  above comments for the computation in this paper. 
 For an internal co-exact 1-form $v_1$
satisfying $\Delta_1^{M_7}v_1=\sss\,v_1$ with $\sss>0$,
\begin{equation}
 \big(\Delta_2^{\Hf}+\Delta_1^{M_7}\big)
 (\omega\wedge v_1)=\sss\,\omega\wedge v_1\,.
 \la{257}
\end{equation}
This is a nonzero eigenmode of the Euclidean quadratic operator,
rather than an on-shell zero mode, and it contributes to the
determinant. It is not a pure-gauge 3-form: since
$\|\dd_{M_7}v_1\|^2=\sss\|v_1\|^2>0$, one has
$\dd_{11}(\omega\wedge v_1)=\omega\wedge\dd_{M_7}v_1\ne0$.

The projector onto $\mathcal H^2(\Hf)$ is part of the complete
 spectral resolution of the Hodge Laplacian, in addition to
the continuous Plancherel integral. 
The harmonic 2-form ghost modes were included in the zero-mode calculation of \ci{Bhattacharyya:2012ye}. 
In    ref. \ci{Bobev:2026gir},   in the context of the fixed-$C_3$ (or $\upmu$-) ensemble  discussion, 
different conditions on this ghost field  were assumed, 
including a vanishing tangential pull-back, and as a result there were no such ghost 
zero modes.
These conditions concern only the sector constant on $M_7$.


The present computation includes the mixed 3-form modes built from harmonic 2-forms (in addition to  the harmonic
ghost sector) and this is essential for the agreement with the $A(k)$ term in the ABJM localization result.

\section{1-loop  UV  divergence  in 11d supergravity} 
\la{241}

Considering  the  quadratic fluctuation action of 11d   supergravity expanded near an 
 on-shell Ricci-flat background, it was  shown in \ci{Fradkin:1982kf}  that  the first 4 Seeley coefficients in the small proper time 
  expansion ($t =\varepsilon=\Lambda^{-2} \to 0$)  of the  combined ``bulk'' heat kernel in \rf{35}  vanish
\be {\rm b}_0={\rm b}_2={\rm b}_4={\rm b}_6=0\ . \la{2004}\ee
This implies the absence of 
 $\Lambda^{11},\Lambda^9,\Lambda^7,\Lambda^5$    UV divergences  in the  1-loop  effective action $\Gamma_1$.
 Also, 
  on  a 11d manifold without boundary where  ${\rm b}_{2p+1}=0$  there  are  no $\Lambda^{2n}$ and $\log \Lambda$ 
  divergences   (cf. \rf{35},\rf{305}). 
  The   divergent part of the effective action   may  thus  contain  only  two terms 
\begin{equation}
\Gamma_{1,\infty}=c_8\,\Lambda^3\,{\rm I}_8+c_{10}\,\Lambda\,{\rm I}_{10}\,,\qquad\qquad \qquad 
{\rm I}_n=\int\dd^{11}x\sqrt g\,I_n\,,
\la{242}
\end{equation}
where $I_n$ are local invariants of dimension $n$   built out of curvature, other fields   and covariant derivatives. 

$I_8$   should be an  on-shell supersymmetric invariant
whose purely gravitational part starts with $R^4$  terms, together with
the associated 4-form field-strength and fermionic terms
(see, e.g.,
\ci{Deser:1998jz,Peeters:2000qj,Howe:2003cy,Cederwall:2004cg}).
A putative $I_{10}$ would similarly have a curvature  part 
of the  form $D^2R^4+R^5$. Such an independent dimension 10 
super-invariant is expected to be absent in the maximally
supersymmetric 11d  supergravity.\foot{This   can be seen, in particular,  from 
 the point of view  of the  dimensionally reduced  theory --  $D=4$, $\cN=8$
supergravity. The  linearized  four-graviton  invariants start with
$R^4$, $D^4R^4$, $D^6R^4$ and $D^8R^4$.
The first three have linearized  superspace representations
involving integrations over $16$, $24$ and $28$ odd coordinates
\ci{Kallosh:1980fi,Howe:1980th,
Bossard:2009sy}.
$D^8R^4$ invariant  also exists  despite the vanishing 
of the  full superspace volume \ci{Bossard:2011tq}.
The absence of a dimension 10  invariant should
not, however, be inferred solely from counting superspace
integration coordinates.
A direct on-shell argument follows from locality and the
supersymmetry Ward identities \ci{Elvang:2010jv}:   from  4-point scattering amplitude 
the candidate $D^2R^4$ matrix element is proportional to
$(s+t+u)\,\mathcal A_{R^4}=0$, while at 5-point  level  no independent
local supersymmetric $R^5$-type  matrix element exists.
Dimensional reduction provides a necessary consistency test (cf. \ci{Fradkin:1982kf})
for an 11d  candidate, but its application
requires retaining the complete reduced supermultiplet;
a restriction to 4d external gravitons alone
need not detect every higher-dimensional  invariant.
The absence of the  dimension 10  invariant  is also supported 
by the structure of the  type II string effective action
\ci{Green:2005ba,Green:2006gt,Richards:2008jg}, and also 
 by the explicit 1-loop  amplitude calculations in 10d 
type IIA and IIB supergravity  in  \ci{Edison:2022jln}, which find no 
 logarithmic UV divergences at 5 points 
in the sectors examined.}

The cubic divergence term  is  related, upon
reduction on the 11d circle, to the quadratic divergence  in  type II 10d  supergravity   which itself is related to the $\alpha'\to 0$ limit  of the  finite  1-loop (genus-one)   term  $\alpha'^{-1}{\rm I}_8^{(10)}$ 
 in the type II string theory  effective action \ci{Metsaev:1987ju}.
 The    cutoffs are assumed to be 
$\Lambda \sim \lpp ^{-1} $ and 
$R_{11}\Lambda^3\sim \Lambda_{10}^2\sim \alpha'^{-1}$ ($R_{11}$ is the radius of the 11d circle  which in the 
present \adssz case is  related to ${1\ov k}$).

 Also, in the 11d  1-loop
amplitude on a circle \ci{Green:1997as,Russo:1997mk,Green:1999pu}
 the zero-winding term contains  the
cubic divergence while the non-zero windings are finite and reproduce the tree-level $\zeta(3)R^4$ term
of type IIA string theory.\foot{To recall, in 
  the Schwinger  proper-time representation of the 1-loop amplitude on $\RR^{10}\times S^1$, the
sum over the discrete KK momenta along the 11d circle can be Poisson-resummed into a sum over the winding
number $w\in\ZZ$ of the worldline of the particle in the loop around the circle. The $w=0$ term is the
amplitude of the decompactified 11d theory and contains  its cubic  UV divergence. The terms with
$w\neq0$ are cut off at small proper time by $e^{-\pi^2R^2_{11}w^2/t}$ and are UV finite. Their sum reproduces
the $\zeta(3)R^4$ term in tree-level  part  of the type IIA string effective action, while the $w=0$ term corresponds to the
1-loop 10d term \ci{Green:1997as,Russo:1997mk,Green:1999pu}. 
}  Note that a similar  structure appears in  the present case of the 
$\ZZ_k$ quotient, where the $p=0$ term of the group average  in \rf{43}  is the covering-space contribution and the
$p\ne0$ terms are finite.

Specializing to the \adssz background  \rf{31},
the local 11d  curvature-type   invariants  scale as the 11d  volume, i.e. 
are proportional to $1\ov k$. 
One should thus find  that 
\be 
{\rm I}_8=\tfrac{1}{k} \rr_8 L^3\,,\qquad \qquad  {\rm I}_{10}= 0 \ , 
\la{243}
\end{equation}
where $L$ is the scale in \rf{31}   and   $\rr_8$ is a constant   whose  value depends on  precise  form 
 of the (currently unknown) 
 $F_4=dC_3$  completion of the $R^4$ invariant (cf.  discussion in Appendix C in \ci{Beccaria:2023hhi}). 
 
The  resulting  $\Lambda^3 L^3$ term in the 1-loop 11d supergravity  free energy  has the same form as the 
 ${\rm a}_1k^{-1} \rm L^3 $  term in the ABJM free energy 
\rf{01} provided the 11d  UV cutoff $\Lambda$ is  set to be proportional to  $\lpp ^{-1}$   \ci{Beccaria:2023hhi}.


\section{1-loop Casimir energy  on   $\AdS_4\times S^7/\ZZ_k$}
\la{247}

The 1-loop vacuum energy of supergravity on
$\AdS_4\times S^7$ that was computed in \ci{Gibbons:1984dg,Inami:1984vp} 
 is the Casimir energy associated with
global AdS time (with boundary being $\mathbb R\times S^2$). 
Refs.  \ci{Gibbons:1984dg,Inami:1984vp} evaluate a regulated sum of global AdS energies  
 and find its cancellation
within each  $S^7$ KK multiplet. 
In this
Appendix we review   the spectral representations  involved, extend the vanishing of the Casimir
energy to the case of the  \adsz  quotient, where it again holds level by level, and comment on  the meaning of this result 
 on the dual 
ABJM theory side.


 Let $\omega_i$ be the energies of the one-particle states in global  Lorentzian 
$\AdS_4$  (in units of the inverse $\AdS_4$ radius).
 Let $d_i$ be their multiplicities and $\fF_i$ their fermion numbers. The (regularized)  Casimir energy is given by 
\begin{equation}
E_c=\half\sum_i(-1)^{\fF_i}d_i\,\omega_i=\half\,\zeta_E(-1)\,,\qquad\qquad \ \ \ 
\zeta_E(z)\equiv \sum_i(-1)^{\fF_i}d_i\,\omega_i^{-z}\,,
\la{248}
\end{equation}
with $\zeta_E$  defined  by an  analytic continuation  from large ${\rm Re}\,z$.  
This is the quantity   computed in \ci{Gibbons:1984dg} and \ci{Inami:1984vp} 
and found to vanish 
for each complete $OSp(8|4)$ multiplet.\foot{These papers also examined 
ultraviolet cancellations   directly in  $\AdS_4$. In \ci{Inami:1984vp} the poles of the regulated
energy sum were shown to cancel, at each KK level and, using spin-energy  sum rules, for the complete tower.
In \ci{Gibbons:1984dg} the two logarithmically divergent local terms of the four-dimensional 1-loop
effective action were considered, the Euler density $\int {\cal E}_4$ and the $SO(8)$ gauge-field term
$\int{\rm tr}F^2$, whose coefficient is the 1-loop beta function of the gauged ${\cal N} =8$  supergravity. 
The latter was shown to cancel separately within each $OSp(8|4)$ multiplet, which is a supertrace-type
sum rule. The degeneracies of the KK levels are polynomials of degree $6$ in $n$, so a non-vanishing
$F^2$ coefficient would produce, in a sum over the levels with cutoff $n_{\rm max}\sim\Lambda L$, power
divergences $(\Lambda L)^7,(\Lambda L)^5,\dots$ multiplying $\int{\rm tr}F^2$. These are the reductions on
$S^7$ of the 11d power divergences $\Lambda^{7}\,{\rm b}_4,\Lambda^{5}\,{\rm b}_6,\dots$, since the $SO(8)$
gauge field comes from the 11d metric. The level-by-level cancellation is thus the 4d counterpart of the  vanishing of the 11d Seeley coefficients in \rf{2004}. 
The 4d Euler  density  coefficient  (a polynomial in the level $n$)
 cancels only for the complete tower, with the sum over $n$ defined by
zeta-function regularization, which discards the power-law growth of the partial sums.}

 Let us define the one-particle supercharacter, refined by the $U(1)_m$ fugacity $v$, as 
\begin{equation}
\chi^{(-)}(\tau,v)=\sum_i  (-1)^{\fF_i}d_i\,e^{-\tau\omega_i}\,v^{\,q_{m,i}}={\rm Str}\big(q^{D}\,v^{R_m}\big)\,,\qquad \ \ \ \ 
 q=e^{-\tau}\,.
\la{290}
\end{equation}
It is the character in \rf{64} 
at $x=-1$, since $(-1)^{\fF}=(-1)^{2J_3}$. $\tau$ is the same variable as in \rf{63},\rf{7766} 
conjugate to the global energy $E$.
 In terms of $\chi^{(-)}$ the Casimir energy is  given by  
\begin{equation}
E_c(v)=-\half\,\partial_\tau\chi^{(-)}(\tau,v)\Big|_{\tau \to 0}\,,\qquad \qquad E_c=E_c(1) \ ,
\la{291}
\end{equation}
where taking  the finite part in the limit $\tau\to0$ is in general assumed. In fact, here the 
 supercharacter of each supermultiplet is regular at $\tau=0$ (see \rf{294})  so no subtraction is needed.

 To compare, the   free
energy in $\AdS_4$ with $S^3$ boundary 
 is instead given by  $F'_1= -\frac12\int\dd\tau\,\tau^{-1}\cI(\tau)$   where the integrand is 
built from the same characters  (see \rf{84},\rf{101}).

 The quantities  $E_c$ and   $F'_1$   are different functionals of the same multiplet representation
data. The lowest global energy $E$, spin $j$ and internal $SO(8)$  quantum
numbers specify the global AdS$_4$ representation and the corresponding
Euclidean field with its chosen boundary quantization.
The spectral zeta function of the   operator 
 $\OO_{E,j}$
has the  form  (cf. \rf{80})
$
 \zeta_{{E,j}}(z)\sim 
 \int_0^\infty \dd\lambda\,\mu_j(\lambda)
   [\lambda^2+(E-\tfrac32)^2]^{-z}.
$

The character integral representation rewrites this Euclidean
spectral expression in terms of global AdS characters
\ci{Basile:2018zoy,Sun:2020ame}.\foot{
The Euclidean continuation of global AdS$_4$  with noncompact time
$u\in\RR$ has metric
$
4 L^{-2}\dd s^2
 =\dd\rho^2+\cosh^2\rho\,\dd u^2      +\sinh^2\rho\,\dd\Omega_2^2.
$
This represents  the same complete $\Hf$ as  the metric 
$4L^{-2}\dd s^2=\dd R^2+\sinh^2R\,\dd\Omega_3^2$:
the radial coordinates are related by
$
 \cosh R=\cosh\rho\cosh u.
$
At the conformal boundary,
$
 \dd\Omega_3^2
 =\operatorname{sech}^2u\,
   (\dd u^2+\dd\Omega_2^2),
 \ \cos\theta=\tanh u,
$
where $\theta$ is the polar angle on $S^3$.
In  the  coordinates of the first metric  with  the cylindrical $\mathbb R \times S^2$ boundary 
the  boundary  3-sphere  effectively  appears   with  its two poles removed.
The  determinant is unchanged when its operator,
boundary conditions and regularization prescription
are expressed in these global coordinates.
In particular, a spherical radial cutoff becomes a coupled
cutoff in $(\rho,u)$ (cf.   \ci{Giombi:2014yra}). }
The resulting partition function is the same 
$\Hf$ partition function,  
 the global-energy
character provides the representation-theoretic input to the
Euclidean determinant on $\Hf$ with $S^3$ conformal boundary.
In general, there is no direct relation  between the  $E_c$  and $F_1= \half \log \det \OO$
(cf. \ci{Gibbons:1984dg,Gibbons:2006ij,Giombi:2014yra}).
\foot{For comparison, the thermal partition function of a free  field 
on $S^1_\beta\times S^2$, with   charge fugacity
$v=e^{i\theta}$, is
$
 \log\widehat Z_{}(\beta,v)
 =-\beta E_c(1)+\log Z_{}(\beta,v),
$
where
$
 \log Z_{}(\beta,v)
 =\sum_{\ell\ge1}\, \ell^{-1}
   \big[\chi_B(\ell\beta,v^\ell)
       +(-1)^{\ell+1}\chi_F(\ell\beta,v^\ell)\big].
$
Here $\beta$ is measured in units of the AdS radius, and
$\chi_B$ and $\chi_F$ are the ungraded one-particle traces
over bosonic and fermionic excitations.
Periodic identification of  global AdS Euclidean  time produces
a thermal quotient with $S^1_\beta\times S^2$ boundary,
which is different from the unquotiented $\Hf$ used for
the sphere partition function.}


Let us now  consider the generalization of the $\AdS_4 \times S^7$  result  \ci{Gibbons:1984dg,Inami:1984vp}
 of the  level-by-level vanishing of $E_c$     to the  \adssz  case. 
The $\ZZ_k$-invariant spectrum on $S^7/\ZZ_k$ consists of the states with $q_m\in k\ZZ$, and its
supercharacter is the average over the group, as in \rf{157}
\begin{equation}
\chi^{(-)}_k(\tau)=\frac1k\sum_{p=0}^{k-1}\chi^{(-)}(\tau,\omega^p)\,,\qquad\qquad \omega=e^{2\pi i/k}\,.
\la{293}
\end{equation}
Since \rf{291} is linear in the character, the Casimir energy of the projected spectrum is the
average of the refined Casimir energy over the roots of unity
\be \la{5551}
E^{(k)}_c=\frac1k\sum_pE_c(\omega^p)\ . \ee
Note that for $k\ge3$ the projection breaks
$OSp(8|4)$ to $OSp(6|4)\times U(1)_m$, so the question is whether the $\cN=8$ cancellation of
\ci{Gibbons:1984dg,Inami:1984vp} survives in the $\cN=6$ supersymmetric case.

 Using the decomposition of the $\frac12$-BPS multiplets $\cM_n$ into conformal
multiplets (see Table \ref{211}), with the null-state subtractions at $n=2$, and the $so(8)$ characters
restricted to $U(1)_m$ we computed
$\chi^{(-)}_{\cM_n}(\tau,v)$ for $n=2,\dots,8$ as an exact function of $\tau$ and $v$. 
We found that  the supercharacter of each multiplet is regular at $\tau=0$ (without the poles
$\tau^{-3},\tau^{-2},\tau^{-1}$ that the individual conformal multiplets have)
\begin{equation}
\chi^{(-)}_{\cM_n}(\tau,v)=\n_0^{(n)}(v)+0\cdot\tau+\n_2^{(n)}(v)\,\tau^2+\OO(\tau^3)\,.
\la{294}
\end{equation}
The constant term $\n_0^{(n)}(v) $  is a non-trivial Laurent polynomial in
$v$, for instance,
\begin{equation}
\n_0^{(2)}=4c^2-1\,,\quad \n_0^{(3)}=4c\,(2c^2-1)\,,\quad \n_0^{(4)}=(4c^2-1)^2-4c^2\,,\quad c=\tfrac12(v+v^{-1})\,,\quad \n_0^{(n)}(1)=n+1 \ . 
\la{295}
\end{equation}
 As a result, the refined Casimir energy of each multiplet 
vanishes for all $v$, level by level,  and thus also $E_c^{(k)}$ in \rf{5551}
\begin{equation}
E_{c,\cM_n}(v)=0\,\qquad\to \qquad E^{(k)}_c=0\ ,  \ \qquad  k\ge1\,.
\la{296}
\end{equation}
The $k=1$ case is the result of \ci{Gibbons:1984dg,Inami:1984vp}. 

The new observation  is that the
cancellation does not require the complete $\cN=8$ multiplets: the $\ZZ_k$-invariant part of each
$\cM_n$, which for $k\ge3$ is a sum of $OSp(6|4)$ multiplets, has vanishing Casimir energy by itself.
Eq. \rf{294}  also shows that no regularization of the sum over descendants is needed once the states
are combined into multiplets, so the result is prescription-independent.

The vanishing for all $n$ can be proved  also from the explicit  knowledge of the spectrum. Introduce a level fugacity $s$,
independent of $q$, and consider the generating function
\begin{equation}
G(s,\tau,v)=\sum_{n\ge2}s^n\,\chi^{(-)}_{\cM_n}(\tau,v)\,.
\la{297}
\end{equation}
With the decomposition of $\cM_n$ into conformal multiplets of Table \ref{211}, the null-state
subtractions at $n=2$ and the absent lowest harmonics, $G$ is a rational function of $s$, $q^{1/2}$ and
$c=\frac12(v+v^{-1})$, built from the generating functions $\sum_ms^m\chi^{(b,m)}(v)$ of the $so(8)$
characters restricted to $U(1)_m$. Its expansion in $\tau\to 0$ is  
\begin{equation}
G(s,\tau,v)=\tfrac{s^2\,(4c^2- 2 s c -1)}{1-2cs+s^2}+0\cdot\tau+\OO(\tau^2)\,.
\la{298}
\end{equation}
 Hence the coefficient of $\tau$ in
$\chi^{(-)}_{\cM_n}(\tau,v)$ vanishes for every $n$ and every $v$, which proves \rf{296} for all $k$. The
constant term in \rf{298} generates the $\n_0^{(n)}(v)$ of \rf{295}, and at $c=1$ reduces to
$s^2(3-2s)/(1-s)^2=\sum_{n \geq2} (n+1)s^n$.

\iffa Let us note that the vanishing does not follow from the algebra alone. For a supercharge $Q$ with
$q_m=0$ one may consider ${\rm Str}(\{Q,S\}X)$ with $X=q^Dv^{R_m}$, but since $[D,Q]=\frac12Q$ the
regulator does not commute with $Q$, and graded cyclicity gives
${\rm Str}(\{Q,S\}X)=(1-q^{1/2})\,{\rm Str}(SQX)$ rather than zero.
 The Casimir energy is thus the
residue of  a pole of   ${\rm Str}(SQX)$, whose vanishing is a property of the spectrum, not of
the algebra. The proof \rf{298} uses the $S^7$ spectrum and does not by itself extend to other
quotients.\fi 

\iffa 
Hence for every unitary multiplet $\cM$ and every $v$  one has 
\begin{equation}
{\rm Str}_{\cM}\big(D\,v^{R_m}\big)=0
\la{298}
\end{equation}
The sum over states is made meaningful by the fugacity
$q$, and \rf{298} is the statement that the coefficient of $\tau$ in \rf{294} vanishes.
The argument uses only one supercharge commuting with the fugacity. 
\fi 

Finally,  let us note   that  the vanishing of the 1-loop $\AdS_4$ Casimir energy  has
the expected interpretation on the  dual ABJM side.
In a conformal renormalization scheme, the ordinary vacuum
on $\RR\times S^2$ corresponds under radial quantization
to the identity operator, whose scaling dimension is zero.
In three dimensions there is no Weyl anomaly producing
an additive vacuum-energy shift.\foot{In four dimensions the analogous quantity is non-zero and is 
related to the  conformal anomaly \ci{Assel:2015nca}.}
Thus, 
 the conformal vacuum has zero energy.
This is true  for any CFT$_3$, independently
of supersymmetry (see, e.g., \ci{Giombi:2014yra}).\foot{For example, a conformally coupled scalar on a unit $S^2$
gives
$
 E_c=\frac12\sum_{\ell\ge0}(2\ell+1)(\ell+\tfrac12)
     =\zeta(-2,\tfrac12)=0,
$
where the sum is defined by an analytic continuation and
$\zeta(z,a)$ is the Hurwitz zeta function.}

The cancellation in \rf{296} provides the corresponding
1-loop holographic check at order $N^0$ in the fixed-$k$   large $N$ expansion in ABJM theory. 
Once again, the  vanishing  of $E_c$  is thus unrelated to the  value of  the $S^3$ partition function  and 
$A(k)$ in particular.

\iffa \foot{
A hemisphere path integral  prepares a state on $S^2$,
and gluing the two hemispheres gives the sphere partition
function as its norm. This finite value contains information that is not fixed
by the  Hamiltonian eigenvalue of the vacuum.}
\fi 

\section{1-loop   10d   supergravity on $\AdS_4\times\CP^3$ 
  and    large-$k$ limit}\la{280}

The reduction of \adssz  on 11d circle gives the $\AdS_4\times\CP^3$  solution  of  type IIA  10d supergravity. 
One  may ask  how  the $k\to \infty$  limit of the 1-loop   free energy of 11d  supergravity 
on \adsz is related to the 1-loop correction  computed directly in type IIA supergravity on $\AdS_4\times\CP^3$.

In this Appendix we  show that 
 the   10d   IIA  supergravity  1-loop computation  represents  the $U(1)_m$-neutral part of
the 1-loop expression in 11d  case 
 and that the limit $k\to\infty$ does not
commute with the $\tau$-integral: the charged KK modes, although exponentially suppressed in the
integrand, produce the $k^2$, $\log k$ and $k^{-2g}$ terms of $A(k)$.

The Hopf circle  of $S^7/\ZZ_k$ 
has radius $L/k$. The weakly coupled IIA description
applies when $L/k\ll\lpp $, while the supergravity approximation also
requires $\lambda=N/k\gg1$.
The 10d supergravity fields are the modes with
vanishing $U(1)_m$ charge $q_m$, i.e. the KK towers on $\CP^3$ of the
massless IIA fields \ci{Nilsson:1984bj}. The modes with $q_m=kr$,
$r\ne0$, carry $r$ units of D0 charge. In the Hopf-charge expansion
of Section \ref{155} this separation is explicit. The $k$-dependent
part of the integrand \rf{163},
\begin{equation}
\cC_k(\tau)=\frac{1}{\sinh^2(\frac{1}{2}\tau)}+\frac{2}{\sinh^2(\frac{1}{2}\tau)\,(e^{k\tau/2}-1)}
=\frac{1}{\sinh^2(\frac{1}{2}\tau)}\big[1+2\sum_{r\ge1}e^{-kr\tau/2}\big]\,,
\la{281}
\end{equation}
consists of the neutral term  ($r=0$)  
and of the sum over the charged modes, each weighted by
$e^{-kr\tau/2}$. 
The term $\cR$ in \rf{163} is neutral. 

The harmonic sector also survives
the  dimensional  reduction. Writing an invariant internal 1-form as
$v_1=a_1+f\eta$  (where $a_1$ is a 1-form on $\CP^3$, $f$ is a scalar
and $\eta$ is the Hopf connection 1-form) the modes
$\omega\wedge a_1$ give the $\Hf$  harmonic components of the RR
3-form, while $f\omega\wedge\eta$ gives components of the NSNS
2-form.  
The modes $\omega\otimes1$ remain the 2-form ghost zero modes of the
RR 3-form, and their regularized $\Hf$ count $n_2=1$ is unchanged.

Keeping only the neutral modes, i.e. those with $U(1)_m$ charge $q_m=0$, amounts to dropping the charged
terms in \rf{281}. This replaces $\cC_k$ in \rf{163} by its $k\to\infty$ limit at fixed $\tau$,
\begin{equation}
\cC_k(\tau)=\frac{\coth(\frac14k\tau)}{\sinh^2(\frac12\tau)}\ \ \to\ \ \cC_\infty(\tau)=\frac{1}{\sinh^2(\frac12\tau)}\,, 
\qquad \qquad  \cI_\infty\equiv \cC_\infty+\cR \ . 
\la{9991}
\end{equation}
The elementary term $\cR$ in \rf{161} is neutral and is kept, and so are the harmonic sector and the zero
modes.  In the prescription \rf{171} and in the units of \rf{21}, the
contribution of the non-zero modes to the 1-loop free energy of IIA supergravity is
\begin{equation}
F'_{\rm IIA}=-\half\,\zeta'[\cI_\infty](0)-\half\,\zeta[\cI_\infty](0)\,\log\big(\tfrac12\mu L\big)\,,
\la{747}
\end{equation}
where $\zeta[f](z)$ is the Mellin transform defined in the footnote below.\foot{\la{dee}For a function $f(\tau)$ with a
Laurent expansion at $\tau\to0$ and exponential decay at infinity, $\zeta[f](z)\equiv\frac{1}{\Gamma(z)}
\int_0^\infty\dd\tau\,\tau^{z-1}f(\tau)$ denotes its Mellin transform, continued analytically in $z$ as in
\rf{171}, so that the regularized integral is $\int_0^\infty\frac{\dd\tau}{\tau}f(\tau)=\zeta'[f](0)$. For
$f=\sum_ie^{-\lambda_i\tau}$ it is the zeta function $\sum_i\lambda_i^{-z}$ of the numbers $\lambda_i$. When
$f$ is the integrand that represents an operator through the two linear factors of each eigenvalue, the zeta
function of the operator at zero is $\frac12\zeta[f](0)$. This is the coefficient of the scale term in
\rf{747}.}
Using $\cC_\infty=4\sum_{m\ge1}m\,e^{-m\tau}$ and
$\cR=e^{-2\tau}+e^{-3\tau}-4\sum_{m\ge1}e^{-m\tau}$ one finds
that 
$
\zeta[\cI_\infty](z)=4\zeta(z-1)+2^{-z}+3^{-z}-4\zeta(z)\,$  $
\zeta[\cI_\infty](0)=\tfrac{11}{3}\,,$ \  $
\zeta'[\cI_\infty](0)=4\zeta'(-1)-\log6+2\log(2\pi)\,, $ so that 
\al{F'_{\rm IIA}=-2\zeta'(-1)+\half\log6-\log(2\pi)-\tfrac{11}{6}\log\big(\tfrac12\mu L\big)\,.
\la{282}
}
The constant $-2\zeta'(-1)=-\half\zeta'[\cC_\infty](0)$ is the $k$-independent constant of $-A(k)$ in \rf{6}.

The scale-dependent  term in \rf{282} has two parts, 
\begin{equation}
-\tfrac{11}{6}\log\big(\tfrac12\mu L\big)=-2\log\big(\tfrac12\mu L\big)+\tfrac16\log\big(\tfrac12\mu L\big)\,.
\end{equation}
The first term is the same as in the 11d result \rf{21}. It is due to the removal of the ghost zero modes and
is cancelled by the zero-mode integral, so it is not associated with a UV divergence. The second term
comes from $\zeta[\cC_\infty](0)=-\frac13$. It cannot be the coefficient of a local  non-topological logarithmic  UV 
divergence of IIA supergravity.\foot{Such a divergence, if present, would have to be a local on-shell
supersymmetric invariant of dimension 10. A non-topological invariant of the $D^2R^4+R^5$ type does not
exist (Appendix \ref{241}), in agreement with the absence of logarithmic divergences in the 1-loop 10d
amplitudes. This does not exclude a topological term, i.e. an Euler-type density, whose integral is
proportional to the Euler number of the background and which is invisible to scattering amplitudes and not
forbidden by supersymmetry. In 11d no such term exists. 
We have not checked whether the coefficient
$\frac16$ is of this type.}
The number $4\zeta(-1)$ is the regularized value of the divergent sum $4\sum_{m\ge1}m$ over the levels of
the neutral tower, in the $\tau$-Mellin prescription.

The charged modes remove this term. The difference $\cC_k-\cC_\infty$ is exponentially small at fixed
$\tau>0$, but its Mellin transform is dominated by $\tau\lesssim1/k$, where $\cC_\infty\simeq4/\tau^2$.  One finds, for large $k$,
\begin{equation}
\zeta[\cC_k-\cC_\infty](z)\sim\frac{8(\frac12k)^{2-z}\zeta(z-2)}{(z-1)(z-2)}-\tfrac23\big(\tfrac12k\big)^{-z}\zeta(z)
+\tfrac1{30}\big(\tfrac12k\big)^{-z-2}z(z+1)\zeta(z+2)+\dots\,.
\la{285}
\end{equation}
At $z=0$ the first and third terms vanish, since $\zeta(-2)=0$, and the second gives $+\frac13$, which
cancels the neutral value $-\frac13$, so that $\zeta[\cC_k](0)=0$ and the scale term of the neutral
sector disappears. The first term gives,  using  $\zeta'(-2)=-\zeta(3)/(4\pi^2)$, the leading $k^2$ term in  $A(k)$ in \rf{6}
and the second gives the logarithm,
\begin{equation}
-\half\zeta'[\cC_k](0)+2\zeta'(-1)\ \to \  \tfrac{1}{8\pi^2}\zeta(3)\,k^2+\tfrac16\log\tfrac{k}{4\pi}-\sum_{g\ge1}\tfrac{1}{k^{2g}}\mathcal B_g\,,
\la{286}
\end{equation}
which is $-A(k)+2\zeta'(-1)$ (see  \rf{6}). The coefficient $\frac16$ of the logarithm is the same,
including the sign, as that of the neutral scale-dependent  term $\frac16\log(\frac12\mu L)$. After the sum over the
charged modes the arbitrary scale $\mu$ is thus replaced by $\sim k/L$  (which is natural, since the charged
modes have masses of order $k/L$: the logarithm of the neutral sector runs up to the KK scale).

The limit $k\to\infty$ taken in the integrand  of the $\tau$-integral retains only $-2\zeta'(-1)$ and the neutral scale term.
The limit taken after the evaluation of the regularized integral gives the whole expansion \rf{6}. The two
operations do not commute. A related mechanism appears in the circle compactification of the 11d 1-loop
four-graviton amplitude \ci{Green:1997as,Russo:1997mk}: the sum over the KK momenta on the circle produces
the tree-level $\zeta(3)R^4$ term of IIA string theory, in addition to the 1-loop term.

On the ABJM side the natural comparison is with the expansion at large $N$ and $k$ with fixed
$\lambda=N/k$. Let us  isolate the explicit log\,$k$ term  and  consider (see also \ci{Bobev:2022eus,Beccaria:2023hhi})
\begin{equation}\la{1000}
\rF=\tfrac16\log k+\sum_{g\ge0}k^{2-2g}\hat \rF_g(\hat \lambda)\,, \qquad\ \ \ \  \ \ \  \hat\lambda=\lambda-\tfrac1{24}\ . 
\end{equation}
The expression \rf{7} gives
\begin{align}
\hat \rF_0=\te \tfrac{\pi\sqrt2}{3}\,\hat\lambda^{3/2}+\frac{1}{8\pi^2}\zeta(3)\,,
\qquad \ \ \ \ \hat \rF_1=-\tfrac{\pi}{3\sqrt2}\,\hat\lambda^{1/2}+\tfrac14\log\hat\lambda-2\zeta'(-1)+\tfrac{11}{12}\log2-\tfrac16\log\pi\,.
\la{288}
\end{align}
The constant $\frac{\zeta(3)}{8\pi^2}$ in $\hat \rF_0$ is due to the term $-\frac{\zeta(3)}{8\pi^2}k^2$ of $A(k)$
and is
  interpreted as originating  from  the tree-level $\zeta(3) 
\alpha'^3R^4$ correction in the type II   superstring effective action 
\ci{Drukker:2010nc,Beccaria:2023hhi}. In $\hat \rF_1$ the term $-2\zeta'(-1)$ is the finite part of the
neutral sector contribution  in \rf{282}, the term $\frac14\log\hat\lambda$ is due to the zero modes, the term
$\frac16\log k$ is the neutral scale term converted by the charged modes, 
and the remaining constants are fixed by the normalization of the zero-mode measure (see the end of
Section \ref{155}).

The  $ \hat\lambda^{1/2}$ term  in $\hat \rF_1$   should be the $\AdS_4\times\CP^3$  value of
   the string 1-loop $\alpha'^{-1} \int\dd^{10}x\sqrt g\,(R^4+...)$ term in  type IIA  string theory
\ci{Green:1997as,Russo:1997mk},  which  scales as $L^2\alpha'^{-1} \sim\hat\lambda^{1/2}$.
 It is the IIA
counterpart of the term $ {\rm L}^3/k$ in \rf{01} that originates from  the local 1-loop $R^4$ term of 11d
theory. 

Thus the IIA supergravity 1-loop contribution \rf{282} corresponds to the genus-one term $\hat \rF_1$ in the ABJM free
energy, with corrections suppressed by inverse powers of $\lambda$ (inverse string tension) dropped. The 11d
1-loop computation gives, in addition, the term corresponding to the string tree-level $\zeta(3)R^4$
interaction and all the higher-genus terms, which come from the sum over the D0-brane charges in the loop.\foot{
The IIA supergravity 1-loop computation cannot fix the $k$-independent constant of $A(k)$ by itself, since the
neutral sector depends on how the infinite tower of KK modes on $\CP^3$ is regularized, as seen from the
scale term $\frac16\log(\frac12\mu L)$ in \rf{282}. The 11d computation fixes this by including the charged
modes, for which $\zeta[\cC_k](0)=0$. The $k^2$ term, which on the string side corresponds to a tree-level
$\alpha'^3$ correction, arises in 11d from the 1-loop contribution of the charged KK modes. This is the
supergravity counterpart of the statement that the tree-level $R^4$ term of IIA string theory is generated by
the 11d supergravity 1-loop amplitude. It is also the reason why a 1-loop 11d computation can
reproduce a function, $A(k)$, whose large-$k$ expansion contains the large-tension parts of the contributions
of all higher genera.}


\section{1-loop   free energy   for the  ADHM quotient}\la{258}

The $\mathcal{N}=4$ $U(N)$  3d ADHM theory can be constructed from 
$\mathcal{N}=8$ $U(N)$   3d  SYM by flavouring a single adjoint chiral
multiplet. For $p$ flavours, the geometric branch of the moduli
space of the abelian theory is
$\mathbb{C}^2\times(\mathbb{C}^2/\mathbb{Z}_{p})$, and the dual
holographic description is in terms of M-theory on
$\AdS_{4}\times S^7/\mathbb{Z}_p$.  The
finite group quotient acts on the  4  cone coordinates $w_a$   as
\begin{equation}
(w_1,w_2,w_3,w_4)
\to
(w_1,w_2,\omega w_3,\omega^{-1}w_4)\,,
\qquad
\omega=e^{2\pi i/p}\,.
\la{259}
\end{equation}
For $p>1$, this action is not free: it fixes pointwise the
3-sphere $w_3=w_4=0$. The quotient
$\AdS_4\times S^7/\ZZ_p$ therefore has a transverse
$A_{p-1}$ singularity along $\AdS_4\times S^3$.
In M-theory, the addition of flavours is realised by a stack
of $p$ coincident Kaluza--Klein (KK6) monopoles, with the
M-theory circle fibre shrinking to zero size at their common
core. In the M2-brane near-horizon limit, their common
worldvolume becomes $\AdS_4\times S^3$, coinciding with
the singular locus
\cite{Benini:2009qs,Jafferis:2009th}
(see also \cite{Kurlyand:2026gxi} for a recent discussion).

The perturbative partition function of ADHM on $S^3$ is again
an Airy function of the form \rf{2}, with
$C={2\ov \pi^2p}$, $B={1\ov 2p}- {1\ov 8}p$, and an analog of the  constant  $A$  that was proposed
and tested in \ci{Hatsuda:2014vsa}
($N_f\equiv p$) 
\begin{equation}
A_{\rm ADHM} (p)
=\half\big[A(p)+p^2A(1)\big]\,.
\la{260}
\end{equation}
Here  $A(p)$  is  the same as  the  ABJM function in \rf{4}  with  $k$ replaced by $p$.  
Below we derive \rf{260} as the 1-loop   contribution on the 11d  M-theory (supergravity) 
  side. 

In the notation of Appendix~\ref{206}, the $\mathbb{Z}_{p}$
quotient can be implemented, for instance, along the
$U(1)_{\widetilde m}$ subgroup of $SO(8)$ generated by
\rf{115}:
\begin{align}\label{115aa}
\widetilde{R}_{m}=R_1-R_2\,.
\end{align}
To compute the contribution of the $\mathbb{Z}_p$-invariant
modes in the KK and harmonic sectors, one can follow the
procedure of Sections~\ref{61} and \ref{155}. To implement
the finite group quotient along \rf{115aa}, it is convenient
to retain two fugacities $u$ and $v$ associated with the
$so(8)$ Cartans $2R_1$ and $2R_2$, respectively. Namely, we
consider the two-fugacity extension of the refined free energy
\rf{666},  with the singleton contribution \rf{139} included:
\al{
&F_1'(u,v)
=-\tfrac12\int_0^\infty\frac{\dd\tau}{\tau}\,
\cI(\tau,u,v)\,,
\label{667}
\\ 
\label{667a}
&F_1'
=-\tfrac12\int_0^\infty\frac{\dd\tau}{\tau}\,
\cI_p(\tau)\,,
\qquad\qquad 
\cI_p(\tau)
=\frac1p\sum_{l=0}^{p-1}
\mathcal I\big(\tau,e^{\pi il/p},e^{-\pi il/p}\big)\,.
}
Following the same steps as in Section~\ref{68} and using
the explicit characters of Appendix~\ref{235}, we find for
the KK tower contribution, with the singleton multiplet contribution  in 
\rf{099} added,
\begin{align}
&(\mathcal I_{\cM_1}+\mathcal I_{\mathrm{KK}})(\tau,u,v)
=
-\frac{2(q+1)}{1-q}
+\frac{2q(\mathrm Q_1+\mathrm Q_2)}
{(1-q)\mathrm Q_1\mathrm Q_2}
\nonumber\\
&\qquad\qquad \qquad\qquad \qquad\quad
+\frac{
(1-q)\big[
2(q^2+1)(\mathrm Q_1+\mathrm Q_2)
-(q+1)(2q^2+q+2)
\big]
}{\mathrm Q_1^2\mathrm Q_2^2}\,,
\label{I.5}\\
&\mathrm Q_1(q,u,v)
=
1+q-q^{1/2}\big(uv+u^{-1} v^{-1} \big)\,,
\quad
\mathrm Q_2(q,u,v)
=
1+q-q^{1/2}\big(uv^{-1} +vu^{-1} \big)\,,
\quad q\equiv e^{-\tau}\,.
\label{I.6}
\end{align}
Similarly, following Section~\ref{140}, we find for the
harmonic 3-form sector contribution 
\begin{align}\la{8899}
\cI_H(\tau,u,v)
=
2+q^2+q^3
-
\frac{
(1-q)\big[
2(q^2+1)(\mathrm Q_1+\mathrm Q_2)
-(q+1)(2q^2+q+2)
\big]
}{\mathrm Q_1^2\mathrm Q_2^2}\,.
\end{align}
The cancellation observed  in \rf{154} persists in the two-fugacity
extension, and  we get explicitly
\begin{align}\la{771}
\cI(\tau,u,v)
=\cI_{\cM_1}+\cI_{\mathrm{KK}}+\mathcal I_H
=
2+q^2+q^3
+\frac{2q}{1-q}
\Big(
\frac1{\mathrm Q_1}+\frac1{\mathrm Q_2}-\frac1q-1
\Big)\,.
\end{align}
The $\mathbb{Z}_p$-invariant integrand $\cI_p$ in
\rf{667a} then becomes (cf. \rf{161},\rf{163})\foot{Here  $\cC_1$ denotes the same function $\cC_p$ 
with $p=1$, which  corresponds to 
 the unquotiented $S^7$, i.e. to the ABJM theory at $k=1$. In \rf{161a} it comes from the directions
$w_1,w_2$ on which the $\ZZ_p$ does not act.}
\begin{gather}
\cI_p(\tau)
=\half\big[\cC_p(\tau)+\cC_1(\tau)\big]+\cR(\tau)\,,
\la{161a}\\
\cC_p(\tau)
=\frac{\coth(\frac14p\tau)}{\sinh^2(\frac12\tau)}\,,
\qquad\qquad
\cR(\tau)
=e^{-2\tau}+e^{-3\tau}-\frac4{e^\tau-1}\,.
\la{163a}
\end{gather}
For $p>1$, the invariant part of the singleton is a free
$\mathcal N=4$ hypermultiplet; for $p=1$, the full
$\mathcal N=8$ singleton survives.

Note that the $\cR(\tau)$ part of \rf{161a} 
is the same as in the  $S^7/\ZZ_k$ 
case \rf{163}, while, with the $\zeta$-function
regularization \rf{171}, the $\cC_p+ \cC_1 $ part contributes
$-\half[A(p)+A(1)]$ to the 1-loop free energy. 
Thus  the projected contribution misses the
$\half(p^2-1)A(1)$ piece of the ADHM  function in 
\rf{260}.

This is not surprising. The missing piece results from the part of the 
KK spectrum localised on the M2-KK6 worldvolume
$\AdS_4\times S^3$. On the dual 3d gauge theory side, these modes
correspond to gauge-invariant operators in the
twisted sector and contribute to both the superconformal
index and the character \cite{Choi:2008za,Imamura:2009hc}.
On the  M-theory  side, these modes are described by a  7d 
 $\mathcal N=2$ $SU(p)$ vector multiplet
and, at the linearised level, can be treated as $p^2-1$
$U(1)$ vector multiplets \cite{Imamura:2010ei}.\foot{Reduction along the transverse Taub--NUT circle maps an 11d KK
monopole to a type IIA D6-brane
\ci{Townsend:1995kk,Hull:1997kt,Bergshoeff:1997gy}. For separated centres, the
normalisable harmonic two-forms of multi-centre Taub--NUT give
$U(1)^p$ worldvolume gauge fields through the reduction of $C_3$.
When the centres coincide, M2-branes wrapped on the vanishing
2-cycles become massless and  give  the off-diagonal states of
the $U(p)$ gauge theory \ci{Sen:1997js}.} The fields
are organised in terms of the irreducible representations
of the 7d  R-symmetry $SU(2)_r$, as well as
the $S^3$ isometry $SU(2)'_r\times SU(2)'_f$.

Upon KK compactification on $S^3$, the fluctuations decompose
into representations of $osp(4|4)\oplus su(2)'_f$, where the
bosonic subalgebra of $osp(4|4)$ is
$so(3,2)\oplus so(4)$ and
$so(4)=su(2)_r\oplus su(2)'_r$.
The contribution of these modes to the 1-loop partition
function can be obtained following the construction of
Section~\ref{68}. In the case of the ADHM quotient,
$\mathbb Z_p$ acts in the transverse directions to KK6,
and the localised tower does not require finite
group projection.  
It is therefore sufficient to keep track
of the conformal dimensions, spins and internal
multiplicities of the conformal multiplets.

The 4d  fields form a tower of $\frac12$-BPS
$osp(4|4)$ vector multiplets with scalar
superconformal primaries of dimension $E_0=n/2$,
where $n\geq2$ \cite{Imamura:2010ei}. The lowest level
is the massless $\mathcal N=4$ vector multiplet.
At each level, there is also a commuting $su(2)'_f$
representation of spin $s_f=(n-2)/2$, whose dimension
$n-1$ is included in the multiplicities listed below.
After tracing over the internal representations,
the conformal multiplets have the following dimensions
$E$, spins $j$ and multiplicities $d$
\cite{Imamura:2010ei}:
\begin{table}[H]
\centering
\begin{tabular}{cccccc}
\toprule
spin $j$ & energy $E$ & $s_r$ & $s_r'$ & $s_f$
& multiplicity $d$\\
\midrule
$1$ & $(n+2)/2$ & $0$ & $(n-2)/2$ & $(n-2)/2$
& $(n-1)^2$\\
$1/2$ & $(n+1)/2$ & $1/2$ & $(n-1)/2$ & $(n-2)/2$
& $2n(n-1)$\\
$1/2$ & $(n+3)/2$ & $1/2$ & $(n-3)/2$ & $(n-2)/2$
& $2(n-1)(n-2)$\\
$0$ & $n/2$ & $0$ & $n/2$ & $(n-2)/2$
& $(n-1)(n+1)$\\
$0$ & $(n+2)/2$ & $1$ & $(n-2)/2$ & $(n-2)/2$
& $3(n-1)^2$\\
$0$ & $(n+4)/2$ & $0$ & $(n-4)/2$ & $(n-2)/2$
& $(n-1)(n-3)$\\
\bottomrule
\end{tabular}
\caption{\small The conformal multiplets in the localized KK spectrum.
The spins $(s_r,s_r',s_f)$ label representations of
$SU(2)_r\times SU(2)'_r\times SU(2)'_f$ in the notation of
\cite{Imamura:2010ei}. The internal multiplicity is
$d=(2s_r+1)(2s_r'+1)(2s_f+1)$.
Representations with a negative spin are absent for $n=2,3$.
At $n=2$, the spin-one conformal multiplet is a
conserved-current multiplet, whose scalar null descendant
has $(E,j)=(3,0)$. This descendant must be subtracted.}
\la{211a}
\end{table}

The contribution $\cJ_{E,j}$ of each conformal multiplet to
the free-energy integrand is obtained from \rf{101}
(with $so(3,2)$ characters $\cC_{E,j}$ as in \rf{332}),
weighted by its multiplicity $d$.
The contribution of the modes localised on KK6 to the
1-loop free energy takes the form
\begin{align}\la{660}
F'_{\mathrm{KK}6}
=
-\tfrac12 ({p^2-1})
\int_0^\infty\frac{\dd\tau}{\tau}\,
\mathcal I_{\mathrm V}(\tau)\,,\qquad \ \ 
\qquad
\mathcal I_{\mathrm V}(\tau)
=
\sum_{n=2}^{\infty}\mathcal I_{\mathrm V,n}(\tau)\,,
\end{align}
defined using  the $\zeta$-function regularization  as in \rf{171}. 
The contribution at each KK level $n$ can be written as%
\footnote{For $n=2$, the last term equals $-\cJ_{3,0}$
and implements the null-state subtraction.}
\begin{align}
\mathcal I_{\mathrm V,n}(\tau)
&=
(n-1)\Big[
(n+1)\cJ_{\frac n2,0}(\tau)
+2n\cJ_{\frac{n+1}{2},\frac12}(\tau)
+3(n-1)\cJ_{\frac{n+2}{2},0}(\tau)
\nonumber\\
&\qquad\qquad\ \  
+(n-1)\cJ_{\frac{n+2}{2},1}(\tau)
+2(n-2)\cJ_{\frac{n+3}{2},\frac12}(\tau)
+(n-3)\cJ_{\frac{n+4}{2},0}(\tau)
\Big]\,.
\end{align}
Evaluating \rf{101} for each conformal multiplet and
combining the contributions, the integrand at each level
simplifies to
\begin{align}
\mathcal I_{\mathrm V,n}(\tau)
=
\frac{2(n-1) \, e^{-\frac12 n \tau}}{1-e^{-\tau}}\,.
\end{align}
Summing over the KK levels gives
\begin{align}\la{885}
\mathcal I_{\mathrm V}(\tau)
&=
\sum_{n=2}^{\infty}    \mathcal I_{\mathrm V,n}(\tau)
=\tfrac12\cC_1(\tau)
=
\frac{\coth(\frac14\tau)}{2\sinh^2(\frac12\tau)}\,.
\end{align}
Adding the $p^2-1$ localized towers, one for each generator of $SU(p)$, the complete integrand is, for
$p\ge2$,
\begin{equation}
\cI^{\rm ADHM}_p
=
\half\big[\cC_p(\tau)+p^2\cC_1(\tau)\big]+\cR(\tau)\,.
\la{271}
\end{equation}
With the $\zeta$-function regularization of Section~\ref{155} and the zero-mode contribution \rf{200},
in which $V_7=\vol(S^7/\ZZ_p)=\frac{\pi^4}{3p}L^7$, we obtain for the total 1-loop free energy
\begin{equation}
F^{\rm ADHM}_{1}(p)
=
\tfrac32\log({\lpp }^{-1} L)
-\half\big[A(p)+p^2A(1)\big]
-\half\log p-\half\log\pi\,.
\la{272}
\end{equation}
The zero-mode factor depends on the quotient only through the volume of the internal space, so the
constant is the same as in \rf{201}.

\iffa
With the $\zeta$-function regularization of 
Section~\ref{155} and including the same zero-mode   contribution 
as in \rf{200}, we obtain for the total  1-loop free energy
\begin{equation}
F^{\rm ADHM}_{1}(p)
=
\tfrac32\log({\lpp }^{-1} L)
-\half\big[A(p)+p^2A(1)\big]
-\half\log p+c_0\,,
\la{272}
\end{equation}
with the same constant $c_0$ as in \rf{201}.
\fi 

On the ADHM  theory side, the Airy function with $C={2\ov \pi^2p}$
and the leading-order relation as in \rf{9}, i.e. 
$L^6=32\pi^2pN\lpp ^6$,  give  the following $N^0$  terms 
in the large-$N$ expansion of the gauge theory  free energy on $S^3$: 
\be \la{7755}
{\cal F}_1= \tfrac32\log(\lpp^{-1} L )
-\tfrac12\log(\pi p) -  A_{\rm ADHM}(p)  \  . \ee
With \rf{260} and the same zero-mode normalization as in
\rf{304}, this agrees with \rf{272} for every $p\geq2$,
including the constant term.
For $p=1$, the computation reduces to
the ABJM case at $k=1$, with the
$\mathcal N=8$ singleton included.


\small 
\bibliographystyle{JHEP-v2_9}
\bibliography{Ak}

\end{document}